\documentclass[12pt]{article}
\usepackage{xcolor}
\usepackage{graphicx}
\usepackage{float}
\usepackage{amsthm,comment}
\usepackage{dsfont}
\usepackage{url}
\usepackage[ruled,vlined]{algorithm2e}
\usepackage{enumitem}
\RequirePackage{amsthm,amsmath,amsfonts,amssymb}
\RequirePackage[authoryear]{natbib}
\usepackage[ruled,vlined]{algorithm2e}
\usepackage{setspace}
\usepackage{tabularx} 
\usepackage{booktabs}
\usepackage{authblk}
\usepackage{bbm}
\usepackage{caption,subcaption}
\usepackage{wrapfig}

\newtheorem{theorem}{Theorem}[section]
\newtheorem{lemma}[theorem]{Lemma}
\theoremstyle{remark}

\allowdisplaybreaks

\makeatletter
\newcommand*{\rom}[1]{\expandafter\@slowromancap\romannumeral #1@}
\makeatother

\DeclareMathOperator*{\argmin}{argmin} 

\RequirePackage[colorlinks,citecolor=blue,urlcolor=blue]{hyperref} 
\newcommand{\blind}{1}

\begin{document}

\def\spacingset#1{\renewcommand{\baselinestretch}%
{#1}\small\normalsize} \spacingset{1}

\if1\blind
{
	 \begin{center} 
	\spacingset{1.5} 	{\LARGE\bf  A Bayesian Approach for Selecting Relevant External Data (BASE): Application to a Study of Long-Term Outcomes in a Hemophilia Gene Therapy Trial (HOPE-B)} \\ \bigskip \bigskip
		\spacingset{1} 
		{\large Tianyu Pan$ ^1 $, Yiyao Shi$ ^2 $, Xiang Zhang$ ^3 $, Weining Shen$ ^2 $, and Ting Ye$ ^4 $ } \\ \bigskip
	 {$ ^1 $Department of Pathology, Stanford University\\  
    $ ^2 $Department of Statistics, University of California, Irvine\\
    $ ^3 $Medical Affairs and HTA Statistics, Quantitative Clinical Sciences and Reporting, CSL Behring\\
    $ ^4 $Department of Biostatistics, University of Washington\\
    }
	\end{center}
} \fi

\if0\blind
{
  \bigskip
  \bigskip
  \bigskip
  \begin{center}
  \spacingset{1.5}
    {\LARGE\bf A Bayesian Approach for Selecting Relevant External Data (BASE): Application to a Study of Long-Term Outcomes in a Hemophilia Gene Therapy Trial (HOPE-B)}
\end{center}
  \medskip
} \fi

\bigskip

\begin{abstract}

Gene therapies aim to address the root causes of diseases, particularly those stemming from rare genetic defects that can be life-threatening or severely debilitating. Although an increasing number of gene therapies have received regulatory approvals in recent years, understanding their long-term efficacy in trials with limited follow-up time remains challenging. To address this critical question, we propose a novel Bayesian framework  designed to selectively integrate relevant external data with internal trial data to improve the inference of the durability of long-term efficacy. We proved that the proposed method has desired theoretical properties, such as identifying and favoring external subsets deemed relevant, where the relevance is defined as the similarity, induced by the marginal likelihood, between the generating mechanisms of the internal data and the selected external data. We also conducted comprehensive simulations to evaluate its performance under various scenarios. Furthermore, we apply this  method to predict and infer the endogenous factor IX (FIX) levels of patients who receive Etranacogene dezaparvovec over the long-term.   
Our estimated long-term FIX levels, validated by recent trial data, indicate that Etranacogene dezaparvovec induces sustained FIX production. Together, the theoretical findings, simulation results, and successful application of this framework underscore its potential to address similar long-term effectiveness estimation and inference questions in real world applications.

\end{abstract}
\noindent%
\vfill

\spacingset{1.7} 
\newpage
\section{Introduction}\label{sec: intro}


\vspace{-5mm}
\subsection{ {Hemophilia B and Etranacogene dezaparvovec}}\label{sec: introGene}\vspace{-2mm}

 {Hemophilia B is an X-linked bleeding disorder which is caused by a partial or complete deficiency of circulating factor IX activity due to mutations in the gene \citep{srivastava2020wfh}. The severity of Hemophilia B is classified based on the levels of clotting factor IX (FIX) in the blood. Severe Hemophilia B is defined by FIX levels less than 1\% of normal, leading to frequent spontaneous bleeding, including into joints and muscles, which can be very painful and result in long-term complications. Moderate Hemophilia B involves FIX levels ranging from 1\% to 5\% of normal, where bleeding can occur after minor injuries and may include spontaneous bleeding episodes \citep{blanchette2014definitions}.}

Gene therapy holds great promise as a one-time treatment for life-threatening, severe-debilitating diseases such as Hemophilia B, with demonstrated increases in FIX expression post-treatments and substantial reductions in both bleeds and utilization of factor replacement therapy in treating breakthrough bleeding \citep{nathwani2011adenovirus, nathwani2014long, george2017hemophilia,leebeek2021gene}.  {However, given the small sample size and limited follow-up data of those trials which aim to investigate the effect of gene therapies on patients with Hemophilia B, there exists uncertainty in the long-term effectiveness of those gene therapies. Because gene therapies are a potentially curative option, it is vital to understand their long-term effectiveness, in order to help the patients and health care professionals to make the optimal treatment decisions. In addition,} the high  {cost of gene therapy and the uncertainty about the lasting effect of gene therapies prompts due diligence from} payers,  {including regional/national Health Technology Assessment (HTA) entities,} who base the price accounting for long-lasting effects \citep{kee2019value}.

 {Etranacogene dezaparvovec} is the first FDA-approved gene therapy designed for treating adults with severe  {or moderately severe} Hemophilia B.  {There have been several clinical and economic evaluations conducted by Payer authorities for Etranadez dezaparvovec.}  {For instance, during pre-/early marketing period, a} flourishing body of assessment reports from entities  {evaluated its durable effect}, including the Institute for Clinical and Economic Review  \citep{Tice2022} in the United States and the National Institute for Health and Care Excellence \citep{Farmer2023} in the United Kingdom.  {However, these authorities have noted a lack of long term treatment effect data posed challenges to assessing the value for Etranacogene dezaparvovec, which necessitates} the development of appropriate  {statistical} methods to infer long-term effectiveness and  {to} bridge the evidence gap.    

\vspace{-5mm}
\subsection{ {Main Clinical Trials in Etranacogene dezaparvovec Clinical Development}}\label{sec: introHemoB}\vspace{-5mm}

\begin{figure}[ht]\centering
      \includegraphics[width=.8\linewidth]{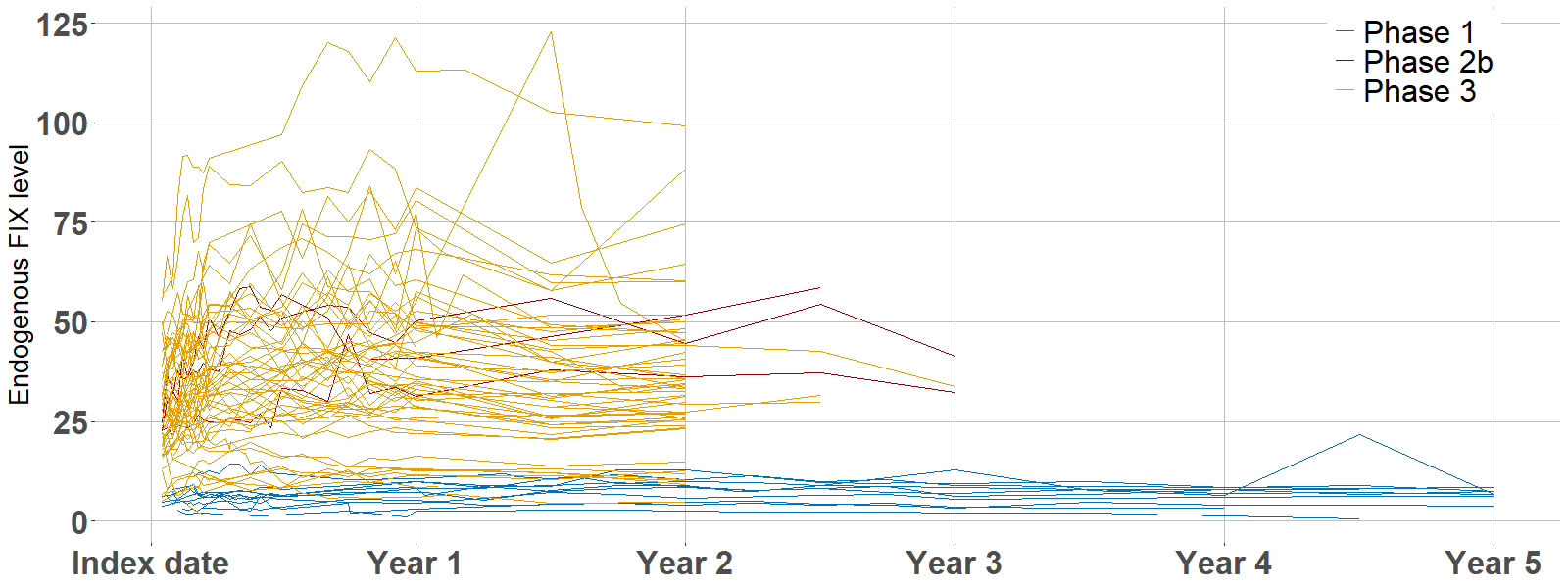}
        \caption{\label{fig: Spaghetti} FIX expression over time in clinical trials. The index date refers to the date when patients received Etranacogene dezaparvovec.}
\end{figure}

The data for this analysis were obtained from three studies (i.e., phase 1, phase 2b, and phase 3 studies). Specifically, the phase 3 study (HOPE-B) is an  {open-label, single-dose, multicenter study that evaluates the efficacy and safety of Etranacogene dezaparvovec for adults with severe or moderately severe Hemophilia B} \citep[ClinicalTrials.gov number, NCT03569891]{pipe2023gene}, with a total of 54 male participants. At the time of our data analysis, information spanning up to 3 years post-treatment was accessible. We opted to utilize only the data from the initial 2 years, reserving the remaining dataset for the purpose of assessing predictions. The ongoing collection of data  {in HOPE-B trial} will continue for a duration of up to 15 years.  {The phase 2b is an open-label, single-dose study (ClinicalTrials.gov number, NCT03489291) that evaluated Etranacogene dezaparvovec in 3 adults with severe or moderately severe Hemophilia B. The phase 1 study is a multinational, open label, dose-escalation study (ClinicalTrials.gov number, NCT02396342) that evaluated gene therapy with Etranacogene dezaparvovec in 10 adults with severe or moderately severe Hemophilia B.}

At the time of this analysis,  {the phase 1 study has a 5 year follow-up \citep{Miesbach2021} and the phase 2b study has }2.5-years follow-up \citep{Gomez2021} data available,  {which are crucial for helping infer the long-term effect of Etranacogene dezaparvovec.}  {However, it is important to note that the therapeutic genetic material differs between the phase 1 and phase 2b/3 studies. Specifically, the phase 2b/3 studies employed the gain of function Padua variant of the factor IX gene, while the phase 1 study used a wild-type factor \rom{9} gene}. These disparities could introduce systematic differences in the FIX level across studies. Indeed, as shown in Figure \ref{fig: Spaghetti}, the phase 1 study displays considerable variation in FIX levels and their longitudinal trajectories in comparison to the other two studies. This highlights the potential for biased estimates if data across the three studies are directly pooled. Throughout the rest of the article, we designate the combined phase 2b and phase 3 study (which share substantial similarities) as our \emph{internal study}, for which we aim to infer the long-term effectiveness, and we consider the phase 1 study as our \emph{external study}, from which we seek to borrow information. Of note, there are other gene therapy trials in Hemophilia B area which has long follow-up data. Those trials, had we have access to their patient-level data,  could be considered as external studies to borrow information.

 {To provide additional interpretations for Figure \ref{fig: Spaghetti}, the ``Index date" denotes the timing of Etranacogene dezaparvovec infusion. Because of the trial designs, the FIX levels of trial participants could be contaminated by previous FIX replacement therapies during first 2 weeks after index date. Therefore, all trajectories are recorded starting at 3 weeks post-treatment to minimize the influence of exogenous sources of FIX.}

\vspace{-5mm}
\subsection{Prior work and our contributions}\label{sec: introDataInt}\vspace{-2mm}

As mentioned in Section \ref{sec: introHemoB}, combining the external and internal datasets can be perceived as a data integration procedure or, more generally, as an effort to combine information from multiple data sources for an integrative and efficient inference. 

Over the past decades, data integration has been an active research area. To list a few of  {data integration methods}, \textit{data fusion} methods combine multiple datasets by assuming a shared latent variable \citep{liu2022b} or sufficient conditional overlapping support \citep{li2021efficient} for various purposes such as integrated epigenetic index estimation or average treatment effect estimation.  {However, when} datasets are in matrix format,  {data fusion methods} can sometimes lead to block-wise missingness. To tackle this, \textit{data integration} methods consider low-rank matrix recovery \citep{cai2016structured}, spectral clustering \citep{park2021integrating}, and multiple block-wise imputation \citep{xue2021integrating}. 
\textit{Multimodal data analysis}, where the collected data come in different types, can also be treated as a form of data selection. For example, to regress the outcome on the covariates from multiple datasets, people have considered the conventional linear regression \citep{li2022integrative} and the non-linear regression \citep{dai2022orthogonalized}. From a philosophical standpoint, Bayesian methods 
are natural data integration approaches. Among these, the \textit{power prior} \citep{chen1999prior,chen2000power} is a notable example designed for data integration. 
It incorporates an uncertain discounting factor $\alpha\in [0,1]$ into the historical likelihood to downgrade its importance while combining it with the current likelihood \citep{neuenschwander2009note,ibrahim2015power}. However, these methods downweight external data equally and  
do not effectively address the task of selecting relevant information from external sources. 
Numerous other data integration methods have emerged in diverse directions, including methods for  heterogeneous treatment effect estimation \citep{yang2020elastic}, long-term treatment effect estimation \citep{athey2019surrogate, imbens2022long}, doubly robust estimation for non-probability samples \citep{yang2020doubly, chen2020doubly}, and adaptive shrinkage strategy 
 \citep{chen2021minimax, oberst2022bias, hector2022turning}.
We refrain from an exhaustive discussion on this rapidly evolving literature 
and refer interested readers to the following reviews \citep{ritchie2015methods,ibrahim2015power,hassler2023data}.

Despite the comprehensive developments in data integration, all the aforementioned methods either rely on incorporating the entire external information or on applying weight scaling to the whole external data to alleviate the inferential bias when conducting the data integration. We take a distinct perspective, inspired by the within-study heterogeneity. We propose a \textit{data selection procedure} that selects external subsets that are similar to the internal data generating mechanism. This approach recognizes  {heterogeneity of the external datasets, with a subset of data} potentially being more relevant to the internal dataset. To the best of our knowledge, the selection of relevant external subsets remains an area  {worth more investigation.}

The main contribution of  {our work} is a novel data integration procedure, termed data selection, which has been largely overlooked till now. Our proposal is a general-purpose Bayesian data selection procedure by assigning a prior to all possible external data subsets and using the marginal likelihood as the criteria to favor specific external subsets in the sampling process.  {This method has desired theoretical properties} that the relevant external subset can be consistently selected with high posterior probability.  {In the application}, we introduce a novel spline model for the trend of  {endogenous FIX level and use the proposed BASE method}, producing promising results that are both statistically valid and scientifically interpretable. In comparison to the results given by the direct combination of external and internal data, and the method without external information incorporation \citep{shah2023comprehensive}, our method  {yields less variable prediction} about the long-term effect  {effect of Etranacogene dezaparvovec} by selecting relevant external subset.


\vspace{-6mm}
\section{A general Bayesian strategy for data selection}
\label{sec: method}
\vspace{-5mm}
\subsection{Method}\label{sec: BayesDatSelect}
\vspace{-2mm}

In this section, we present a general BAyesian approach for SElecting relevant
external data, which  {is termed} BASE.  {In short, BASE selects subsets from the external data that are similar to the internal data to help infer the long-term outcomes. We will use the motivating example to describe our idea in details.}

 {Suppose $Y_{sit}$ is the outcome (i.e., endogenous FIX level, abbreviated as “factor level” below) measured for patient $i$ at time $t$ in study $s$. The data collected from the studies are represented as $\{Y_{sit}, i = 1,\ldots, N_s, t \in \mathcal{T}_i^{(s)}, s = 0,1\}$, where $N_s$ is the number of patients in study $s$, $\mathcal{T}_i^{(s)}$ denotes the collection of the time schedule indices ordered temporally for the $i$-th subject in study $s$, where we let $s = 1$ refer to the internal study, while $s = 0$ refer to the external study. We use the term “trajectory” to describe the longitudinal nature of $Y_{sit}$. In our application, the internal trajectories are censored at the end of 2nd year post-treatment, while most external trajectories are observed up to 5 years after treatment. We assume that in the external study, there exists a subset $\mathcal{C}_0 \subseteq \{1,\ldots, N_0\}$, where the trajectories $\{Y_{0it}, i \in \mathcal{C}_0, t \in \mathcal{T}_i^{(s)}\}$ are generated under the same data generating process (DGP) as the internal trajectories, up to and beyond 2 years after treatment. The remaining external trajectories behave differently from the internal ones, up to and beyond 2 years after treatment. Our primary object is to integrate the relevant external information by incorporating only the external trajectories whose subject indices are in $\mathcal{C}_0$.}

 {To this end, suppose we model the trajectories using $L(\cdot\mid \boldsymbol{\theta}_1)$, which is parameterized by $\boldsymbol{\theta}_1$, our data selection procedure treats the correct external subset $\mathcal{C} \subseteq \{1,\ldots, N_0\}$ as an unknown parameter, and assigns a prior $\Pi(\mathcal{C})$ to it. The external subset is then selected by drawing posterior samples of $\mathcal{C}$, from the following distribution,
\begin{align}
        & \Pi\left(\mathcal{C}\mid \{Y_{sit}, i = 1,\ldots, N_s, t \in \mathcal{T}_i^{(s)}, s = 0,1\}\right)\label{posteriorC}\\
        & \propto \underbrace{\left(\int \prod_{i=1}^{N_1} L(\{Y_{1it}\}_{t\in\mathcal{T}_i^{(1)}}\mid \boldsymbol{\theta}_1) \times \underbrace{\pi(\boldsymbol{\theta}_1\mid \{Y_{0it}, i \in \mathcal{C}, t \in \mathcal{T}_i^{(0)}\})}_{[2].~\text{the posterior density of $\boldsymbol{\theta}_1$ given $\mathcal{C}$}}d \boldsymbol{\theta}_1\right)}_{[3].~\text{the marginal likelihood}}\times \underbrace{\Pi(\mathcal{C})}_{[1].~\text{the prior of $\mathcal{C}$}}, \nonumber
\end{align}
where $\pi(\boldsymbol{\theta}_1)$ and $\pi(\boldsymbol{\theta}_1\mid \cdot)$ denote the prior and posterior density functions of $\boldsymbol{\theta}_1$, and $L(\{Y_{1it}\}_{t\in\mathcal{T}_i^{(1)}}\mid \boldsymbol{\theta}_1)$, parameterized by $\boldsymbol{\theta}_1$, refers to the joint likelihood of the $i$-subject's longitudinal data in the internal study over its time schedule.}

 {We provide further interpretations of \eqref{posteriorC} to clarify the rationale behind this selection procedure, which is governed by the posterior distribution defined in \eqref{posteriorC}. For clarity, we break down the posterior distribution into three components and interpret each. For part [1], $\Pi (\mathcal{C})$ represents our prior belief about external subsets $\mathcal{C}$'s. Typically, a uniform prior is assigned to all possible $\mathcal{C}$'s, indicating our equal prior preference over these subsets. This specification introduces uncertainty, ensuring valid posterior inference on $\mathcal{C}$. In part [2], the selected external data (with indices assigned to $\mathcal{C}$) are used to form a posterior distribution of $\boldsymbol{\theta}_1$, which captures the intrinsic trends within the selected data. Part [3] represents the marginal likelihood, which evaluates the similarity between the data-generating mechanisms of the external subset $\mathcal{C}$ and the internal data. Unlike the traditional marginal likelihood, which integrates over the prior of $\boldsymbol{\theta_1}$, our definition is more general, as it considers the posterior of $\boldsymbol{\theta_1}$ given $\mathcal{C}$. The original definition is a special case when $\mathcal{C}$ is an empty set. In particular, if the selected external subset $\mathcal{C}$ follows the same data-generating mechanism as the internal data, the marginal likelihood will be large, and vice versa. We will provide a theoretical justification for this in Section \ref{sec 2.2: theory} under i.i.d settings. In our real data application, most internal trajectories are censored at the end of 2 years post-treatment, while most external trajectories are observed up to 5 years post-treatment. As a result, the marginal likelihood (Part [3]) will automatically prioritize external subsets whose early-stage trends match that of the internal trajectories. This behavior will be further explored through simulation studies in Section \ref{sec: Simul}.}

Since the marginal likelihood plays an important role in our proposed method, we provide more discussions here. In Bayesian studies, the marginal likelihood is commonly used to quantify the goodness-of-fit for multiple competing models (or priors). The models that assign higher probability masses around the true parameter values (i.e., fit the data better) in general yield higher marginal likelihood values \citep{robert2007bayesian}.  {In the context of \eqref{posteriorC}, the posterior density functions induced by different subset indices $\mathcal{C}$'s (Part [2]) can be interpreted as the candidate models. The prior $\Pi(\mathcal{C})$ serves as a weighting average for these candidate models, implying that our approach is essentially a model selection procedure through model averaging. This further justifies adopting a uniform prior over $\mathcal{C}$, as mentioned in the previous paragraph, since it reflects that we treat all possible models equally within the framework of model selection. During the data selection procedure, the posterior densities induced by specific external subsets, as defined in the parentheses in \eqref{posteriorC}, that exhibit higher marginal likelihood values are considered better at recovering the internal data patterns preceding 2 years post-treatment. As a result, these subsets are more likely to be drawn during the sampling process}. Note that it is possible that the selection process ends up with selecting all patients in the external dataset. However, instead of naively utilizing all external data, our approach uses a data-driven way to determine which of the external data to incorporate, making it more flexible and robust.
\vspace{-5mm}
\subsection{A theoretical perspective}\label{sec 2.2: theory}
\vspace{-2mm}

We consider a toy example to provide theoretical insights into BASE.  {Suppose the external and internal trajectories, $\{X_{ik}\}_{k=1}^{T_i^{(x)}}$ and $\{Y_{jl}\}_{l=1}^{T_j^{(y)}}$, for $i=1,\ldots,N_0$ and $j = 1,\ldots,N_1$, denoted by $\left\{X_{ik}\right\}_{\substack{i=1,k=1}}^{N_0,T_i^{(x)}}$ and $\left\{Y_{jl}\right\}_{\substack{j=1,l=1}}^{N_1,T_j^{(y)}}$, are both i.i.d. between and within subject. For clarity and succinctness, we slightly abuse the notations by using $\{X_{ik}\}_{k=1}^{T_i^{(x)}}$ to represent the first through $T_i^{(x)}$-th observations collected for the $i$-th subject, ordered temporally. Specifically, we assume,
\begin{equation}\label{iidDist}
    \begin{split}
        & X_{11},\ldots,X_{1T_1^{(x)}},\ldots,X_{N_01},\ldots,X_{N_0T_{N_0}^{(x)}}\stackrel{i.i.d.}{\sim} p_{\theta_0},\\
        & Y_{11},\ldots,Y_{1T_1^{(y)}},\ldots,Y_{N_11},\ldots,Y_{N_1T_{N_1}^{(y)}}\stackrel{i.i.d.}{\sim} p_{\theta_1},
    \end{split}
\end{equation}
where $p_{\theta}$ is a known  probability density function with parameter $\theta$. In the following theorem, we show that by adopting the marginal likelihood in \eqref{posteriorC} as a criterion for Bayesian estimation with a uniform prior on external subset indices $\mathcal{C}$'s, with $\mathcal{C}\subseteq \{1,\ldots,N_0\}$, our method can correctly estimate $\mathcal{C}$ as the entire external subset if $\theta_0$ and $\theta_1$ are sufficiently close, or as empty set otherwise.}

\begin{theorem}\label{postExample}
Suppose external and internal trajectories are generated following \eqref{iidDist} and Assumptions (A1)-(A4) detailed in the Supplementary Materials  hold.  If $\|p_{\theta_0}-p_{\theta_1}\|_1\lesssim \epsilon_{N_0^*}$, the expected Bayes factor can be controlled as follows,
\begin{equation}\label{postExampleEq1}
    \begin{split}
        & \frac{\int \prod_{\substack{j=1, l=1}}^{N_1,T_j^{(y)}} p_{\theta}(Y_{jl})d\pi(p_{\theta})}{\int_{\|p_{\theta}-p_{\theta_0}\|_1\leq K \epsilon_{N_0^*}} \prod_{\substack{j=1, l=1}}^{N_1,T_j^{(y)}} p_{\theta}(Y_{jl})d\pi(p_{\theta}\mid \left\{X_{ik}\right\}_{\substack{i=1,k=1}}^{N_0,T_i^{(x)}})}\stackrel{p}{\to} 0,~~\text{as}~N_0^*,N_1^*\to\infty.
    \end{split}
\end{equation}
On the other hand, if $\|p_{\theta_0}-p_{\theta_1}\|_1\gtrsim \epsilon_{N_0^*}\times \psi_{N_0^*}$ instead, for $\psi_{N_0^*}$ diverging to $\infty$ at any rates, it yields that
\begin{equation}\label{postExampleEq2}
    \begin{split}
        & \frac{\int_{\|p_{\theta}-p_{\theta_0}\|_1\leq K \epsilon_{N_0^*}} \prod_{\substack{j=1, l=1}}^{N_1,T_j^{(y)}} p_{\theta}(Y_{jl})d\pi(p_{\theta}\mid \left\{X_{ik}\right\}_{\substack{i=1,k=1}}^{N_0,T_i^{(x)}})}{\int \prod_{\substack{j=1, l=1}}^{N_1,T_j^{(y)}} p_{\theta}(Y_{jl})d\pi(p_{\theta})} \stackrel{p}{\to} 0,~~\text{as}~N_0^*,N_1^*\to\infty,
    \end{split}
\end{equation}
where $N_0^*\equiv \sum_{i=1}^{N_0} T_i^{(x)}$ and $N_1^*\equiv \sum_{j=1}^{N_1} T_j^{(y)}$ respectively denote the total observations of the external and internal trajectories, $a\lesssim b$ denotes $a\leq C \times b$ for a universal positive constant $C$, $K > 0$ is defined in the Supplementary File, $\pi(\cdot)$ is a prior on $p_{\theta}$, $\epsilon_n =\left(\frac{M_n}{n}\right)^{1/2}$ refers to the posterior contraction rate and $M_n = o(\log n)$ is a sequence of numbers going to $\infty$ as $n \rightarrow \infty$.

\end{theorem}
 {Theorem \ref{postExample} can be interpreted as a Bayesian testing (model selection) consistency result. When the external trajectories $\left\{X_{ik}\right\}_{\substack{i=1,k=1}}^{N_0,T_i^{(x)}}$ are similar to the internal trajectories $\{Y_{jl}\}_{\substack{j=1, l=1}}^{N_1,T_j^{(y)}}$ (i.e., $p_{\theta_0}$ and $p_{\theta_1}$ are sufficiently close), the marginal likelihood of the internal trajectories, obtained by integrating over the posterior density given the external trajectories, dominates the likelihood integrated over the prior. In such cases, it is beneficial to integrate the entire external trajectories, since they provide additional useful information. On the other hand, when the external and internal trajectories are generated from significantly different mechanisms, we prefer to discard the external trajectories. In this situation, the marginal likelihood of the internal trajectories, obtained by integrating over the prior will dominate. In summary, Theorem \ref{postExample} suggests these two situations, namely, whether we should incorporate or discard the entire external trajectories, are correctly determined by the difference in the magnitude of the corresponding marginal likelihoods.} 

 {For brevity, we provide the details of Assumptions (A1)-(A4) in Theorem \ref{postExample} in the Supplementary File. Here, we briefly describe and interpret them in the context of integrating external trajectories: (A1) The underlying trend is modeled by a parametric model $p_{\theta}$, to which the prior assigned is non-degenerate (non-informative); (A2) The underlying trends of both external and internal trajectories can be correctly estimated using the model in (A1), with the true models denoted by $p_{\theta_0}$ and $p_{\theta_1}$, respectively; (A3) The number of external observations $N_0^*\equiv \sum_{i=1}^{N_0} T_i^{(x)}$ is smaller than the internal ones $N_1^*\equiv \sum_{j=1}^{N_1} T_j^{(y)}$ at a specific rate; and (A4) Both $p_{\theta_0}$ and $p_{\theta_1}$ are within the support of the prior defined in (A1), and receive non-trivial prior concentration. Assumption (A1) is mild if the model and the prior is properly designed. Assumptions (A2) and (A4) are testable and widely used in Bayesian asymptotics literature; and they can be easily verified following the steps of Lemma 8.1 and Theorem 2.1 in \citet{ghosal2000convergence} and Lemma B2 in \citet{shen2013adaptive}. Assumption (A3) requires the internal data size to be larger than the external data size, which is reasonable given that we would like to only borrow information from the external set while treating the internal data as the main source for inference.} 

 {Next, we consider a more realistic scenario where the external trajectories are generated by two mechanisms (i.e., $\left\{X_{ik}\right\}_{\substack{i=1,k=1}}^{N_0^{(x)},T_i^{(x)}}$ and $\{Y_{jl}\}_{\substack{j=N_0^{(x)} + 1, l=1}}^{N_0^{(x)} + N_0^{(y)},T_j^{(y)}}$), where one of these mechanisms matches the the internal trajectories (i.e., $\{Z_{rs}\}_{\substack{r=1, s=1}}^{N_1,T_r^{(z)}}$), while the other does not. Specifically, we assume,
\begin{equation}\label{hybridExample}
    \begin{split}
        & X_{11},\ldots,X_{1T_1^{(x)}},\ldots,X_{N_0^{(x)}1},\ldots,X_{N_0^{(x)}T_{N_0^{(x)}}^{(x)}}\stackrel{i.i.d.}{\sim} p_{\theta_1},\\
        & Y_{(N_0^{(x)} + 1)1},\ldots,Y_{(N_0^{(x)} + 1)T_{(N_0^{(x)} + 1)}^{(y)}},\ldots,Y_{(N_0^{(x)} + N_0^{(y)})1},\ldots,Y_{(N_0^{(x)} + N_0^{(y)})T_{(N_0^{(x)} + N_0^{(y)})}^{(y)}}\stackrel{i.i.d.}{\sim} p_{\theta_0},\\
        & Z_{11},\ldots,Z_{1T_1^{(z)}},\ldots,Z_{N_1^{(z)}1},\ldots,Z_{N_1^{(z)}T_{N_1^{(z)}}^{(z)}}\stackrel{i.i.d.}{\sim} p_{\theta_1}.\\
    \end{split}
\end{equation}
In Theorem \ref{theoremFinal}, we demonstrate that our method can correctly identify the correct external subset $\mathcal{C}_0\equiv\{1,\ldots,N_0^{(x)}\}$ using the selected external subset from our model $\mathcal{C}\subseteq \{1,\ldots,N_0^{(x)},N_0^{(x)} + 1,\ldots,N_0^{(x)} + N_0^{(y)}\}$. This comes with the cost of at most including a small amount of data from the incorrect external trajectories, with indices forming a subset of $\{N_0^* + 1,\ldots,N_0^* + N_0'\}$. The key observation is that if a non-negligible number of external observations $\{Y_{jl}\}_{\substack{j\in\mathcal{C}', l=1}}^{~~~~~T_j^{(y)}}$, where $\mathcal{C}'$ is a subset of $\{N_0^{(x)} + 1,\ldots, N_0^{(x)} + N_0^{(y)}\}$, are included, the induced density function (a mixture of $p_{\theta_0}$ and $p_{\theta_1}$) diverges from $p_{\theta_0}$ at a detectable distance, allowing our model to identify it.}

 {Theorem \ref{theoremFinal} holds under Assumption (A1), (A3), (B1), (C1) and (C2). Briefly, (B1) and (C2) are modifications of (A2) and (A4) for pooled samples; (C1) assumes that the correct external observations $N_0^*\equiv \sum_{i=1}^{N_0^{(x)}}T_{i}^{(x)}$ tends to infinity as the pooled external observations $N_0\equiv N_0^* + N_0'$ approaches infinity, where $N_0'\equiv \sum_{i=1}^{N_0^{(y)}} T_{(N_0^{(x)} + i)}^{(y)}$ denotes the number of incorrect external observations. Theorem \ref{theoremFinal} compares the marginal likelihoods of two models based on two different external subsets: $\mathcal{C}_0 ',\mathcal{C}_1'\subseteq \{1,\ldots,N_0^{(x)},N_0^{(x)} + 1,\ldots,N_0^{(x)} + N_0^{(y)}\}$. Specifically, the model using $\mathcal{C}_1'$ effectively selects the correct external trajectories while including only a small amount of incorrect external observations. That is, it holds that $\{1,\ldots,N_0^{(x)}\} \subseteq \mathcal{C}_1'$, and for $m' \equiv\sum_{i\in\mathcal{C}_1^{''}}T_{i}^{(y)}$, where $\mathcal{C}_1''$ is the subset of $\mathcal{C}_1'$ excluding the correct external indices, the incorrect observations are sufficiently small compared to the correct external observations $N_0^*\equiv \sum_{i=1}^{N_0^{(x)}}T_{i}^{(x)}$. In contrast, the other model $\mathcal{C}_0'$ either includes too few pooled external observations or contains a non-negligible amount of incorrect external observations. Specifically, for $\mathcal{C}_0^{'*}\equiv \mathcal{C}_0'\cap \{1,\ldots,N_0^{(x)}\}$ and $\mathcal{C}_0^{''}\equiv \mathcal{C}_0^{'}\cap\{N_0^{(x)} + 1,\ldots,N_0^{(x)} + N_0^{(y)}\}$, the correct and incorrect external observation numbers $u\equiv\sum_{i\in\mathcal{C}_0^{'*}}T_{i}^{(x)}$ and $v\equiv\sum_{i\in\mathcal{C}_0^{''}}T_{i}^{(y)}$ must satisfy certain conditions. The details are presented as follows,
\begin{theorem}\label{theoremFinal}
Suppose Assumptions (A1), (A3), (B1), (C1), and (C2) in the Supplementary File hold. If $m'$ satisfies $\|p_{\theta_1}-p_{\theta_0}\|_1 = o\left(\frac{m' + N_0^*}{m'}\times\left(\sqrt{\frac{M_{m' + N_0^*}}{m' + N_0^*}} - \sqrt{\frac{M_{N_1}}{N_1}}\right)\right)$, we have
\begin{equation}\label{concludeBF}
    \begin{split}
        & \frac{\int_{\|p_{\theta}-p_{u,v}^*\|_1\leq K^* \epsilon_{u+v}} \prod_{\substack{r=1, s=1}}^{N_1^{(z)},T_r^{(z)}} p_{\theta}(Z_{rs})d\pi(p_{\theta}\mid \left\{X_{ik}\right\}_{\substack{i\in\mathcal{C}_0^{'*},k=1}}^{~~~~~T_i^{(x)}},\left\{Y_{jl}\right\}_{\substack{j\in\mathcal{C}_0^{''},l=1}}^{~~~~~T_j^{(y)}})}{\int_{\|p_{\theta}-p_{N_0^*,m'}^*\|_1\leq K^* \epsilon_{N_0^*+m'}} \prod_{\substack{r=1, s=1}}^{N_1^{(z)},T_r^{(z)}} p_{\theta}(Z_{rs})d\pi(p_{\theta}\mid \left\{X_{ik}\right\}_{\substack{i = 1,k=1}}^{N_0^{(x)},T_i^{(x)}},\left\{Y_{jl}\right\}_{\substack{j\in\mathcal{C}_1^{''},l=1}}^{~~~~~T_j^{(y)}})}\stackrel{p}{\to}0,
    \end{split}
\end{equation}
$\text{as}~N_0^*,~N_0,~u,~v \to\infty$ if either of the following two conditions holds (i) $\frac{u + v}{v}\times\left(\sqrt{\frac{M_{N_1}}{N_1}} + \sqrt{\frac{M_{u+v}}{u+v}}\right) = o\left(\|p_{\theta_1}-p_{\theta_0}\|_1\right)$ or (ii) $u + v = o(N_0^* + m')$, where $N_1\equiv\sum_{i=1}^{N_1^{(z)}}T_i^{(z)}$ denotes the number of internal observations, $M_n = o(\log n)$ is a sequence of numbers going to $\infty$ as $n \rightarrow \infty$, $\epsilon_n =\left(\frac{M_n}{n}\right)^{1/2}$, $p^*_{n,m} \equiv \frac{n}{n+m}p_{\theta_1} + \frac{m}{n+m}p_{\theta_0}$ refers to the weighted average of $p_{\theta_0}$ and $p_{\theta_1}$ given $n\leq N_0^*$, $m\leq N_0'$, and $n,m\in \mathbbm{Z}^+$.
\end{theorem}}

 {Theorem \ref{theoremFinal} states that models induced by $\mathcal{C}_1'$, which select the entire correct external observations $\left\{X_{ik}\right\}_{\substack{i=1,k=1}}^{N_0^{(x)},T_i^{(x)}}$, along with $m'$ observations from the incorrect external observations $\{Y_{jl}\}_{\substack{j=N_0^{(x)} + 1, l=1}}^{N_0^{(x)} + N_0^{(y)},T_j^{(y)}}$, are preferred over alternative models induced by $\mathcal{C}_0'$ in terms of marginal likelihood values.
The proportion of incorrect selections, $\frac{m'}{N_0^*}$, is determined by the discrepancy between $p_{\theta_0}$ and $p_{\theta_1}$. For instance, if the discrepancy is small (e.g., 0), $m'$ can be as large as $N_0^*$ because the incorrectness is minimal. On the other hand, when the discrepancy is large, $m'$ becomes negligible compared to $N_0^*$. The number of correctly and incorrectly selected external observations, $u$ and $v$, is assumed to satisfy one of the following conditions: (i) $p^*_{u,v}\equiv \frac{u}{u+v}p_{\theta_1} + \frac{v}{u+v}p_{\theta_0}$ is recognizably distant from $p_{\theta_1}$, or (ii) the information provided by $\left\{X_{ik}\right\}_{\substack{i\in\mathcal{C}_0^{'*},k=1}}^{~~~~~~T_i^{(x)}},\left\{Y_{jl}\right\}_{\substack{j\in\mathcal{C}_0^{''},l=1}}^{~~~~~~T_j^{(y)}}$ is less informative that provided by $\left\{X_{ik}\right\}_{\substack{i=1,k=1}}^{N_0^{(x)},T_i^{(x)}}$, $\left\{Y_{jl}\right\}_{\substack{j\in\mathcal{C}_1^{''},l=1}}^{~~~~~~T_j^{(y)}}$. Notably, when $u + v$ is finite, the result still holds by slightly modifying the proof of Theorem \ref{postExample}.}
 
 {Theorem \ref{postExample} and \ref{theoremFinal} provide theoretical justification for the data selection procedure based on the marginal likelihood criterion. Specifically, the model can correctly select the relevant external subset, where the external observations $\left\{X_{ik}\right\}_{\substack{i=1,k=1}}^{N_0^{(x)},T_i^{(x)}}$ are on the order of $O(N_0^*)$, while discarding the wrong external observations $\{Y_{jl}\}_{\substack{j=N_0^{(x)} + 1, l=1}}^{N_0^{(x)} + N_0^{(y)},T_j^{(y)}}$ except for a negligible amount. Motivated by these theoretical results under i.i.d. assumptions, we apply the selection approach in \eqref{posteriorC} to explore the relevant external subset in the presence of temporal correlation. This approach improves early-stage parameter estimation and long term prediction by reducing the relative error (RE) compared to directly combining the internal and external data. The validity and performance of this procedure will be further evaluated in Section \ref{sec: Simul} through extensive simulations.}

\vspace{-8mm}
\section{Model specification for the  {trend of endogenous FIX levels}}
\label{sec: model}
\vspace{-5mm}
\subsection{Concatenated Cubic Hermite spline (CCHs)}\label{sec: ConCubHerSpline} \vspace{-2mm}
 {In this section, we introduce a parametric model to characterize the mean factor level. Motivated by a shared pattern among similar gene therapy products -- where factor levels increase after treatment but may decrease over time and eventually reach a plateau \citep{Nathwani2018,shah2023comprehensive} -- we propose modeling the mean factor level using a concatenation of two Cubic Hermite splines. Specifically, we model the mean factor level for both the internal trajectories and the selected external trajectories using $\psi(t;\boldsymbol{\theta})$, for $t \in [0, T]$,
where the parameter vector is $\boldsymbol{\theta}\equiv (\mu_{0},m_{0},\mu_{1},m_{1},\mu_{2})$. The function $\psi(t;\boldsymbol{\theta})$ is defined by two Cubic Hermite splines over the intervals $[0,\alpha]$ and $(\alpha, T]$, concatenated at a turning point $\alpha$:
\begin{equation}\label{meanFunc}
  \psi(t;\boldsymbol{\theta})=\begin{cases}
    h_{00}(\frac{t}{\alpha})\mu_{0} + h_{10}(\frac{t}{\alpha})\alpha m_{0} + h_{01}(\frac{t}{\alpha})\mu_{1} + h_{11}(\frac{t}{\alpha})\alpha m_{1}, & t \in [0,\alpha],\\
    h_{00}(\frac{t - \alpha}{T - \alpha})\mu_{1} + h_{10}(\frac{t - \alpha}{T - \alpha})(T - \alpha) m_{1} + h_{01}(\frac{t - \alpha}{T - \alpha})\mu_{2}, & t \in (\alpha,T],
  \end{cases}
\end{equation}
where $T$, the ``plateau time point", is a pre-specified value at which the mean factor level is expected to stabilize and remain constant thereafter. In real data analysis, $T$ could be chosen with flexibility with other knowledge such as clinical pharmacology or experts’ opinions. For instance, it may be set to 6 to represent the mean factor level approaches its plateau at 6 years post-treatment. Sensitive analyses by choosing different $T$ values are detailed in Section \ref{sec: Real} and the Supplementary File.} 

\begin{figure}[ht]\centering
      \includegraphics[width=.8\linewidth]{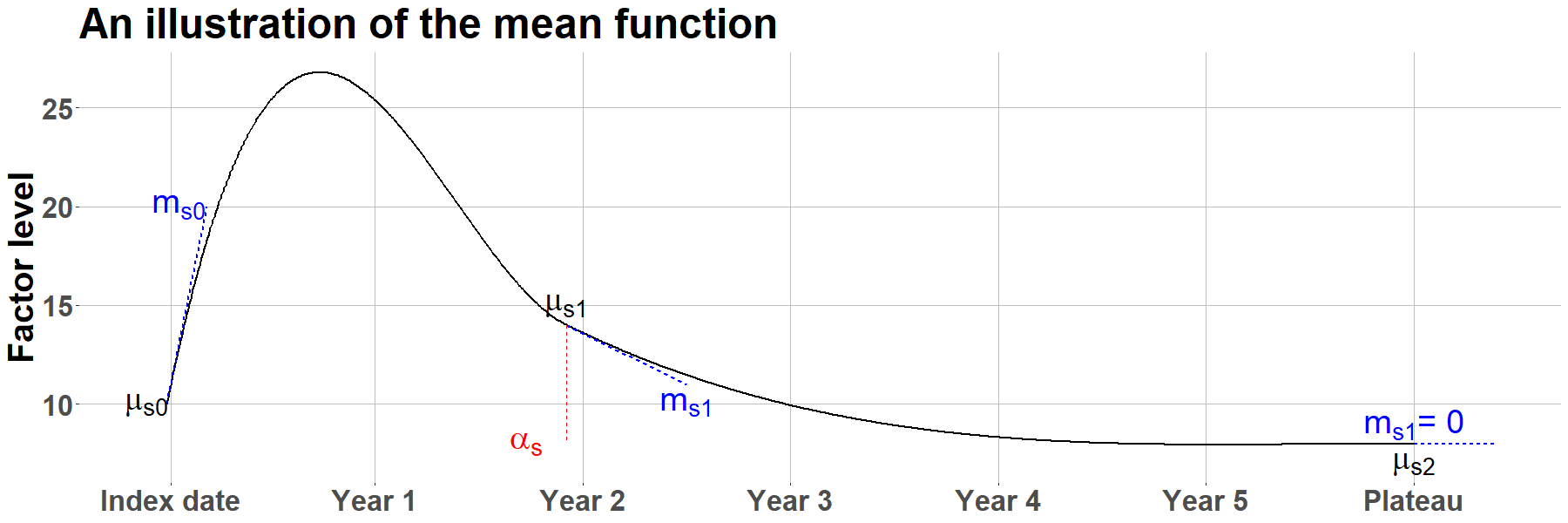}
        \caption{\label{fig: MeanFunc} The blue dashed lines represent the derivatives at specific time points, while the black solid line illustrates the mean value. The turning point is indicated by the red dashed line.}
\end{figure}

 {We provide further interpretation of the mean factor level $\psi(t;\boldsymbol{\theta})$. For $\mu_{0}$ and $m_{0}$, they represent the starting value and the derivative of the mean factor level at $t = 0$, respectively. Similarly, $\mu_{1}$ and $m_{1}$ correspond to the mean factor level and its derivative at the turning point $\alpha$. The mean factor level at the plateau time point $T$ is denoted by $\mu_{2}$, with the derivative set to 0, as per the definition of the ``plateau time point". The Cubic Hermite basis functions involved in \eqref{meanFunc}, $h_{00},h_{10},h_{01},h_{11}$, are defined as $h_{00}(t)  = 2t^3 - 3t^2 + 1, h_{10}(t) = t^3 - 2t^2 + t,  h_{01}(t)  = -2t^3 + 3t^2, $ and $ h_{11}(t)  = t^3 - t^2.$ In Figure \ref{fig: MeanFunc}, we provide an example of $\psi(t;\boldsymbol{\theta}_s)$ with annotations for illustration.}

 {Our choice of using a CCHs offers three-fold benefits. First, the model can incorporate relevant constraints, such as the mean factor levels increasing after treatment and potentially decreasing over time \citep{Nathwani2018}. This is achieved by constraining $m_{0}$ to be positive and $m_{1}$ to be negative, respectively, reflecting the post-treatment dynamics. Second, the CCHs can be expressed as a linear combination of the basis functions of a Cubic Hermite spline. From a Bayesian perspective, this allows for the use of a truncated multivariate normal prior on $\boldsymbol{\theta}$, providing conjugacy, improving computational efficiency given limited sample sizes, and ensuring the satisfaction of the specified constraints. Third, the parametric nature of $\psi(t;\boldsymbol{\theta})$ aligns well with the selection procedure outlined in Section \ref{sec: method}. This enables the model to identify relevant external trajectories (factor levels) and leverage their long-term information to predict the long-term behavior of the internal trajectories.}

\vspace{-5mm}
\subsection{Full model specification}\label{sec: Model}
\vspace{-2mm}

 {We present our model before applying the data selection strategy outlined in Section \ref{sec: method}. Our proposed model comprises two main steps: Step (1). Sample external subsets; Step (2). Sample CCHs parameters given the most representative external subset $\hat{\mathcal{C}}$ obtained from Step (1). The details are outlined as follows,}
\begin{equation}\label{Model}
    \begin{split}
        & \textbf{Step (1).}\\
        & \Pi\left(\mathcal{C}\mid \{Y_{sit}, i = 1,\ldots, N_s, t \in \mathcal{T}_i^{(s)}, s = 0,1\}\right)\\
        & \propto \Pi\left(\mathcal{C}\right) \times \int \prod_{i=1}^{N_1}L(\{Y_{1it}\}_{t\in\mathcal{T}_i^{(1)}}\mid \boldsymbol{\theta},\alpha,\rho,\sigma_0^2,\sigma_1^2)\times\\
        & ~~~~\pi(\boldsymbol{\theta},\alpha,\rho,\sigma_0^2,\sigma_1^2\mid \{Y_{0it}, i \in \mathcal{C}, t \in \mathcal{T}_i^{(0)}\})d \boldsymbol{\theta}d\alpha d\rho d\sigma_0^2 d\sigma_1^2,\\
        & \textbf{Step (2).}\\
        & Y_{1it} = \psi(t;\boldsymbol{\theta}) + \epsilon_{1it},~\text{for}~i=1,\ldots,N_1,~ t\in \mathcal{T}_i^{(1)},\\
        & Y_{0jt} = \psi(t;\boldsymbol{\theta}) + \epsilon_{0jt},~\text{for}~j\in \hat{\mathcal{C}},~t\in \mathcal{T}_j^{(0)},\\
        & \psi(t;\boldsymbol{\theta}) = \begin{cases}
			h_{00}(\frac{t}{\alpha})\mu_{0} + h_{10}(\frac{t}{\alpha})\alpha m_{0} + h_{01}(\frac{t}{\alpha})\mu_{1} + h_{11}(\frac{t}{\alpha})\alpha m_{1}, & t\in[0,\alpha],\\
            h_{00}(\frac{t-\alpha}{T-\alpha})\mu_{1}+h_{10}(\frac{t-\alpha}{T-\alpha})(T-\alpha)m_{1} + h_{01}(\frac{t-\alpha}{T-\alpha})\mu_{2}, & t\in (\alpha,T],
		 \end{cases}\\
        & \boldsymbol{\theta}\equiv(\mu_{0},m_{0},\mu_{1},m_{1},\mu_{2})^T\sim N^{(2^+,4^-)}(\beta_0,\Psi_0),\\
        & (\epsilon_{sit})_{t\in \mathcal{T}_s}\mid\sigma_s^2,\rho \sim N(0,\sigma_s^2\times \Sigma_{\rho}),~ \sigma_s^{-2} \sim \text{Gamma}(a_0,b_0),\\
        & \alpha \sim tN(\text{mean}=2,\text{sd} = 1;\text{lb} = 0, \text{ub} = T),~ {\log(\rho) \sim N(0,\psi_0^2)},\\
        & \hat{\mathcal{C}} = \argmin_{\mathcal{C}\in \text{Post}(\mathcal{C})}\|\boldsymbol{Z_{\mathcal{C}}} - \boldsymbol{\bar{Z}_{\mathcal{C}}}\|_2, \\
        & \boldsymbol{\bar{Z}_{\mathcal{C}}} = \frac{1}{|\text{Post}(\mathcal{C})|}\sum_{\mathcal{C}\in \text{Post}(\mathcal{C})}\boldsymbol{Z_{\mathcal{C}}}, 
    \end{split}
\end{equation}
 {where $\|\cdot\|_2$ denotes the $\ell^2$ vector norm, $\boldsymbol{Z_{\mathcal{C}}}$ denotes a vector of length $N_0$ (i.e., the number of external trajectories), with the $j$-th entry being 1 if the $j$-th external trajectory is selected ($j\in\mathcal{C}$) and 0 otherwise, $\text{Post}(\mathcal{C})$ represents the collection of the posterior samples of $\mathcal{C}$ collected from Step (1), $\boldsymbol{\theta}$ refers to the CCHs parameter shared by both external and internal trajectories, respectively. Additionally, $\alpha$ is shared by both external and internal trajectories to increase the external information borrowing ratio, $N^{(2^+,4^-)}$ refers to a multivariate normal distribution that is truncated above 0 for the second entry and below 0 for the fourth entry, as discussed in Section \ref{sec: ConCubHerSpline}. The function $L(\cdot\mid \boldsymbol{\theta},\alpha,\rho,\sigma_0^2,\sigma_1^2)$ denotes the joint likelihood function for a longitudinal trajectory, given all model parameters. The posterior distribution of the parameters, $\pi(\boldsymbol{\theta},\alpha,\rho,\sigma_0^2,\sigma_1^2\mid \{Y_{0it}, i \in \mathcal{C}, t \in \mathcal{T}_i^{(0)}\})$ is constructed using the priors defined in Step (2) and $L(\cdot\mid \boldsymbol{\theta},\alpha,\rho,\sigma_0^2,\sigma_1^2)$ as the likelihood function. The covariance matrix $\Sigma_{\rho}$ is based on an exponential kernel, with the $(k,l)$ entry expressed as $\exp\{-|k - l|/\rho \}$ for any $k,l\in\mathcal{T}_i^{(1)}~\text{or}~\mathcal{T}_j^{(0)}$, and $i=1,\ldots,N_1$ and $j = 1,\ldots,N_0$, and $tN(\text{mean}=2,\text{sd} = 1;\text{lb} = 0, \text{up} = T)$ refers to a normal distribution $N(2,1)$ truncated between 0 and $T$, which is derived from a previous study \citep{samelson2021follow}. The truncated normal distribution suggests that the turning point likely occurs around 2 years post-treatment, with approximately a one-year standard deviation as uncertainty.}

 {We proceed by interpreting our approach \eqref{Model}, which operates as a ``two-step procedure". In Step (1), the external subset $\mathcal{C}$ is sampled according to the procedure described in Section \ref{sec: BayesDatSelect}. Since $\mathcal{C}$ is treated as a random variable in this step, summarizing the results can be challenging due to the numerous posterior samples of $\mathcal{C}$. To address this, we select the most representative external subset $\hat{\mathcal{C}}$ for reporting and use it in Step (2). In Step (2), the internal trajectories and the selected (i.e., those with indices assigned to $\hat{\mathcal{C}}$) external trajectories are combined and used as the inputs for the CCHs model. It is crucial to note that Step (1) and Step (2) are intrinsically consistent because they are derived from a joint posterior distribution $\Pi^*$ of $\left(\mathcal{C},\boldsymbol{\zeta}\right)$. This posterior is constructed using the priors defined in Step (2) times $\Pi(\mathcal{C})\times \pi(\boldsymbol{\zeta}\mid \{Y_{0it}, i \in \mathcal{C}, t \in \mathcal{T}_i^{(0)}\})$, resulting in:
\begin{equation}\label{jointParams}
    \begin{split}
        & \Pi^*\left(\mathcal{C},\boldsymbol{\zeta}\mid \{Y_{sit}, i = 1,\ldots, N_s, t \in \mathcal{T}_i^{(s)}, s = 0,1\}\right) \\
        & \propto \Pi\left(\mathcal{C}\right) \times \prod_{i=1}^{N_1}L(\{Y_{1it}\}_{t\in\mathcal{T}_i^{(1)}}\mid \boldsymbol{\zeta})\times \pi(\boldsymbol{\zeta}\mid \{Y_{0it}, i \in \mathcal{C}, t \in \mathcal{T}_i^{(0)}\}),
    \end{split}
\end{equation}
where $\boldsymbol{\zeta}\equiv \left(\boldsymbol{\theta},\alpha,\rho,\sigma_0^2,\sigma_1^2\right)$ represents all the parameters involved in Step (2). The posterior samples of $\boldsymbol{\zeta}$ given $\hat{\mathcal{C}}$, collected in Step (2), are drawn from the conditional distribution of $\boldsymbol{\zeta}$ given $\hat{\mathcal{C}}$ under \eqref{jointParams}. This is a valid Bayesian step, as $\hat{\mathcal{C}}$ itself is a valid draw from the joint distribution \eqref{jointParams}, obtained by integrating out $\boldsymbol{\zeta}$.}

 {For both simulation and real data analyses, we consider the following hyper-parameter settings for the priors, $\beta_0 = (0,0,0,0,0)^T$, $\nu_0 = 0.01$, $\Psi_0 = 100\times \text{diag}(1,1,1,1,1)$, $\psi_0^2 = 100$, $a_0 = 0.01$, $b_0 = 0.01$ and $\Pi(\mathcal{C})\propto 1$. For the real data analysis, we choose $T = 6$, assuming the factor levels approach the plateau value by 6 years post-treatment. The simulation settings are briefly introduced in Section \ref{sec: Simul} and detailed in the Supplementary File. We will also conduct sensitivity analyses for the real data analysis to assess whether increasing $T$ significantly affects the inference results.}

 {For the sampling procedure, in Step (1), we use the harmonic mean estimator \citep{neton1994approximate} to approximate the marginal likelihood, taking 4000 samples after discarding the first 1,000 samples for burn-in. To calculate $\hat{\mathcal{C}}$, we take 1,000 posterior samples of $\mathcal{C}$ from the categorical distribution based on the estimated marginal likelihood. In Step (2), we apply the Gibbs sampler with the full conditional distribution to take samples, retaining one sample every 10 iterations from 10,000 MCMC iterations after 1,000 burn-in samples. Empirically, the marginal likelihood approximation in Step (1) achieves bearable stochastic error, while the chosen priors in Step (2) are sufficiently non-informative. Additionally, the number of iterations is sufficient to approach the stationary posterior distribution with good mixing. The most time-consuming task, when deployed on a server that operates at 3.80 GHz, takes approximately 16 hours. This duration is manageable for a Bayesian method given such iteration settings. The efficiency of the task is attributed to the conjugacy achieved by adopting the multivariate-normal inverse-Gamma distribution, as defined in \eqref{Model}. In Section \ref{sec: Simul}, we will further investigate the performance of our proposed model.}

\vspace{-8mm}
\section{Simulation}\label{sec: Simul}
\vspace{-4mm}

 {In this section, we conduct simulation studies to assess the performance of our approach across three main settings. Setting 1 is discussed in the main article, while Settings 2 and 3 are detailed in the Supplementary File. In Setting 1, we compare our selection procedure (BASE; abbreviated as SP) to two alternative methods: the direct combination approach (DC) and the no-external-information-borrowing approach (NB). The primary objective of Setting 1 is to validate that our approach improves the estimation of CCHs parameters by effectively prioritizing external trajectories generated from the same mechanism as the internal trajectories (hereafter referred to as ``correct external trajectories") over those generated differently, termed ``incorrect external trajectories". Setting 2 compares our approach to a Subset-base Synthetic Control Method (SSCM; \citet{abadie2010synthetic, doudchenko2016balancing}), focusing on identifying the conditions where our approach may outperform SSCM and vice versa. Setting 3 aims to stress-test our approach by evaluating its performance against the DC and NB methods when the correct and incorrect external trajectories are generated with much closer similarity than in Setting 1. All generated trajectories are subject to the censoring patterns observed in the real data to mimic the real-data setting.}

 {Throughout all simulation settings, we assume that both the external and internal trajectories are generated using our model. Specifically, we assume that the turning point $\alpha$ occurs at 1.15 years post-treatment (or 60 weeks post-treatment) and that the expected outcome stabilizes at 6 years post-treatment. The subject level variances for external and internal studies, that is, $\sigma_0^2$ and $\sigma_1^2$ are chosen to be 1.5. The time schedule for the external and internal trajectories vary across the three settings, with further details provided in the Supplementary File. To replicate the sample sizes observed in the real data, we assume 10 external trajectories and 57 internal trajectories. Multiple data-generating processes (DGPs) are designed for each setting to evaluate the methods' performance, and 100 Monte Carlo replications are performed to obtain the averaged performance for each candidate method under each DGP.}

\vspace{-5mm}
\subsection{Setting 1}
\vspace{-2mm}

 {In Setting 1, we assume the external trajectories consist of a mix of $K_1$ correct external trajectories and $K_2$ incorrect external trajectories ($K_1 + K_2 = 10$). The DGP used in this section, termed DGP 1, has a true CCHs parameter for the internal trajectories of $\boldsymbol{\theta}^* = (20,1,35,-0.05,28)$. For the external trajectories, the CCHs parameters $\boldsymbol{\theta}^*$'s are a mixture of $(20,1,35,-0.05,28)$ and $(20,2,65,-0.2,16)$. This setup in DGP 1 is designed to demonstrate that our approach can effectively prioritize correct external trajectories over incorrect ones when the two processes are significantly different. Additionally, we explore various combinations of $K_1$ and $K_2$, as well as different values of $\rho$. In Figure \ref{fig: Simul12}, we present 10 trajectories for each generating process, setting $\rho$ to be 50, and $K_1 = K_2 = 5$ for illustration. This implies that the correlation between two observations separated by 52 weeks is approximately $\exp\{-52/50\} \approx 0.353$.}

 {To evaluate the performance of the three methods in estimating the true parameters, we use relative error metrics to assess the accuracy of both the trend estimation up to the turning point (denoted as $\ell_{\text{S}}$) and the plateau value estimation (denoted as $\ell_{\text{P}}$) for each of the three methods (i.e., SP, DC, and NB). These metrics are intended to demonstrate whether incorporating the relevant external information, specifically the correct external trajectories, enhances early-stage and long-term estimation accuracy. Additionally, we define $p_{\text{s}}$ as an indicator function, taking value 1 if the $\ell_{\text{S}}$ or $\ell_{\text{P}}$ of the current method is smaller than that of SP, allowing us to compare our approach against the two alternatives.}

\begin{figure}[ht]\centering
      \includegraphics[width=.8\linewidth]{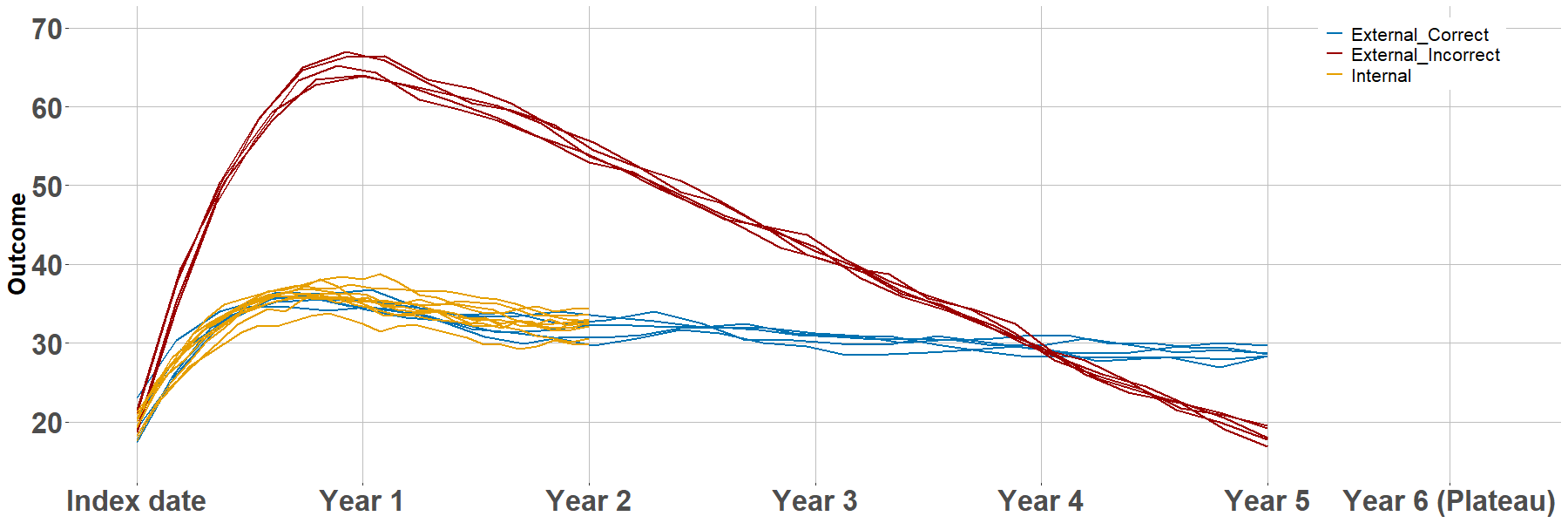}
        \caption{\label{fig: Simul12} Simulated sample data under DGP 1. The trajectories are generated with the real-data censoring patterns.}
\end{figure}

 {We also examine the empirical coverage of the $95\%$ credible intervals of each method, focusing on the frequency with which these credible intervals cover the true outcome value at 3 years post-treatment. For each method, the upper and lower bounds of the $95\%$ credible interval are determined by the 2.5th and 97.5th percentiles of the corresponding 1,000 posterior samples. The indicator $Cvr_3$ denotes whether the true outcome value at 3 years post-treatment is covered by the $95\%$ credible intervals, with a value of 1 if the true value is covered and 0 otherwise. Additionally, we let $len_3$ represent the median length of the $95\%$ credible intervals across 100 Monte Carlo replications. To facilitate method comparisons, we introduce $p_{\text{small};3}$, an indicator function that takes a value of 1 if the $len_3$ of the current method is shorter than that of SP.}

 {For our proposed approach, we further evaluate the proportions of correct and incorrect selections, denoted as $p_{[1]}$ and $p_{[2]}$, along with the preference level for favoring the correct external trajectories, denoted by $p_{\geq}$. Here, $p_{\geq}$ is an indicator function that takes a value of 1 if $p_{[1]}$ exceeds $p_{[2]}$.  These three metrics are intended to measure our approach's accuracy in selecting relevant external trajectories. Detailed definitions of these metrics can be found in Section 2.2 of the Supplementary File. The results for DGP 1 are presented in Table \ref{tab 1: setting1} and \ref{tab 2: setting1}, with reported statistics annotated in the footnotes, based on 100 Monte Carlo replications.}
\begin{table}[h]
    \centering
        \caption{Performance of the three methods in parameter estimation across 100 Monte Carlo replications for DGP 1, evaluated under varying values of $K_1$, $K_2$ and $\rho$.}
    \label{tab 1: setting1}
   \begin{tabular}{l|l|l|l|l|l|l|l|l}
   \toprule
    \toprule
  \multicolumn{3}{c|}{}& \multicolumn{2}{c|}{SP} & \multicolumn{2}{c|}{DC} & \multicolumn{2}{c}{NB} \\
    \midrule
 $\rho$ & $K_1$ & $K_2$  & $\ell_{\text{S}}$ & $\ell_{\text{P}}$ & $\ell_{\text{S}}$ & $\ell_{\text{P}}$ & $\ell_{\text{S}}$ & $\ell_{\text{P}}$ \\
\midrule
 10 & 2 & 8 & 0.32 & 0.17 & 0.51 (0.13) & 0.39 (0.00) & 0.55 (0.12) & 0.37 (0.08)\\
  & 5 & 5 & 0.17 & 0.09 & 0.34 (0.25) & 0.24 (0.09) & 0.44 (0.09) & 0.34 (0.01)\\
 & 8 & 2 & 0.13 & 0.01 & 0.20 (0.43) & 0.09 (0.27) & 0.46 (0.10) & 0.33 (0.02) \\
 30 & 2 & 8 & 0.16 & 0.04 & 0.48 (0.10) & 0.37 (0.04) & 0.32 (0.32) & 0.21 (0.22)\\
  & 5 & 5 & 0.11 & 0.01 & 0.30 (0.17) & 0.24 (0.08) & 0.25 (0.31) & 0.18 (0.07) \\
  & 8 & 2 & 0.11 & 0.01 & 0.15 (0.37) & 0.10 (0.12) & 0.25 (0.27) & 0.18 (0.10)\\
  50 & 2 & 8 & 0.20 & 0.06 & 0.46 (0.12) & 0.34 (0.09) & 0.22 (0.58) & 0.13 (0.42)\\
   & 5 & 5 & 0.13 & 0.02 & 0.28 (0.23) & 0.22 (0.14) & 0.16 (0.34) & 0.12 (0.15)\\
   & 8 & 2 & 0.11 & 0.02 & 0.14 (0.42) & 0.10 (0.16) & 0.16 (0.39) & 0.12 (0.11) \\ 
  \bottomrule
  \bottomrule
    \end{tabular}
    \newline
    \footnotesize\textsuperscript{*} Median values (with mean $p_{\text{s}}$) for $\ell_{\text{S}}$ and $\ell_{\text{P}}$
\end{table}

 {The results in Table \ref{tab 1: setting1} indicate our approach (SP) outperforms the DC approach in estimating the overall trend in DGP 1, where incorrect external trajectories differ significantly from correct external trajectories. This is evidenced by the smaller values of $\ell_{\text{S}}$ and $\ell_{\text{P}}$ and further supported by the low $p_{\text{s}}$ values (in the parentheses). When incorporating incorrect external trajectories introduces substantial error in the CCHs parameter estimation, filtering out these trajectories yields notable improvements. As a result, our approach achieves meaningful gains through selection compared to directly combining all entire external trajectories (DC).}

 {It is noteworthy that our approach does not always select all correct external trajectories, as indicated by the $p_{[1]}$ values in Table \ref{tab 2: setting1} being less than 1. This discrepancy may be derived from the inherent stochastic error in marginal likelihood approximation due to the limited samples (i.e., 4000 samples as introduced in Section \ref{sec: Model}). Such errors are more pronounced when $\rho$ is small, corresponding to a higher true marginal likelihood, as the approximation becomes less reliable with the limited samples. Consequently, our approach may occasionally exclude relevant or include irrelevant external trajectories if these actions do not significantly alter the true marginal likelihood. This also explains the selection of a small number of incorrect external trajectories, as shown by the low $p_{[2]}$ values in Table \ref{tab 2: setting1}.}

\begin{table}[h]
    \centering
        \caption{Preference level of our approach (SP) and coverage level of the true outcome value at 3 years post-treatment for the three methods across 100 Monte Carlo replications for DGP 1, with varying values of $K_1$, $K_2$ and $\rho$.}
    \label{tab 2: setting1}
   \begin{tabular}{l|l|l|l|l|l|l|l|l|l|l|l}
   \toprule
    \toprule
  \multicolumn{3}{c|}{}& \multicolumn{5}{c|}{SP} & \multicolumn{2}{c|}{DC} & \multicolumn{2}{c}{NB} \\
    \midrule
 $\rho$ & $K_1$ & $K_2$ & $p_{[1]}$ & $p_{[2]}$ & $p_{\geq}$ & $Cvr_3$ & $len_3$ & $Cvr_3$ & $len_3$ & $Cvr_3$ & $len_3$ \\
    \midrule
 10 & 2 & 8 & 0.50 & 0.13 & 0.82 & 0.71 & 2.15 & 0.00 & 2.66 (0.23) & 0.58 & 5.28 (0.00)\\
 & 5 & 5 & 0.4 & 0.2 & 0.78 & 0.79 & 1.93 & 0.29 & 2.33 (0.22) & 0.62 & 5.32 (0.00)\\
 & 8 & 2 & 0.38 & 0.00 & 0.69 & 0.95 & 0.68 & 1.00 & 1.78 (0.24) & 0.63 & 5.22 (0.00) \\
  30 & 2 & 8 & 0.50 & 0.00 & 0.83 & 0.75 & 1.20 & 0.00 & 2.15 (0.39) & 0.75 & 4.19 (0.01) \\
 & 5 & 5 & 0.60 & 0.00 & 0.90 & 0.83 & 0.83 & 0.01 & 1.86 (0.19) & 0.74 & 4.06 (0.00) \\
  & 8 & 2 & 0.50 & 0.00 & 0.88 & 0.90 & 0.80 & 0.88 & 1.49 (0.12) & 0.78 & 4.10 (0.00)\\
50 & 2 & 8 & 0.50 & 0.06 & 0.82 & 0.63 & 1.68 & 0.00 & 1.95 (0.50) & 0.84 & 3.49 (0.01)\\
 & 5 & 5 & 0.60 & 0.00 & 0.85 & 0.76 & 0.86 & 0.01 & 1.71 (0.22) & 0.82 & 3.40 (0.00)\\
 & 8 & 2 & 0.50 & 0.00 & 0.84 & 0.86 & 0.84 & 0.72 & 1.33 (0.19) & 0.88 & 3.40 (0.00)\\ 
  \bottomrule
  \bottomrule
    \end{tabular}
    \newline
    \footnotesize\textsuperscript{*} Median values for $p_{[1]}$, $p_{[2]}$; Mean values for $p_{\geq}$ and $\text{Cvr}_3$; Median value (with mean $p_{\text{s};3}$) for $len_3$.
\end{table}

 {It is important to note that with the real-data censoring patterns, none of the evaluated approaches achieves a satisfactory coverage level, suggested by the $Cvr_3$ columns being not close to 0.95 for all approaches. This limitation arises primarily because the credible interval is constructed for a time point beyond the observed timeframe of the internal trajectories. Nonetheless, our approach provides credible intervals with relatively higher coverage levels than the DC approach, and narrower intervals compared to the NB approach.}

 {Selection-wise, our approach exhibits a tendency to correctly prioritize the correct external trajectories over the incorrect ones, as indicated by the $p_{\geq}$ values, which exceeds 0.5 in most cases in Table \ref{tab 2: setting1}. This finding suggests the robustness of our approach in identifying relevant external trajectories in practice, with censoring patterns similar to those in Setting 1.}

\vspace{-5mm}
\subsection{Main findings from Settings 2 and 3}
\vspace{-2mm}

 {We begin by briefly describing the DGPs used in Settings 2 and 3. In Settings 2, two DGPs (DGPs 2 and 3) are designed to compare our approach with SSCM, where external trajectories are generated from a mixture of three generating processes, labeled as processes [1], [2] and [3]. The number of external trajectories corresponding to each process are denoted by $K_1$, $K_2$ and $K_3$, respectively. Specifically, process [1] trajectories are generated identically to the internal trajectories, whereas the trajectories from processes [2] and [3] follow distinct generating mechanisms. In DGP 2, the trajectories from processes [2] and [3] differ only slightly from the internal trajectories, allowing us to demonstrate that both our approach and SSCM can identify the correct external trajectories. However, in DGP 3, the trajectories from processes [2] and [3] are generated by slightly shifting the mean trend of process [1] vertically, causing SSCM to misidentify these trajectories as they closely resemble the internal trend. In Setting 3, additional DGPs are constructed by generating external trajectories that align with the internal ones at the early stage but diverge after the turning point. We use these DGPs to compare our approach with the DC and NB approaches, particularly demonstrating that our approach retains its ability to select the correct external trajectories under more challenging conditions.}

 {For conciseness, we summarize the key findings from Settings 2 and 3, with detailed results provided in the Supplementary File. In Setting 2, it is noteworthy that the SSCM approach primarily aims at prediction by flexibly weighting selected external trajectories following criteria that minimize the error between the synthesized external trend and the internal trend. While this approach generally yields better estimates and predictions, as demonstrated by the results in DGPs 2 and 3, it is statistically unreliable for identifying relevant external trajectories. Specifically, results from DGP 3 suggest that SSCM can be misled by certain generating processes, where it prioritizes external trajectories that coincidentally mimic the internal trend despite originating from a different generating process. Our main conclusion from DGPs 2 and 3 is that SSCM cannot effectively capture the similarity in progression trends between internal and selected external trajectories. Consequently, the subsets identified by SSCM are less informative for future studies. Our approach can achieve this by selecting based on progression trend similarity.}

 {In Setting 3, we stress-test our approach by generating external trajectories similar to the internal ones to evaluate its robustness in predictions and inferences. Compared to the DC and NB approaches, our approach demonstrates superior early-stage trend estimation and long-term value prediction. Additionally, the coverage level of the true outcome value at the 3 years post-treatment remains comparable to that of the NB approach, which serves as a benchmark in our setting. Notably, our approach benefits from leveraging the long-term information from external trajectories, enabling a reduction in credible interval length while preserving coverage.}

\vspace{-8mm}
\section{Long-term outcome after a hemophilia gene therapy}\label{sec: Real}

\vspace{-5mm}
\subsection{Outcome data before transformation}
\vspace{-2mm}

\begin{figure}[ht] \centering
      \includegraphics[width=.8\linewidth]{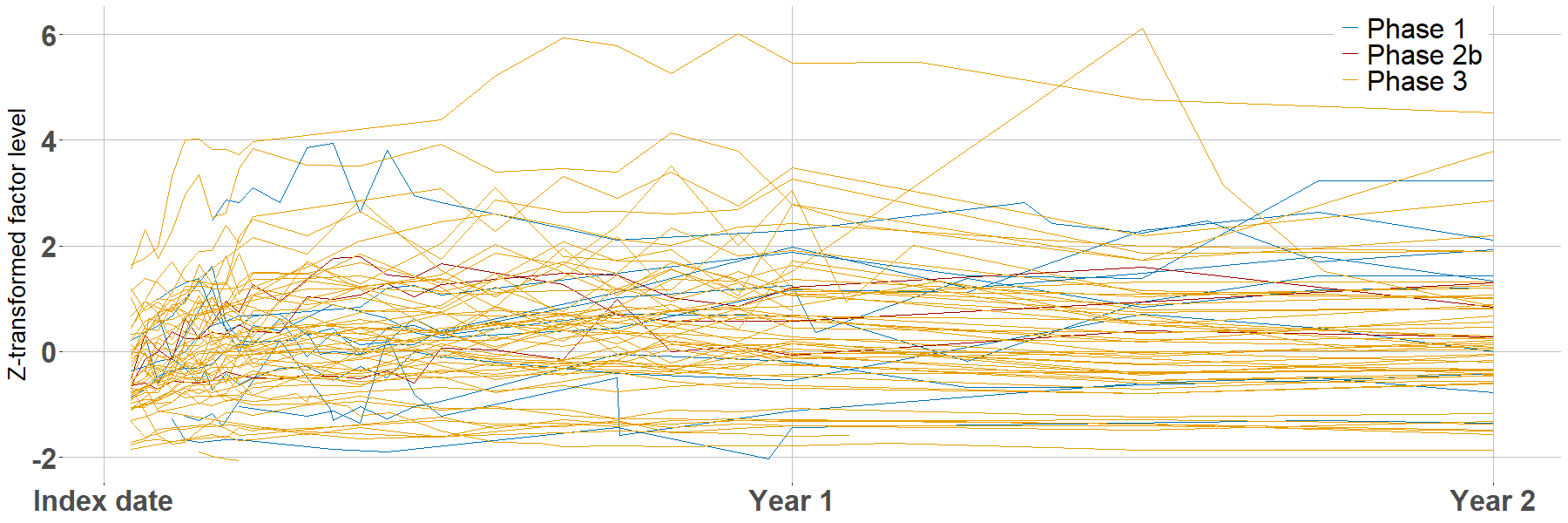}
        \caption{\label{fig: ZTrans} The trajectories of z-transformed factor level up to the 2nd year after index date.}
\end{figure}

 {Motivated by the data introduced in Section \ref{sec: intro}, our analysis begins with a z-transformation  {of endogenous FIX levels} in both the external and internal studies. This transformation uses the respective mean values and standard deviations calculated from the data up to week 10 visit post-treatment of each study. The standardized data, shown in Figure \ref{fig: ZTrans}, reveals a closer alignment between the external and internal data following the transfromation.}

 {To further justify the use of the z-transformation from a modeling perspective, it is important to note that this procedure standardizes both the initial values and the corresponding variances across studies. Specifically, under our model specification \eqref{Model}, the z-transformation forces the first entry of $\beta^*$ to be zero and normalizes both $\sigma_0^2$ and $\sigma_1^2$ to be one for both studies. This normalization potentially increases the proportion of external data that can be effectively borrowed by the internal data within Model \eqref{Model}. Moreover, this transformation allows our selection procedure to focus specifically on differences in trend progression rather than baseline disparities. This provides more interpretable results, as the external trajectories selected for borrowing will be those that exhibit similarity in trend with the internal trajectories.}

\vspace{-5mm}
\subsection{Primary analysis}
  \vspace{-2mm}
  
 {We apply BASE method to those clinical trials mentioned above, in order to predict factor levels of adult patients with Hemophilia B who received Etranacogene dezaparvovec in the long-term. To obtain posterior samples of external subsets, we run 100 chains using the marginal likelihood approximation and the same hyper-parameter settings described in Section \ref{sec: Model}. For each chain, we draw 1,000 posterior samples of $\mathcal{C}$ from the categorical distribution based on the estimated marginal likelihood (Step (1) in Section \ref{sec: Model}). The most representative external subset, $\hat{\mathcal{C}}$, is selected according to the definition in  \eqref{Model} using these 1,000 posterior samples. Next, we obtain posterior samples of the CCHs parameter $\boldsymbol{\zeta}$ (defined in \eqref{jointParams}), conditional on the most representative external subset $\hat{\mathcal{C}}$ (Step (2)). To reduce within-chain auto-correlation, we retain one sample every 10 iterations from the 10,000 MCMC iterations after 1,000 burn-in samples, resulting in a final collection of 1,000 posterior samples of $\boldsymbol{\zeta}$ for each chain. Our final analysis is based on 100 samples of $\hat{\mathcal{C}}$ and 100,000 samples of $\boldsymbol{\zeta}$ conditional on $\hat{\mathcal{C}}$, by pooling 1,000 samples across the 100 chains.}

\begin{figure}[ht]
  \centering
      \includegraphics[width=.8\linewidth]{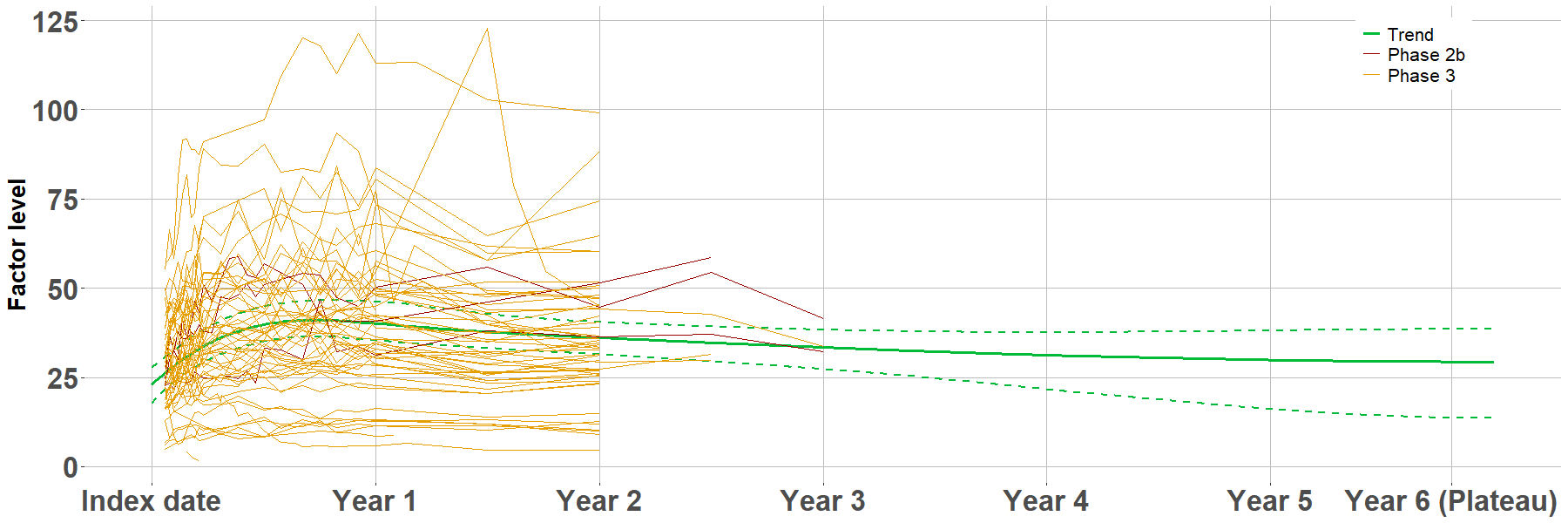}
        \caption{\label{fig: Real} The spaghetti plot of outcomes from internal study and the estimated trend of outcome using BASE. The green dashed curves represent the $95\%$ credible interval.}
\end{figure}

 {We begin by presenting the result of the estimated mean factor levels using BASE method. Based on the 100,000 pooled samples, we conclude that the median value of the initial factor level at index date is 22.93, with a standard deviation of 2.53. The median value of the plateau factor level (i.e., the stabilized factor level at 6 years post-treatment) is 29.37, with a standard deviation of 6.57. In Figure \ref{fig: Real}, we visualize the estimated factor level trend using the posterior median of $\boldsymbol{\theta}$, as defined in \eqref{Model}. The figure also includes a $95\%$ credible region, banded by the $2.5\%$ and $97.5\%$ quantiles of the 100,000 pooled samples. Additionally, the detailed annual results, along with the $95\%$ credible region for the factor level trend, are presented in Table \ref{tab: annual}.}
\begin{table}[h]
    \centering
        \caption{The posterior median (MD), lower bound (LB), upper bound (UB) of the $95\%$ credible interval, and standard deviation (SD) for the annual post-treatment factor levels.}
    \label{tab: annual}
   \begin{tabular}{l|l|l|l|l|l|l|l}
   \toprule
    \toprule
& Year 0 & Year 1 & Year 2  & Year 3 & Year 4 & Year 5 & Year 6\\
    \midrule
MD & 22.93 & 40.10 & 36.11 & 33.42 & 31.23 & 29.89 & 29.37\\
LB & 17.90 & 35.30 & 31.48 & 27.29 & 21.67 & 16.13 & 13.57\\
UB & 27.79 & 46.17 & 40.46 & 38.34 & 37.66 & 38.10 & 38.51\\
SD & 2.53 & 2.74 & 2.31 & 2.80 & 4.11 & 5.77 & 6.57\\
  \bottomrule
  \bottomrule
    \end{tabular}
\end{table}

 {Prediction-wise, we obtain a 95\% credible interval for the factor level at 3-years post-treatment of $[27.29,38.34]$, which includes the actual median factor level (36.0) reported in the Phase 3 study (available at \url{https://ash.confex.com/ash/2023/webprogram/Paper187624.html}). This provides further evidence that the $95\%$ credible interval produced by the BASE approach has a plausible coverage probability in practice. We also investigate to which extent each external trajectory is preferred by the internal dataset. Specifically, we report the posterior probability of selection, which is given by averaging $Z_{\hat{\mathcal{C}}}$ over the 100 $\hat{\mathcal{C}}$ obtained from the 100 chains. The entry-wise mean and standard deviation (SD) of $Z_{\hat{\mathcal{C}}}$ over the 100 chains are given in Table \ref{tab: pseudoIDs} with pseudo IDs.}
\begin{table}[h]
    \centering
        \caption{The entry-wise mean and standard deviation (SD) of $Z_{\hat{\mathcal{C}}}$ over 100 chains.}
    \label{tab: pseudoIDs}
   \begin{tabular}{l|l|l|l|l|l|l|l|l|l|l}
   \toprule
    \toprule
Pseudo ID & 1 & 2 & 3 & 4 & 5 & 6 & 7 & 8 & 9 & 10\\
    \midrule
Mean & 0.01 & 0.24 & 0.61 & 0.38 & 0.95 & 0.85 & 1.00 & 0.07 & 1.00 & 0.88\\
SD & 0.10 & 0.43 & 0.49 & 0.49 & 0.22 & 0.36 & 0.00 & 0.26 & 0.00 & 0.33\\
  \bottomrule
  \bottomrule
    \end{tabular}
\end{table}

 {Additionally, the median (MD) and interquartile range (IQR) of the selected subjects, calculated as $\sum_{i \in Z_{\hat{\mathcal{C}}}} i/|Z_{\hat{\mathcal{C}}}|$ across the 100 chains, are 0.6 and 0.2, respectively. This result indicates that more than half of the external trajectories can be leveraged for long-term inference, based on the similarity in early-stage trends between external and internal trajectories after Z-transformation.  {From a scientific perspective, the proportion of external trajectory utilization suggests that the mechanism of action of gene therapies in affecting factor level trend may remain relatively consistent between AMT-060 (external data) and etranacogene dezaparvovec, which differ by 1 single amino acid resulting in naturally occurring highly active FIX Padua variant (FIX-R338L).} Furthermore, the result provides statistical evidence for the hypothesis that this evolution (1 single amino acid-difference) is more influential on the baseline and plateau factor levels post-treatment, while the overall trend information is likely preserved. Moreover, by identifying the study subjects that are frequently selected, potential external sub-populations exhibiting similar trends to the internal population prior to the follow-up endpoint can be uncovered. These insights could be valuable for guiding future research endeavors. More discussions on the results in Table \ref{tab: pseudoIDs} are provided in Section 3 of the Supplementary File.}

\vspace{-5mm}
\subsection{Secondary analysis}\label{sec: secondary}
\vspace{-2mm}

 {We conduct a two-fold secondary analysis. First, we compare the results given by our proposed selection procedure (SP) with those from the direct combination (DC) and no-information-borrowing (NB) from the external dataset. Second, we evaluate the robustness of our model by considering different prior hyper-parameter settings.}

\begin{figure}[ht]
  \centering
      \includegraphics[width=\linewidth]{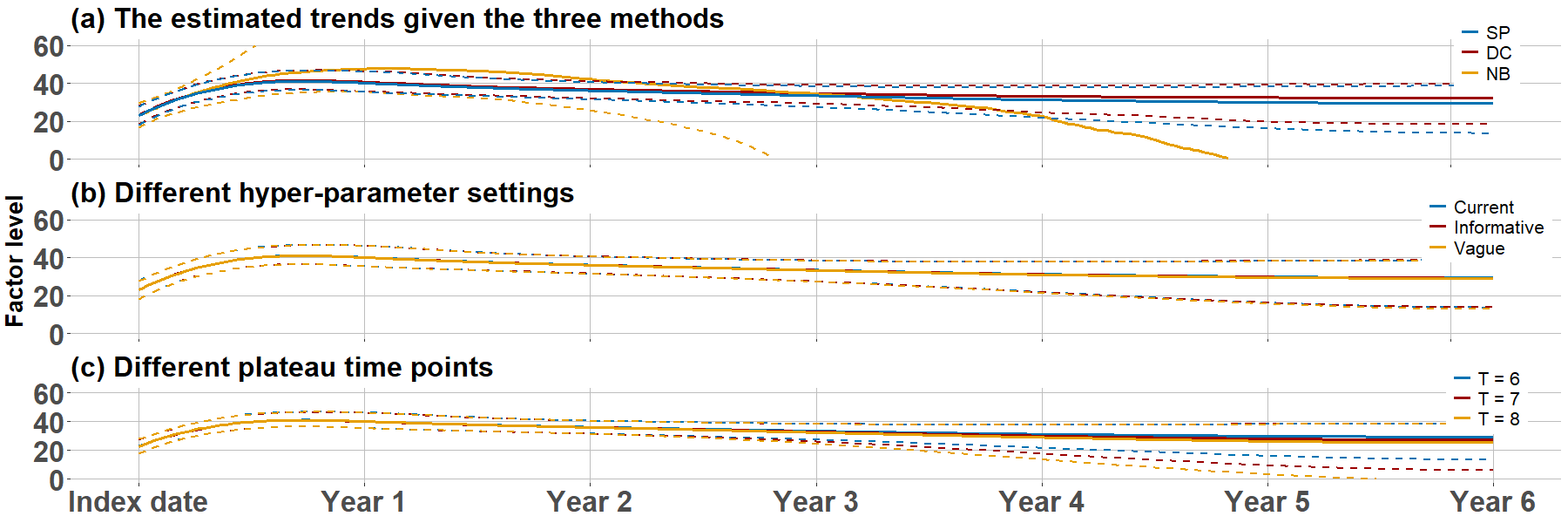}
        \caption{\label{fig: ThreeTrends} (a) The estimated trends (median) along with their $95\%$ credible intervals obtained by the three methods; (b) The sensitivity analysis results under different hyper-parameter settings; (c) The sensitivity analysis results with different plateau time points.}
\end{figure}

 {Figure \ref{fig: ThreeTrends}(a) and (b) display the estimated trends along with their credible intervals obtained by the three methods under the same MCMC settings. Additionally, Table \ref{tab: CIresults} provides details on the estimated annual factor levels and the turning point. Both the proposed selection procedure (SP) and the direct combination (DC) methods yield similar results in the early stage, particularly before 2 years post-treatment, largely due to the dominance of the internal trajectories. In contrast, the curve obtained without external information (NB) exhibits high variance, even in the early stage, due to the absence of long-term data. Specifically, this is because accurately estimating the CCHs parameters requires data observed beyond $\alpha$. Without borrowing external information, this estimation is challenging, as most internal trajectories are censored at 2 years post-treatment, while $\alpha$ has a median of a median of 2.66 years and a 95$\%$ credible interval of $[1.12,3.75]$ years. This is further reflected in the large standard deviations for the estimated factor levels in the fourth and fifth years, as shown in Table \ref{tab: CIresults}. It is important to emphasize that the results for the NB approach in Table \ref{tab: CIresults} are primarily intended to illustrate the excessively wide confidence intervals. Notably, by definition, it is not possible to observe negative factor level values in practice by its definition. Both our method (SP) and the direct combination method (DC) also produce similar estimated trends and plateau values. Two potential factors likely contribute to this similarity. First, our method selects approximately $60\%$ of the external trajectories for inference, capturing the majority of the external information. As a result, any differences in the inferential results between the two methods are expected to be small, given this high selection proportion. Second, the generating processes for both the selection and direct combination strategies are likely inherently similar. In other words, the discarded external trajectories are excluded mainly due to the stochastic nature of the sampling procedure and do not differ significantly from those that are selected. This scenario is more likely when the temporal correlation $\rho$ is high, as indicated in Table \ref{tab: CIresults}. For example, two observations separated by a 1-year time interval would have a temporal correlation of 0.41 if $\rho$ is 1.12 (i.e., the median value of $\rho$ for the proposed selection method).} 

\begin{table}[h]
    \centering
        \caption{The posterior median value (MD), the lower (LB) and upper (UB) bounds of the 95 $\%$ credible interval for the annual factor levels, the turning point $\alpha$ and the temporal correlation $\rho$.}
    \label{tab: CIresults}
   \begin{tabular}{l|l|l|l|l|l|l|l|l|l}
   \toprule
    \toprule
\multicolumn{2}{c|}{\textbf{Method}} & Year 0 & Year 1  & Year 2 & Year 3 & Year 4 & Year 5 & $\alpha$ & $\rho$ \\
    \midrule
\textbf{Selection} & MD & 22.93 & 40.10 & 36.11 & 33.42 & 31.23 & 29.89 & 1.29 & 1.12\\
                   & LB & 17.90 & 35.30 & 31.48 & 27.29 & 21.67 & 16.13 & 1.01 & 0.89\\
                   & UB & 27.79 & 46.17 & 40.46 & 38.34 & 37.66 & 38.10 & 2.08 & 1.41\\
\midrule
\textbf{Combination} & MD & 23.20 & 40.68 & 36.90 & 34.70 & 33.17 & 32.50 & 1.32 & 1.03\\
                     & LB & 18.20 & 35.59 & 32.02 & 29.22 & 24.50 & 19.61 & 1.01 & 0.85\\
                     & UB & 28.04 & 46.20 & 41.06 & 39.05 & 38.60 & 39.23 & 2.27 & 1.34\\
\midrule
\textbf{No-borrow} & MD & 22.94 & 47.58 & 42.15 & 34.53 & 22.63 & -4.24 & 2.66 & 1.60\\
                   & LB & 16.58 & 35.32 & 25.77 & -9.37 & -90.29 & -204.47 & 1.12 & 1.21\\
                   & UB & 29.62 & 96.06 & 173.33 & 213.74 & 197.22 & 221.27 & 3.75 & 1.72\\
  \bottomrule
  \bottomrule
    \end{tabular}
\end{table}

 {To investigate the robustness of our results, we perform a sensitivity analysis using alternative hyper-parameter values $\nu_0,\Psi_0,a_0,b_0$ defined in \eqref{Model} and plateau time points ($T$). Specifically, we consider a more informative prior (Informative) and a vaguer prior (Vague) in comparison to the current setting, and larger plateau time points ($T=7$ and $8$). Further details on these setting are provided in Table 1 of the Supplementary File. As shown in Figure \ref{fig: ThreeTrends}(b) and (c), the results exhibit a general agreement in the median trends and, to some extent, the credible bands. This confirms that the results are robust and not highly sensitive to the choice of the hyper-parameter values and plateau time points. It is important to note that while the trend estimates given different $T$ values are largely similar, Figure \ref{fig: ThreeTrends}(c) indicates slight evidence that increasing $T$ leads to wider credible interval. This finding is further supported by Figure 2 in the Supplementary File, which shows similar results when using $T = 8,~9,~10$. One possible explanation is that additional long-term observations per trajectory are necessary to improve the accuracy of long-term inferences. Specifically, with a longer plateau time, the information beyond $\alpha$ is used to infer behavior over a longer time frame, which can reduce the average effect sample size for these longer-term estimates. It is important to note that increasing plateau time $T$ has a greater impact on long-term inference, while early-stage inference remains relatively unaffected. This is evident from the negligible changes in the credible intervals prior to 2 years post-treatment, as $T$ varies. This is likely because most internal trajectories, which inform early-stage trend inferences, are censored at 2 years post-treatment. Therefore, we recommend choosing a relatively small value for $T$ given the current internal data collection. A plateau time value of $T = 6$ or $7$ seems plausible.}

\vspace{-8mm}
\section{Discussion}\label{sec: Discuss}
\vspace{-5mm}
In this paper, we propose BASE, a Bayesian data selection procedure to enhance understanding of the long-term effects of gene therapies. Our work stands out as the first work that selects subsets from external datasets with theoretical justification.  {Specifically, our method provides a robust framework for justifying the procedure of selecting external data subsets in longitudinal studies with well-defined asymptotic properties. The selected external trajectories are clearly interpreted as being generated in a manner consistent with the internal trajectories. Our method offers deeper scientific insights and has the potential to inspire future clinical investigations, particularly when compared to existing methods. Notably, our method is not confined to hemophilia studies, it can be applied broadly to analyze data from any longitudinal study with similar censoring patterns.} Of note, our work does not intend to provide a definite answer on the long term effectiveness of Etranacogene dezaparvovec, but rather showcase a conceptually novel procedure that manages to refine the long-term efficacy inference. As more data from ongoing clinical trials are available, the predicted values may be updated with the same Bayesian approach. For instance, the factor level on the 15-th year post-treatment can be predicted if the trials continue, rendering more available durability data. 

Despite these achievements, the current model  {leaves several unanswered questions for future research interests}. Firstly, the current MCMC sampling scheme relies on the proposal of the external subset $\mathcal{C}$. Although the current proposal has a satisfactory empirical performance in simulations and real data analysis, e.g., the standard deviation of the external proportion is moderate compared to the point estimate, it remains unexplored whether this proposal is effective when the external data size is large, for instance, over 100 observations. Intuitively, a more effective proposal should be devised in such scenarios, given that the number of possible external subsets increases at an exponential order with the external data size. However, since the size of the hemophilia external data is manageable (10 observations), this question is beyond the scope of this work. Additionally, the estimation of the marginal likelihood can be further investigated. The current strategy \citep{pajor2017estimating} is proven to be effective under the parametric setting, whilst there is limited evidence that this method provides efficient Monte Carlo estimation to the marginal likelihood under the nonparametric setting. Moreover, when the sample size is sufficiently large, one can consider data-driven strategies that select models from a family of parametric models. Criteria such as the Akaike Information Criterion (AIC; \citet{akaike1974new}) 
and marginal likelihood can be employed to explore if any alternatives are superior in capturing the mean trend compared to the Hermite spline specification. Another intriguing future topic is how to embed the spirit of data selection into non-likelihood problems, e.g., treatment effects that are given by estimating equations and non-probability samples \citep{yang2020doubly,chen2020doubly}. In cases where likelihood functions are not well-defined, it would be impossible to use the marginal likelihood as a guideline, rendering our current pipeline unsuitable. Furthermore, it is also challenging to determine whether the internal estimation benefits from the external data selection.  {This approach relies on the key assumption that similar initial trajectories can predict similar long-term outcomes. Extending our method to incorporate additional covariate information for data selection and data integration is another important future work direction.}

\section{Proofs}\label{appenx: proofExamp1}
  {To begin with, we reiterate the notations and definitions of Theorem \ref{postExample} provided in the main article. Suppose the external and internal trajectories, $\{X_{ik}\}_{k=1}^{T_i^{(x)}}$ and $\{Y_{jl}\}_{l=1}^{T_j^{(y)}}$, for $i=1,\ldots,N_0$ and $j = 1,\ldots,N_1$, are both i.i.d. between and within subject. For clarity and succinctness, we slightly abuse the notations by using $\{X_{ik}\}_{k=1}^{T_i^{(x)}}$ to represent the first through $T_i^{(x)}$-th observations collected for the $i$-th subject, ordered temporally. Specifically, we assume,
\begin{equation}\label{iidDist}
    \begin{split}
        & X_{11},\ldots,X_{1T_1^{(x)}},\ldots,X_{N_01},\ldots,X_{N_0T_{N_0}^{(x)}}\stackrel{i.i.d.}{\sim} p_{\theta_0},\\
        & Y_{11},\ldots,Y_{1T_1^{(y)}},\ldots,Y_{N_11},\ldots,Y_{N_1T_{N_1}^{(y)}}\stackrel{i.i.d.}{\sim} p_{\theta_1},
    \end{split}
\end{equation}
where $p_{\theta}$ is a known  probability density function with parameter $\theta$. Additionally, we let $N_0^*\equiv \sum_{i=1}^{N_0} T_i^{(x)}$ and $N_1^*\equiv \sum_{j=1}^{N_1} T_j^{(y)}$ respectively be the total observations of the external and internal trajectories. We prove Theorem \ref{postExample} with the following assumptions,
\begin{enumerate}
    \item[(A1)] It holds $\pi(p_{\theta}: \|p_{\theta}-p_{\theta_i}\|_1\lesssim \epsilon_{N_i^*})\propto \epsilon_{N_i^*}^d$ for $i = 0,1$, where $\|\cdot\|_1$ denotes the $\ell_1$ distance, $\pi$ denotes a prior density function assigned to $\theta$ and $d$ refers to the dimensionality of $\theta$, and $a\gtrsim b$ denotes $a\geq C^*\times b$ for a positive constant $C^*$.
    \item[(A2)] Assume
    $$\mathbbm{E}_{\left\{X_{ik}\right\}_{\substack{i=1,k=1}}^{N_0,T_i^{(x)}}}\left[\pi\left(p_{\theta}: \|p_{\theta}-p_{\theta_0}\|_1\geq K \epsilon_{N_0^*}\mid \left\{X_{ik}\right\}_{\substack{i=1,k=1}}^{N_0,T_i^{(x)}}\right)\mathbbm{1}_{E_{N_0^*}}\right]\leq \exp\{-K^*\times N_0^*\epsilon_{N_0^*}^2\},$$ and $$\mathbbm{E}_{\left\{Y_{jl}\right\}_{\substack{j=1,l=1}}^{N_1,T_j^{(y)}}}\left[\pi\left(p_{\theta}: \|p_{\theta}-p_{\theta_1}\|_1 \geq K \epsilon_{N_1^*}\mid \left\{Y_{jl}\right\}_{\substack{j=1,l=1}}^{N_1,T_j^{(y)}}\right)\mathbbm{1}_{F_{N_1^*}}\right]\leq \exp\{-K^*\times N_1^*\epsilon_{N_1^*}^2\},$$ where $\mathbbm{E}$ denotes the expectation with respect the product density given by the subscript, $E_{N_0^*}$ and $F_{N_1^*}$ are the events defined in \eqref{(A5)}, $\epsilon_n^2 = \frac{M_n}{n}$ is the posterior contraction rate, $K^*$ and $K$ are two positive constants with $K$ depending on $K^*$ and $M_n = o(\log(n))$ approaches to $\infty$ as $n$ increases.
    \item[(A3)] The external and internal data sizes satisfy $\log(N_0^*) \leq C' \times M_{N_1^*}$ for a positive $C'$.
    \item[(A4)] Define $B_{n,i} = \left\{p_{\theta}: -\int p_{\theta_i}(x)\log\left(\frac{p_{\theta}(x)}{p_{\theta_i}(x)}\right)dx\leq \epsilon_n^2, \int p_{\theta_i}(x)\left(\log\left(\frac{p_{\theta}(x)}{p_{\theta_i}(x)}\right)dx\right)^2\leq \epsilon_n^2\right\}$, for $i=0,1$, it follows that
    $$\left\{p_{\theta}: \|p_{\theta}-p_{\theta_i}\|_1\leq C\times \epsilon_{N_i^*}\right\}\subseteq B_{N_i^*,i},$$
    for a positive constant $C$.
\end{enumerate}
We further let $E_{N_0^*}$ denote the event that happens when the following inequality holds
\begin{equation}\label{(A5)}
    \begin{split}
        & \int_{B_{N_0,0}} {\prod}_{\substack{i=1, k=1}}^{N_0,T_i^{(x)}}\frac{p_{\theta}(X_{ik})}{p_{\theta_0}(X_{ik})}\frac{d\pi(p_{\theta})}{\pi(B_{N_0,0})}\geq \exp\{-(1 + K_0)M_{N_0^*}\},
    \end{split}
\end{equation}
where $K_0 > 0$ is a universal constant. Event $F_{N_1^*}$ is defined analogously.}

\begin{theorem}\label{postExample}
Suppose external and internal trajectories are generated following \eqref{iidDist} and Assumptions (A1) - (A4) are satified. If it holds $\|p_{\theta_0}-p_{\theta_1}\|_1\lesssim \epsilon_{N_0^*}$, the expected Bayes factor can be controlled by
\begin{equation}\label{postExampleEq1}
    \begin{split}
        & \mathbbm{E}_{\left\{X_{ik}\right\}_{\substack{i=1,k=1}}^{N_0,T_i^{(x)}},\left\{Y_{jl}\right\}_{\substack{j=1,l=1}}^{N_1,T_j^{(y)}}}\left(\frac{\int \prod_{\substack{j=1, l=1}}^{N_1,T_j^{(y)}} p_{\theta}(Y_{jl})d\pi(p_{\theta})}{\int_{\|p_{\theta}-p_{\theta_0}\|_1\leq K \epsilon_{N_0^*}} \prod_{\substack{j=1, l=1}}^{N_1,T_j^{(y)}} p_{\theta}(Y_{jl})d\pi(p_{\theta}\mid \left\{X_{ik}\right\}_{\substack{i=1,k=1}}^{N_0,T_i^{(x)}})}\mathbbm{1}_{E_{N_0^*}}\mathbbm{1}_{F_{N_1^*}}\right) \\
        & \lesssim \exp\{(1 + K_0)M_{N_0^*} - \frac{d}{2}\log(N_0^*)\}.
    \end{split}
\end{equation}
On the other hand, if it holds $\|p_{\theta_0}-p_{\theta_1}\|_1\gtrsim \epsilon_{N_0^*}\times \psi_{N_0^*}$ instead, for $\psi_{N_0^*}$ diverging to $\infty$ at any rates, it yields that
\begin{equation}\label{postExampleEq2}
    \begin{split}
        & \mathbbm{E}_{\left\{X_{ik}\right\}_{\substack{i=1,k=1}}^{N_0,T_i^{(x)}},\left\{Y_{jl}\right\}_{\substack{j=1,l=1}}^{N_1,T_j^{(y)}}}\left(\frac{\int_{\|p_{\theta}-p_{\theta_0}\|_1\leq K \epsilon_{N_0^*}} \prod_{\substack{j=1, l=1}}^{N_1,T_j^{(y)}} p_{\theta}(Y_{jl})d\pi(p_{\theta}\mid \left\{X_{ik}\right\}_{\substack{i=1,k=1}}^{N_0,T_i^{(x)}})}{\int \prod_{\substack{j=1, l=1}}^{N_1,T_j^{(y)}} p_{\theta}(Y_{jl})d\pi(p_{\theta})}\mathbbm{1}_{E_{N_0……*}}\mathbbm{1}_{F_{N_1……*}}\right)\\
        & \lesssim \exp\{(1+K_0)M_{N_0^*}+\frac{d}{2}\log(N_0^*) - K^*\times M_{N_1^*}\},
    \end{split}
\end{equation}
where $K^*$ satisfies $K^* > 1 + K_0 + \frac{d}{2} \times C'$.

\begin{proof}
    For brevity, since both external and internal observations are i.i.d. samples, without loss of generality, we can reform the generating processes in \eqref{iidDist} into,
    \begin{equation}\label{iidDistModify}
        \begin{split}
            & X_1,\ldots,X_{N_0^*}\stackrel{i.i.d.}{\sim} p_{\theta_0},\\
            & Y_1,\ldots,Y_{N_1^*}\stackrel{i.i.d.}{\sim} p_{\theta_1}.\\
        \end{split}
    \end{equation}
    Our goal is then to prove
    \begin{equation}\label{postExampleEq1Modified}
    \begin{split}
        & \mathbbm{E}_{\left\{X_i\right\}_{i=1}^{N_0^*},\left\{Y_j\right\}_{j = 1}^{N_1^*}}\left(\frac{\int \prod_{\substack{j=1}}^{N_1^*} p_{\theta}(Y_{j})d\pi(p_{\theta})}{\int_{\|p_{\theta}-p_{\theta_0}\|_1\leq K \epsilon_{N_0^*}} \prod_{\substack{j=1}}^{N_1^*} p_{\theta}(Y_{j})d\pi(p_{\theta}\mid \left\{X_{i}\right\}_{i=1}^{N_0^*})}\mathbbm{1}_{E_{N_0^*}}\mathbbm{1}_{F_{N_1^*}}\right) \\
        & \lesssim \exp\{(1 + K_0)M_{N_0^*} - \frac{d}{2}\log(N_0^*)\},
    \end{split}
    \end{equation}
    and
    \begin{equation}\label{postExampleEq2Modified}
    \begin{split}
        & \mathbbm{E}_{\left\{X_{i}\right\}_{i=1}^{N_0^*},\left\{Y_{j}\right\}_{j=1}^{N_1^*}}\left(\frac{\int_{\|p_{\theta}-p_{\theta_0}\|_1\leq K \epsilon_{N_0^*}} \prod_{\substack{j=1}}^{N_1^*} p_{\theta}(Y_{j})d\pi(p_{\theta}\mid \left\{X_{i}\right\}_{\substack{i=1}}^{N_0^*})}{\int \prod_{\substack{j=1}}^{N_1^*} p_{\theta}(Y_{jl})d\pi(p_{\theta})}\mathbbm{1}_{E_{N_0……*}}\mathbbm{1}_{F_{N_1……*}}\right)\\
        & \lesssim \exp\{(1+K_0)M_{N_0^*}+\frac{d}{2}\log(N_0^*) - K^*\times M_{N_1^*}\}.
    \end{split}
    \end{equation}
    Note by the definition of $\pi(p_{\theta}\mid \left\{X_i\right\}_{i=1}^{N_0})$ provided in the main article, it holds that
    \begin{equation}\label{postDen1}
        \begin{split}
            & \frac{\int \prod_{j=1}^{N_1^*} p_{\theta}(Y_j)d\pi(p_{\theta})}{\int_{\|p_{\theta}-p_{\theta_0}\|_1\leq K\epsilon_{N_0^*}} \prod_{j=1}^{N_1^*} p_{\theta}(Y_j)d\pi(p_{\theta}\mid \left\{X_i\right\}_{i=1}^{N_0^*})}\\
            = & \frac{\int \prod_{i=1}^{N_0^*} p_{\theta}(X_i)d\pi(p_{\theta})\int \prod_{j=1}^{N_1^*} p_{\theta}(Y_j)d\pi(p_{\theta})}{\int_{\|p_{\theta}-p_{\theta_0}\|_1\leq K \epsilon_{N_0^*}} \prod_{i=1}^{N_0^*}p_{\theta}(X_i) \prod_{j=1}^{N_1^*} p_{\theta}(Y_j) d\pi(p_{\theta})},\\
            = & \frac{\int \prod_{i=1}^{N_0^*} p_{\theta}(X_i)d\pi(p_{\theta})}{\int_{\|p_{\theta}-p_{\theta_0}\|_1\leq K \epsilon_{N_0^*}} \prod_{i=1}^{N_0^*}p_{\theta}(X_i)d\pi(p_{\theta}\mid \{Y_j\}_{j=1}^{N_1^*})},
        \end{split}
    \end{equation}
    and
    \begin{equation}\label{postDen2}
        \begin{split}
            & \frac{\int_{\|p_{\theta}-p_{\theta_0}\|_1\leq K\epsilon_{N_0^*}} \prod_{j=1}^{N_1^*} p_{\theta}(Y_j)d\pi(p_{\theta}\mid \left\{X_i\right\}_{i=1}^{N_0^*})}{\int \prod_{j=1}^{N_1^*} p_{\theta}(Y_j)d\pi(p_{\theta})}\\
            = & \frac{\int_{\|p_{\theta}-p_{\theta_0}\|_1\leq K\epsilon_{N_0^*}} \prod_{i=1}^{N_0^*}p_{\theta}(X_i)d\pi(p_{\theta}\mid \{Y_j\}_{j=1}^{N_1^*})}{\int \prod_{i=1}^{N_0^*} p_{\theta}(X_i)d\pi(p_{\theta})}.
        \end{split}
    \end{equation}
    We proceed by proving the first part. Suppose it holds that $\|p_{\theta_0}-p_{\theta_1}\|_1\lesssim \epsilon_{N_0^*}$, the following upper bounded can be derived,
    \begin{equation}\label{1stPart1}
        \begin{split}
            & \mathbbm{E}_{\{X_i\}_{i=1}^{N_0^*},\{Y_j\}_{j=1}^{N_1^*}}\left(\frac{\int \prod_{i=1}^{N_0^*} p_{\theta}(X_i)d\pi(p_{\theta})}{\int_{\|p_{\theta}-p_{\theta_0}\|_1\leq K \epsilon_{N_0^*}} \prod_{i=1}^{N_0^*}p_{\theta}(X_i)d\pi(p_{\theta}\mid \{Y_j\}_{j=1}^{N_1^*})}\mathbbm{1}_{E_{N_0^*}}\mathbbm{1}_{F_{N_1^*}}\right)\\
            & \stackrel{(a)}{\propto} \mathbbm{E}_{\{X_i\}_{i=1}^{N_0^*},\{Y_j\}_{j=1}^{N_1^*}}\left(\frac{\int_{\|p_{\theta}-p_{\theta_0}\|_1\leq K\epsilon_{N_0^*}} \frac{\prod_{i=1}^{N_0^*} p_{\theta}(X_i)}{\prod_{i=1}^{N_0^*} p_{\theta_0}(X_i)}d\pi(p_{\theta})}{\int_{\|p_{\theta}-p_{\theta_0}\|_1\leq K\epsilon_{N_0^*}} \frac{\prod_{i=1}^{N_0^*}p_{\theta}(X_i)}{\prod_{i=1}^{N_0^*}p_{\theta_0}(X_i)}d\pi(p_{\theta}\mid \{Y_j\}_{j=1}^{N_1^*})}\mathbbm{1}_{E_{N_0^*}}\mathbbm{1}_{F_{N_1^*}}\right),\\
            & \stackrel{(b)}{\lesssim} \exp\{(1 + K_0)M_{N_0^*}\} \mathbbm{E}_{\{X_i\}_{i=1}^{N_0^*}}\left(\int_{\|p_{\theta}-p_{\theta_0}\|_1\leq K\epsilon_{N_0^*}} \frac{\prod_{i=1}^{N_0^*} p_{\theta}(X_i)}{\prod_{i=1}^{N_0^*} p_{\theta_0}(X_i)}d\pi(p_{\theta})\right),\\
            & \lesssim \exp\{(1 + K_0)M_{N_0^*}\} \pi(p_{\theta}: \|p_{\theta}-p_{\theta_0}\|_1\leq K \epsilon_{N_0^*}),\\
            & \lesssim \exp\{(1 + K_0)M_{N_0^*}\}\epsilon_{N_0^*}^d,
        \end{split}
    \end{equation}
    where inequality (a) is given by Assumption (A2) that $$\mathbbm{E}_{\{X_i\}_{i=1}^{N_0^*}}\pi\left(p_{\theta}: \|p_{\theta}-p_{\theta_0}\|_1\leq K \epsilon_{N_0^*}\mid \{X_i\}_{i=1}^{N_0^*}\right) \to 1,$$ and the definition that
    $$\int {\prod}_{i=1}^{N_0^*} p_{\theta}(X_i)d\pi(p_{\theta}) = \frac{\int_{\|p_{\theta}-p_{\theta_0}\|_1\leq K\epsilon_{N_0^*}} \prod_{i=1}^{N_0^*} p_{\theta}(X_i)d\pi(p_{\theta})}{\pi\left(p_{\theta}: \|p_{\theta}-p_{\theta_0}\|_1\leq K \epsilon_{N_0^*}\mid \{X_i\}_{i=1}^{N_0^*}\right)}.$$ Inequality (b) is given by applying Lemma 8.1 of \citet{ghosal2000convergence}, with $K_0$ being any positive value. Note that the conditions of Lemma 8.1 are met because Assumption (A4), $\|p_{\theta_0}-p_{\theta_1}\|_1\lesssim \epsilon_{N_0^*}$ and $E_{\{Y_j\}_{j=1}^{N_1^*}}\pi\left(p_{\theta}: \|p_{\theta}-p_{\theta_0}\|_1 \leq K \epsilon_{N_0^*}\mid \{Y_j\}_{j=1}^{N_1^*}\right)\to 1$. The last display of \eqref{1stPart1} approaches 0 by Assumption (A2) and (A3).

    To prove the second part, we apply Lemma 8.1 of \citet{ghosal2000convergence} on the denominator, which gives,
    \begin{equation}\label{2ndPart1}
        \begin{split}
            & \mathbbm{E}_{\{X_i\}_{i=1}^{N_0^*},\{Y_j\}_{j=1}^{N_1^*}}\left(\frac{\int_{\|p_{\theta}-p_{\theta_0}\|_1\leq K \epsilon_{N_0^*}} \prod_{i=1}^{N_0^*}p_{\theta}(X_i)d\pi(p_{\theta}\mid \{Y_j\}_{j=1}^{N_1^*})}{\int \prod_{i=1}^{N_0^*} p_{\theta}(X_i)d\pi(p_{\theta})}\mathbbm{1}_{E_{N_0^*}}\mathbbm{1}_{F_{N_1^*}}\right)\\
            & \lesssim \exp\{(1+K_0)M_{N_0^*}\}\epsilon_{N_0^*}^{-d}\times E_{\{Y_j\}_{j=1}^{N_1^*}}\pi(p_{\theta}:\|p_{\theta}-p_{\theta_0}\|_1\leq K \epsilon_{N_0^*}\mid \{Y_j\}_{j=1}^{N_1^*}),\\
            & \stackrel{(a)}{\lesssim} \exp\{(1+K_0)M_{N_0^*}+\frac{d}{2}\log(N_0^*) - \frac{d}{2}\log(\log(N_0^*))\}\times \\
            & E_{\{Y_j\}_{j=1}^{N_1^*}}\pi(p_{\theta}:\|p_{\theta}-p_{\theta_1}\|_1\geq K \epsilon_{N_1^*}\mid \{Y_j\}_{j=1}^{N_1^*}),\\
            & \lesssim \exp\{(1+K_0)M_{N_0^*}+\frac{d}{2}\log(N_0^*) - K^*\times M_{N_1^*}\},
        \end{split}
    \end{equation}
    where inequality (a) is given by $\|p_{\theta_0}-p_{\theta_1}\|_1\gtrsim \epsilon_{N_0^*}\times \psi_{N_0^*}$. By Assumption (A3), the last display of \eqref{2ndPart1} approaches 0 as both $N_0$ and $N_1$ increase.
\end{proof}
\end{theorem}
It is important to note that in the proof of Theorem \ref{postExample}, we also demonstrate that $\mathbbm{P}_{\theta_0}^{N_0^*}[\mathbbm{1}_{E_{N_0^*}}]$ and $\mathbbm{P}_{\theta_1}^{N_1^*}[\mathbbm{1}_{F_{N_1^*}}]$ converge to 1, where $\mathbbm{P}_{\theta}^{n}$ denotes the product of $n$ probability measures induced by $p_{\theta}$. Together with the results of Theorem \ref{postExample}, this leads to the conclusion that the Bayes factors in \eqref{postExampleEq1} and \eqref{postExampleEq2} converge to 0 in probability.

To prepare for Theorem \ref{theoremFinal}, we first prove a lemma that establishes a posterior contraction result where observations are pooled from two random samples. Specifically, we show that the posterior of $\pi(\cdot)$, given $X_1,\ldots,X_{n}\stackrel{i.i.d.}{\sim} p_{\theta_0}$ and $Y_1,\ldots,Y_m\stackrel{i.i.d.}{\sim}p_{\theta_1}$, contracts at $p_{n,m}^*\equiv \frac{n}{n + m}p_{\theta_0} + \frac{m}{n + m}p_{\theta_1}$, under a moderate assumption,
\begin{enumerate}
    \item[(B1)] For $\forall n,m \in \mathbbm{Z}^+$, suppose $p_{n,m}^*$ is in the support of $\pi(\cdot)$, given a random sample $Z_1,\ldots,Z_{n+m}\stackrel{i.i.d}{\sim} p_{n,m}^*$, it holds that $$E_{\{Z_{\ell}\}_{\ell=1}^{n+m}}\left[\pi\left(p_{\theta}: \|p_{\theta}-p_{n,m}^*\|_1\geq K \epsilon_{n+m}\mid \{Z_{\ell}\}_{\ell=1}^{n+m}\right)\mathbbm{1}_{G_{n+m}}\right]\leq \exp\{-K^*(n+m)\epsilon_{n+m}^2\},$$ 
    
    where $\mathbbm{1}_{G_{n+m}}$ is the event defined similarly as the ones given in \eqref{(A5)}, $K^* > 0$ and $K > 0$ are two universal constants with $K$ depending on $K^*$ and $\epsilon_{n+m}^2=\frac{M_{n+m}}{n+m}$ denotes the posterior contraction rate with $M_{n+m} = o(\log(n+m))$ approaching to $\infty$ as $n + m$ increases.
\end{enumerate}

In addition, we let $H_{n,m}$ be the event such that the following inequality holds given random variables $Z_1,\ldots,Z_{n+m}\stackrel{i.i.d}{\sim} p^*_{n,m}$,
\begin{equation}\label{Hevent}
    \begin{split}
        & \prod_{i=1}^n\frac{p_{\theta_0}(Z_i)}{p^*_{n,m}(Z_i)}\prod_{j=n+1}^{n+m}\frac{p_{\theta_1}(Z_j)}{p^*_{n,m}(Z_j)}\leq r_{n+m},
    \end{split}
\end{equation}
where $r_{n+m}$ is a sequence that diverges to infinity as $n + m\to\infty$. The lemma is presented as follows,

\begin{lemma}\label{lemma1.2.}
    Suppose $X_1,\ldots,X_n\stackrel{i.i.d.}{\sim}p_{\theta_0}$ and $Y_1,\ldots,Y_m\stackrel{i.i.d.}{\sim}p_{\theta_1}$, and Assumption (B1) holds and $r_{n+m}\to\infty$ at a rate such that $r_{n+m} = o\left(\exp\left\{K^*(n+m)\epsilon_{n+m}^2\right\}\right).$
    Then it holds that
    \begin{equation}\label{concludeLem}
        \begin{split}
            & E_{\{X_j\}_{j=1}^{n},\{Y_j\}_{j=1}^m}\left(\pi\left(p_{\theta}: \|p_{\theta}-p^*_{n,m}\|_1\geq K \epsilon_{n+m}\mid \{X_j\}_{j=1}^{n},\{Y_j\}_{j=1}^m\right)\mathbbm{1}_{G_{n+m}}\mathbbm{1}_{H_{n,m}}\right)\\
            & \leq r_{n+m}\exp\{-K^*(n+m)\epsilon_{n+m}^2\},
        \end{split}
    \end{equation}
    with $\left[\mathbbm{P}_{n,m}^*\right]^{n+m}(H_{n,m})$ tends to 1 as $n+m$ approaches $\infty$, where $\left[\mathbbm{P}_{n,m}^*\right]^{n+m}$ denotes the product of the probability measures induces by $p^*_{n,m}$.
    \begin{proof}
        By Fubini's theorem, we have
        \begin{equation}\label{fracSplit}
            \begin{split}
                & E_{\{X_j\}_{j=1}^{n},\{Y_j\}_{j=1}^m}\left(\pi\left(p_{\theta}: \|p_{\theta}-p^*_{n,m}\|_1\geq K \epsilon_{n+m}\mid \{X_j\}_{j=1}^{n},\{Y_j\}_{j=1}^m\right)\mathbbm{1}_{G_{n+m}}\mathbbm{1}_{H_{n,m}}\right) \\
                & = \int \prod_{i=1}^n \prod_{j=1}^m p^*_{n,m}(x_i) p^*_{n,m}(y_j)\prod_{i=1}^n\prod_{j=1}^m\frac{p_{\theta_0}(x_i)}{p^*_{n,m}(x_i)}\frac{p_{\theta_1}(y_j)}{p^*_{n,m}(y_j)}\times\\
                & \pi\left(p_{\theta}: \|p_{\theta}-p^*_{n,m}\|_1\geq K \epsilon_{n+m}\mid \{X_j\}_{j=1}^{n},\{Y_j\}_{j=1}^m\right)\times\\
                & \mathbbm{1}_{G_{n+m}}\mathbbm{1}_{H_{n,m}} dx_1,\ldots,dx_n,dy_1,\ldots,dy_m\\
                & = E_{\{Z_{\ell}\}_{\ell=1}^{n+m}}\left[U_{n,m}\times V_{n,m}\right],
            \end{split}
        \end{equation}
        where $U_{n,m}$ and $V_{n,m}$ are defined as
        \begin{equation}\label{UVDef}
            \begin{split}
                & U_{n,m} = \prod_{i=1}^n\frac{p_{\theta_0}(Z_i)}{p^*_{n,m}(Z_i)}\prod_{j=1}^m\frac{p_{\theta_1}(Z_j)}{p^*_{n,m}(Z_j)}\times \mathbbm{1}_{H_{n,m}},\\
                & V_{n,m} = \pi\left(p_{\theta}: \|p_{\theta}-p^*_{n,m}\|_1\geq K \epsilon_{n+m}\mid \{Z_{\ell}\}_{\ell=1}^{n+m}\right)\mathbbm{1}_{G_{n+m}}.
            \end{split}
        \end{equation}
        The result in \eqref{concludeLem} can be directly obtained following the definition of $H_{n,m}$ and Assumption (B1). The remainder is to show $\left[\mathbbm{P}_{n,m}^*\right]^{n+m}(H_{n,m})$ tends to 1 as $n+m$ approaches $\infty$. This result is given by a direct application of Markov's inequality,$$\left[\mathbbm{P}_{n,m}^*\right]^{n+m}\left(\prod_{i=1}^n\frac{p_{\theta_0}(Z_i)}{p^*_{n,m}(Z_i)}\prod_{j=n+1}^{n+m}\frac{p_{\theta_1}(Z_j)}{p^*_{n,m}(Z_j)}\geq r_{n+m}\right)\leq \frac{1}{r_{n+m}}.$$
        This concludes the proof.
    \end{proof}
\end{lemma}

To prove our main result (Theorem \ref{theoremFinal}), the following lemma is required for lower bounding the model evidence, which is a modification of Lemma 8.1. of \citet{ghosal2000convergence}. The key difference in our result is that it applies to a setting where observations are pooled from two random samples generated from two different generating mechanisms.

\begin{lemma}\label{KLLow}
    For $\forall \epsilon > 0$ and a prior $\pi(\cdot)$ of $p_{\theta}$ on the set
    \begin{equation}\label{KLNeigh}
        \begin{split}
            & \left\{p_{\theta}: \mathbbm{P}_{n,m}^*\log\left(\frac{p_{n,m}^*}{p_{\theta}}\right)\leq \epsilon^2,~ \mathbbm{P}_{n,m}^*\left(\log\left(\frac{p_{n,m}^*}{p_{\theta}}\right)\right)^2\leq \epsilon^2\right\},
        \end{split}
    \end{equation}
    and $X_1,\ldots,X_n\stackrel{i.i.d.}{\sim} p_{\theta_0}$, $Y_1,\ldots,Y_m\stackrel{i.i.d.}{\sim}p_{\theta_1}$, with sufficiently large $n$ and $m$ and $\forall K_0 > 0$, it holds that
    \begin{equation}\label{KLprob}
        \begin{split}
            & \mathbbm{P}_{\theta_0}^n \mathbbm{P}_{\theta_1}^m\left(\int \prod_{i=1}^n\frac{p_{\theta}}{p_{n+m}^*}(X_i)\prod_{j=1}^m\frac{p_{\theta}}{p_{n+m}^*}(Y_j)d \pi(p_{\theta})\leq \exp(-(1 + K_0)(n+m)\epsilon^2)\right),\\
            & \leq \frac{1}{K_0^2(n+m)\epsilon^2}
        \end{split}
    \end{equation}
    where $p_{n,m}^* \equiv \frac{n}{n+m}p_{\theta_0}+\frac{m}{n + m}p_{\theta_1}$ is the weighted summation of $p_{\theta_0}$ and $p_{\theta_1}$ and $\mathbbm{P}_{n,m}^*\log\left(\frac{p_{n,m}^*}{p_{\theta}}\right)$ refers to $\int p_{n,m}^*(x)\log\left(\frac{p_{n,m}^*(x)}{p_{\theta}(x)}\right)dx$.
\begin{proof}
    By applying Jensen's inequality to the following inequality, we have
    \begin{equation}\label{Jensen}
        \begin{split}
            & \log\left(\int \prod_{i=1}^n\frac{p_{\theta}(X_i)}{p_{n,m}^*(X_i)}\prod_{j=1}^m\frac{p_{\theta}(Y_j)}{p_{n,m}^*(Y_j)}d \pi(p_{\theta})\right)\leq -(1 + K_0)(n+m)\epsilon^2,\\
            & \sum_{i=1}^n\int \log\left(\frac{p_{\theta}(X_i)}{p_{n,m}^*(X_i)}\right) + \sum_{j=1}^m\int \log\left(\frac{p_{\theta}(Y_j)}{p_{n,m}^*(Y_j)}\right)\leq -(1 + K_0)(n+m)\epsilon^2.
        \end{split}
    \end{equation}
    The probability \eqref{KLprob} can hence be upper bounded by,
    \begin{equation}\label{probUp}
        \begin{split}
            & \mathbbm{P}_{\theta_0}^n \mathbbm{P}_{\theta_1}^m\left(\int \prod_{i=1}^n\frac{p_{\theta}(X_i)}{p_{n,m}^*(X_i)}\prod_{j=1}^m\frac{p_{\theta}(Y_j)}{p_{n,m}^*(Y_j)}d \pi(p_{\theta})\leq \exp(-(1 + K_0)(n+m)\epsilon^2)\right)\\
            & \leq \mathbbm{P}_{\theta_0}^n \mathbbm{P}_{\theta_1}^m\left(\sqrt{\frac{n}{n+m}}\mathbbm{P}_{\theta_0,n}\int\log\left(\frac{p_{\theta}}{p_{n,m}^*}\right)d\pi(p_{\theta}) + \sqrt{\frac{m}{n+m}}\mathbbm{P}_{\theta_1,m}\int\log\left(\frac{p_{\theta}}{p_{n,m}^*}\right)d\pi(p_{\theta}) \leq \right.\\
            & \left. -(1 + K_0)\sqrt{n+m}\epsilon^2 - \sqrt{n+m}\mathbbm{P}_{n,m}^* \int\log\left(\frac{p_{\theta}}{p_{n,m}^*}\right)d\pi(p_{\theta})\right),\\
            & \leq \mathbbm{P}_{\theta_0}^n \mathbbm{P}_{\theta_1}^m\left(\sqrt{\frac{n}{n+m}}\mathbbm{P}_{\theta_0,n}\int\log\left(\frac{p_{\theta}}{p_{n,m}^*}\right)d\pi(p_{\theta}) + \sqrt{\frac{m}{n+m}}\mathbbm{P}_{\theta_1,m}\int\log\left(\frac{p_{\theta}}{p_{n,m}^*}\right)d\pi(p_{\theta}) \leq \right.\\
            & \left. -K_0\sqrt{n+m}\epsilon^2\right),
        \end{split}
    \end{equation}
    where $\mathbbm{P}_{\theta_0,n}$ and $\mathbbm{P}_{\theta_0,m}$ are the empirical processes $\sqrt{n}(\mathrm{P}_{\theta_0,n} - \mathbbm{P}_{\theta_0})$ and $\sqrt{m}(\mathrm{P}_{\theta_1,m} - \mathbbm{P}_{\theta_1})$, $\mathrm{P}_{\theta_0,n}$ is the probability measure induced by $F(x)\equiv \frac{1}{n}\sum_{i=1}^n\mathbbm{1}_{X_i \leq x}$ and $\mathrm{P}_{\theta_1,m}$ is defined in a similar way. The last display of \eqref{probUp} is given by Fubini's theorem and the condition in \eqref{KLNeigh}. By applying Chebyshev's inequality to \eqref{probUp}, the probability \eqref{KLprob} can be further upper bounded by,
    \begin{equation}\label{varIneq}
        \begin{split}
            & \frac{\frac{n}{n+m}\text{var}_{\theta_0}\int \log\left(\frac{p_{\theta}}{p_{n,m}^*}\right)d\pi(p_{\theta}) + \frac{m}{n+m}\text{var}_{\theta_1}\int \log\left(\frac{p_{\theta}}{p_{n,m}^*}\right)d\pi(p_{\theta})}{K_0^2 (n+m)\epsilon^4}\leq \frac{\mathbbm{P}_{n,m}^*\left(\log\left(\frac{p_{n,m}^*}{p_{\theta}}\right)\right)^2}{K_0^2(n+m)\epsilon^4},
        \end{split}
    \end{equation}
    where $\text{var}_{\theta_0}\int \log\left(\frac{p_{\theta}}{p_{n,m}^*}\right)d\pi(p_{\theta})$ refers to $$\mathbbm{P}_{\theta_0}\left(\int \log\left(\frac{p_{\theta}}{p_{n,m}^*}\right)d\pi(p_{\theta})\right)^2 - \left(\mathbbm{P}_{\theta_0}\int \log\left(\frac{p_{\theta}}{p_{n,m}^*}\right)d\pi(p_{\theta})\right)^2,$$
    and $\text{var}_{\theta_1}\int \log\left(\frac{p_{\theta}}{p_{n,m}^*}\right)d\pi(p_{\theta})$ is defined in a similar way. The last display of \eqref{varIneq} is given by Jensen's inequality. By another application of the condition in \eqref{KLNeigh} concludes the result in \eqref{KLprob}.
\end{proof}
\end{lemma}
In particular, when $n$ or $m$ equals to 0, Lemma \ref{KLLow} degenerates to Lemma 8.1. of \citet{ghosal2000convergence} with the assumption that the observations are i.i.d.

  {For our main result (Theorem \ref{theoremFinal}), we show that our method can identify the correct external subset, with a cost of at most including a small amount of data from the incorrect external observations. For clarity, we reiterate the notations and definitions provided in the main article. Suppose the the external trajectories are generated by two mechanisms (i.e., $\left\{X_{ik}\right\}_{\substack{i=1,k=1}}^{N_0^{(x)},T_i^{(x)}}$ and $\{Y_{jl}\}_{\substack{j=N_0^{(x)} + 1, l=1}}^{N_0^{(x)} + N_0^{(y)},T_j^{(y)}}$), where one of these mechanisms matches the the internal trajectories (i.e., $\{Z_{rs}\}_{\substack{r=1, s=1}}^{N_1,T_r^{(z)}}$), while the other does not. Specifically, we assume,
\begin{equation}\label{hybridExample}
    \begin{split}
        & X_{11},\ldots,X_{1T_1^{(x)}},\ldots,X_{N_0^{(x)}1},\ldots,X_{N_0^{(x)}T_{N_0^{(x)}}^{(x)}}\stackrel{i.i.d.}{\sim} p_{\theta_1},\\
        & Y_{(N_0^{(x)} + 1)1},\ldots,Y_{(N_0^{(x)} + 1)T_{(N_0^{(x)} + 1)}^{(y)}},\ldots,Y_{(N_0^{(x)} + N_0^{(y)})1},\ldots,Y_{(N_0^{(x)} + N_0^{(y)})T_{(N_0^{(x)} + N_0^{(y)})}^{(y)}}\stackrel{i.i.d.}{\sim} p_{\theta_0},\\
        & Z_{11},\ldots,Z_{1T_1^{(z)}},\ldots,Z_{N_1^{(z)}1},\ldots,Z_{N_1^{(z)}T_{N_1^{(z)}}^{(z)}}\stackrel{i.i.d.}{\sim} p_{\theta_1}.\\
    \end{split}
\end{equation}
We will demonstrate that our method can correctly identify the correct the external subset $\mathcal{C}_0\equiv\{1,\ldots,N_0^{(x)}\}$ using the selected external subset from our model $\mathcal{C}\subseteq \{1,\ldots,N_0^{(x)},N_0^{(x)} + 1,\ldots,N_0^{(x)} + N_0^{(y)}\}$. This comes with the cost of at most including a small amount of data from the incorrect external trajectories, with indices forming a subset of $\{N_0^* + 1,\ldots,N_0^* + N_0'\}$. We further let the number of the correct external observations be $N_0^*\equiv \sum_{i=1}^{N_0^{(x)}}T_{i}^{(x)}$, the pooled external observations be $N_0\equiv N_0^* + N_0'$ and the incorrect external observations be $N_0'\equiv \sum_{i=1}^{N_0^{(y)}} T_{(N_0^{(x)} + i)}^{(y)}$. The following theorem compares the marginal likelihoods of two models based on two different external subsets: $\mathcal{C}_0 ',\mathcal{C}_1'\subseteq \{1,\ldots,N_0^{(x)},N_0^{(x)} + 1,\ldots,N_0^{(x)} + N_0^{(y)}\}$. Specifically, the model using $\mathcal{C}_1'$ effectively selects the correct external trajectories while including only a small amount of incorrect external observations. That is, it holds that $\{1,\ldots,N_0^{(x)}\} \subseteq \mathcal{C}_1'$, and for $m' \equiv\sum_{i\in\mathcal{C}_1^{''}}T_{i}^{(y)}$, where $\mathcal{C}_1''$ is the subset of $\mathcal{C}_1'$ excluding the correct external indices, the incorrect observations are sufficiently small compared to the correct external observations $N_0^*\equiv \sum_{i=1}^{N_0^{(x)}}T_{i}^{(x)}$. In contrast, the other model $\mathcal{C}_0'$ either includes too few pooled external trajectories or contains a non-negligible amount of incorrect external trajectories. Specifically, for $\mathcal{C}_0^{'*}\equiv \mathcal{C}_0'\cap \{1,\ldots,N_0^{(x)}\}$ and $\mathcal{C}_0^{''}\equiv \mathcal{C}_0^{'}\cap\{N_0^{(x)} + 1,\ldots,N_0^{(x)} + N_0^{(y)}\}$, the correct and incorrect external observation numbers $u\equiv\sum_{i\in\mathcal{C}_0^{'*}}T_{i}^{(x)}$ and $v\equiv\sum_{i\in\mathcal{C}_0^{''}}T_{i}^{(y)}$ must satisfy certain conditions, detailed in Theorem \ref{theoremFinal}. Since the trajectories are generated i.i.d. following $p_{\theta_1}$ or $p_{\theta_0}$, we can simplify the generating process as follows for brevity:
\begin{equation}\label{hybridExampleModify}
    \begin{split}
        & X_1,\ldots,X_{N_0^*}\stackrel{i.i.d.}{\sim}p_{\theta_1},\\
        & Y_1,\ldots,Y_{N_0'}\stackrel{i.i.d.}{\sim}p_{\theta_0},\\
        & Z_1,\ldots,Z_{N_1}\stackrel{i.i.d.}{\sim}p_{\theta_1},\\
    \end{split}
\end{equation}
where $N_1\equiv \sum_{i=1}^{N_1^{(z)}}T_i^{(z)}$ represents the number of internal observations. Note that $u$ and $v$ now refer to the number of correct and incorrect external observations corresponding to their respective subsets.}

  {We proceed by making the following assumptions, based on the generating process \eqref{hybridExampleModify} and the notations above to ensure the validity of our theoretical results,} 
\begin{enumerate}
    \item[(C1)] Suppose data are generated following \eqref{hybridExampleModify} and $N_0\equiv N_0^* + N_0'$ with $N_0^*$ tending to $\infty$ as $N_0$ increases.
    \item[(C2)] Let $p^*_{n,m} \equiv \frac{n}{n+m}p_{\theta_1} + \frac{m}{n+m}p_{\theta_0}$ be the weighted average of $p_{\theta_1}$ and $p_{\theta_0}$ given $n\leq N_0^*$, $m\leq N_0'$, $n,m\in \mathbbm{Z}^+$. We assume 
    $$\left\{p_{\theta}: \|p_{\theta}-p^*_{n,m}\|_1\leq C\times \epsilon_{n+m}\right\}\subseteq B_{n,m},$$
    where $B_{n,m}$ denotes the K-L neighborhood around $p^*_{n,m}$ with radius $\epsilon_{n+m}$, defined as 
    $$B_{n,m} = \left\{p_{\theta}: \mathbbm{P}^*_{n,m}\log\left(\frac{p^*_{n,m}}{p_{\theta}}\right)\leq \epsilon_{n+m}^2, \mathbbm{P}^*_{n,m}\left(\log\left(\frac{p^*_{n,m}}{p_{\theta}}\right)\right)^2\leq \epsilon_{n+m}^2\right\}.$$
\end{enumerate}
Additionally, we let $Q_{u,v}$ represent the event that happens when the following inequality holds,
\begin{equation}
    \begin{split}
        & \int_{B_{u,v}}\prod_{i=1}^{u}\frac{p_{\theta}(X_i)}{p_{u,v}^*(X_i)}\prod_{j=1}^{v}\frac{p_{\theta}(Y_j)}{p_{u,v}^*(Y_j)}\frac{d\pi(p_{\theta})}{\pi(B_{u,v})}\geq \exp\{-(1 + K_0)M_{u+v}\}.
    \end{split}
\end{equation}
Similarly, we let $R_{N_0^*, m'}$ represent the event that happens when the following inequality holds,
\begin{equation}
    \begin{split}
        & \int_{\|p_{\theta}-p_{N_0^*,m'}^*\|_1\leq K\epsilon_{N_0^* + m'}}\prod_{i=1}^{N_0^*}\frac{p_{\theta}(X_i)}{p_{N_0^*,m'}^*(X_i)}\prod_{j=1}^{m'}\frac{p_{\theta}(Y_j)}{p_{N_0^*,m'}^*(Y_j)}d\pi(p_{\theta}\mid \{Z_k\}_{k=1}^{N_1})\\
        & \geq \exp\{-(1 + K_0)M_{N_0^* + m'}\}.
    \end{split}
\end{equation}
Moreover, we let $U_{N_0^*,m'}$ denote the event that happens when the following inequality happens given auxiliary random variables $\Tilde{Z}_1,\ldots,\Tilde{Z}_{N_0^*+m'}\stackrel{i.i.d}{\sim} p_{N_0^*,m'}^*$,
\begin{equation}
    \begin{split}
        & \int_{\|p_{\theta}-p_{N_0^*,m'}^*\|_1\leq K \epsilon_{N_0^* + m'}} \prod_{i=1}^{N_0^*}\frac{p_{\theta}(\Tilde{Z}_i)}{p^*_{N_0^*,m'}(\Tilde{Z}_i)}\prod_{j=1}^{m'}\frac{p_{\theta}(\Tilde{Z}_j)}{p^*_{N_0^*,m'}(\Tilde{Z}_j)}d\pi(p_{\theta})\\
        & \leq r_{N_0^*+m'}\times \pi\left(\left\{p_{\theta}: \|p_{\theta}-p_{N_0^*,m'}^*\|_1\leq K\epsilon_{N_0^* + m'} \right\}\right).
    \end{split}
\end{equation}

We also defined the event $H_{N_0^*,m'}^*$ similarly as the one in \eqref{Hevent}, given auxiliary random variables $\Tilde{Z}_1^*,\ldots,\Tilde{Z}_{N_0^*+m}^*\stackrel{i.i.d.}{\sim} p_{N_0^*,m'}^*$
\begin{equation}
    \begin{split}
        & \prod_{i=1}^{N_0^*}\frac{p_{\theta_0}(\Tilde{Z}_i^*)}{p_{N_0^*,m'}^*(\Tilde{Z}_i^*)}\prod_{j=N_0^*+1}^{N_0^*+m'}\frac{p_{\theta_1}(\Tilde{Z}_j^*)}{p_{N_0^*,m'}^*(\Tilde{Z}_j^*)} \leq r_{N_0^*+m'},
    \end{split}
\end{equation}
and let the event $H_{u,v}'$ happen when the following inequality holds, given auxiliary random variables $\Tilde{Z}_1',\ldots,\Tilde{Z}_{N_0^*+m'}'\stackrel{i.i.d.}{\sim}p_{u,v}^*$
\begin{equation}
    \begin{split}
        & \prod_{i=1}^{N_0^*}\frac{p_{N_0^*,m'}^*(\Tilde{Z}_i')}{p_{u,v}^*(\Tilde{Z}_i')}\prod_{j=N_0^*+1}^{N_0^*+m'}\frac{p_{N_0^*,m'}^*(\Tilde{Z}_j')}{p_{u,v}^*(\Tilde{Z}_j')} \leq r_{N_0^*+m'},
    \end{split}
\end{equation}

Our main result is stated as follows, 
\begin{theorem}\label{theoremFinal}
    Suppose Assumptions (A1), (A3), (B1), (C1) and (C2) hold for $r_{n,m} = o(n + m)$. For any $m'\leq N_0'$ with $m'$ satisfying $\frac{m'}{m' + N_0^*}\|p_{\theta_1}-p_{\theta_0}\|_1 = o\left(\sqrt{\frac{M_{m' + N_0^*}}{m' + N_0^*}} - \sqrt{\frac{M_{N_1}}{N_1}}\right)$, we have
    \begin{equation}\label{concludeBF}
        \begin{split}
            & E_{\{X_i\}_{i=1}^{N_0^*},\{Y_j\}_{j=1}^{N_0'},\{Z_k\}_{k=1}^{N_1}}\left[\frac{\int_{\|p_{\theta}-p_{u,v}^*\|_1\leq K \epsilon_{u+v}} \prod_{k=1}^{N_1} p_{\theta}(Z_k)d\pi(p_{\theta}\mid \left\{X_i\right\}_{i=1}^{u},\left\{Y_j\right\}_{j=1}^{v})}{\int_{\|p_{\theta}-p_{N_0^*,m'}^*\|_1\leq K \epsilon_{N_0^*+m'}} \prod_{k=1}^{N_1} p_{\theta}(Z_k)d\pi(p_{\theta}\mid \left\{X_i\right\}_{i=1}^{N_0^*},\left\{Y_j\right\}_{j=1}^{m'})}\times \right.\\
            & \left.\mathbbm{1}_{Q_{u,v}}\mathbbm{1}_{R_{N_0^*,m'}}\mathbbm{1}_{G_{n+m}}\mathbbm{1}_{U_{N_0^*,m'}}\mathbbm{1}_{H_{N_0^*, m'}^*}\mathbbm{1}_{H_{u,v}'}\right] \\
            & \lesssim \exp\left\{(1+K_0)(M_{u+v} + M_{N_0^*,m'})+\frac{d}{2}\log(u+v) + \right.\\
            & \left. 3 \log(r_{N_0^*,m'}) -\frac{d}{2}\log(N_0^* + m') - \mathbbm{1}_{\text{[1]}}\times K^* \times M_{N_1}\right\},
        \end{split}
    \end{equation}
    where $K^*$ satisfies $K^* > 1 + K_0 + \frac{d}{2} \times C'$, $N_0^*,N_0,u,v$ tend to $ \infty$, given $u\leq N_0^*$, $v\leq N_0'$, and [1] denotes the condition $\sqrt{\frac{M_{N_1}}{N_1}} + \sqrt{\frac{M_{u+v}}{u+v}} = o\left(\frac{v}{u + v}\|p_{\theta_1}-p_{\theta_0}\|_1\right)$.
    \begin{proof}
        By Bayes' theorem,
        \begin{equation}
            \begin{split}
                & \frac{\int_{\|p_{\theta}-p_{u,v}^*\|_1\leq K \epsilon_{u+v}} \prod_{k=1}^{N_1} p_{\theta}(Z_k)d\pi(p_{\theta}\mid \left\{X_i\right\}_{i=1}^{u},\left\{Y_j\right\}_{j=1}^{v})}{\int_{\|p_{\theta}-p_{N_0^*,m'}^*\|_1\leq K \epsilon_{N_0^*+m'}} \prod_{k=1}^{N_1} p_{\theta}(Z_k)d\pi(p_{\theta}\mid \left\{X_i\right\}_{i=1}^{N_0^*},\left\{Y_j\right\}_{j=1}^{m'})}\\
                & = \frac{\int_{\|p_{\theta}-p_{u,v}^*\|_1\leq K \epsilon_{u+v}}\prod_{i=1}^u\frac{p_{\theta}(X_i)}{p_{u,v}^*(X_i)}\prod_{j=1}^v\frac{p_{\theta}(Y_j)}{p_{u,v}^*(Y_j)}d\pi(p_{\theta}\mid \{Z_k\}_{k=1}^{N_1})}{\int_{\|p_{\theta}-p_{N_0^*,m'}^*\|_1\leq K \epsilon_{N_0^*+m'}}\prod_{i=1}^{N_0^*}\frac{p_{\theta}(X_i)}{p_{N_0^*,m'}^*(X_i)}\prod_{j=1}^{m'}\frac{p_{\theta}(Y_j)}{p_{N_0^*,m'}^*(Y_j)}d\pi(p_{\theta}\mid \{Z_k\}_{k=1}^{N_1})}\times\\
                & \frac{\int \prod_{i=1}^{N_0^*}\frac{p_{\theta}(X_i)}{p_{N_0^*,m'}^*(X_i)}\prod_{j=1}^{m'}\frac{p_{\theta}(Y_j)}{p_{N_0^*,m'}^*(Y_j)}d\pi(p_{\theta})}{\int \prod_{i=1}^{u}\frac{p_{\theta}(X_i)}{p_{u,v}^*(X_i)}\prod_{j=1}^{v}\frac{p_{\theta}(Y_j)}{p_{u,v}^*(Y_j)}d\pi(p_{\theta})}.
            \end{split}
        \end{equation}
        With Assumption (A1), (A3), (B1), (C2), by applying Lemma \ref{lemma1.2.} and the fact that 
        $$\{p_{\theta}: \|p_{\theta} - p_{\theta_1}\|_1\leq K\epsilon_{N_1}\}\subseteq \{p_{\theta}: \|p_{\theta} - p_{N_0^*,m'}^*\|_1\leq K\epsilon_{N_0^*+m'}\},$$
        we can lower-bound the model evidence as follows,
        \begin{equation}\label{step1}
            \begin{split}
            & E_{\{X_i\}_{i=1}^{N_0^*},\{Y_j\}_{j=1}^{N_0'},\{Z_k\}_{k=1}^{N_1}}\left[\frac{\int_{\|p_{\theta}-p_{u,v}^*\|_1\leq K \epsilon_{u+v}} \prod_{k=1}^{N_1} p_{\theta}(Z_k)d\pi(p_{\theta}\mid \left\{X_i\right\}_{i=1}^{u},\left\{Y_j\right\}_{j=1}^{v})}{\int_{\|p_{\theta}-p_{N_0^*,m'}^*\|_1\leq K \epsilon_{N_0^*+m'}} \prod_{k=1}^{N_1} p_{\theta}(Z_k)d\pi(p_{\theta}\mid \left\{X_i\right\}_{i=1}^{N_0^*},\left\{Y_j\right\}_{j=1}^{m'})}\times \right.\\
            & \left.\mathbbm{1}_{Q_{u,v}}\mathbbm{1}_{R_{N_0^*,m'}}\mathbbm{1}_{G_{n+m}}\mathbbm{1}_{U_{N_0^*,m'}}\mathbbm{1}_{H_{N_0^*, m'}^*}\mathbbm{1}_{H_{u,v}'}\right] \\
            & = E_{\{X_i\}_{i=1}^{N_0^*},\{Y_j\}_{j=1}^{N_0'},\{Z_k\}_{k=1}^{N_1}}\left[\frac{\int_{\|p_{\theta}-p_{u,v}^*\|_1\leq K \epsilon_{u+v}}\prod_{i=1}^u\frac{p_{\theta}(X_i)}{p_{u,v}^*(X_i)}\prod_{j=1}^v\frac{p_{\theta}(Y_j)}{p_{u,v}^*(Y_j)}d\pi(p_{\theta}\mid \{Z_k\}_{k=1}^{N_1})}{\int_{\|p_{\theta}-p_{N_0^*,m'}^*\|_1\leq K \epsilon_{N_0^*+m'}}\prod_{i=1}^{N_0^*}\frac{p_{\theta}(X_i)}{p_{N_0^*,m'}^*(X_i)}\prod_{j=1}^{m'}\frac{p_{\theta}(Y_j)}{p_{N_0^*,m'}^*(Y_j)}d\pi(p_{\theta}\mid \{Z_k\}_{k=1}^{N_1})}\times\right.\\
            & \left.\frac{\int \prod_{i=1}^{N_0^*}\frac{p_{\theta}(X_i)}{p_{N_0^*,m'}^*(X_i)}\prod_{j=1}^{m'}\frac{p_{\theta}(Y_j)}{p_{N_0^*,m'}^*(Y_j)}d\pi(p_{\theta})}{\int \prod_{i=1}^{u}\frac{p_{\theta}(X_i)}{p_{u,v}^*(X_i)}\prod_{j=1}^{v}\frac{p_{\theta}(Y_j)}{p_{u,v}^*(Y_j)}d\pi(p_{\theta})}\times \mathbbm{1}_{Q_{u,v}}\mathbbm{1}_{R_{N_0^*,m'}}\mathbbm{1}_{G_{n+m}}\mathbbm{1}_{U_{N_0^*,m'}}\mathbbm{1}_{H_{N_0^*, m'}^*}\mathbbm{1}_{H_{u,v}'}\right],\\
            & \leq \exp\left\{(1+K_0)(M_{u+v} + M_{N_0^*,m'})+\frac{d}{2}\log(u + v)\right\}\times \\
            & E_{\{X_i\}_{i=1}^{N_0^*},\{Y_j\}_{j=1}^{N_0'},\{Z_k\}_{k=1}^{N_1}}\left[\int_{\|p_{\theta}-p_{u,v}^*\|_1\leq K \epsilon_{u+v}}\prod_{i=1}^u\frac{p_{\theta}(X_i)}{p_{u,v}^*(X_i)}\prod_{j=1}^v\frac{p_{\theta}(Y_j)}{p_{u,v}^*(Y_j)}d\pi(p_{\theta}\mid \{Z_k\}_{k=1}^{N_1})\right.\\
            & \left. \int \prod_{i=1}^{N_0^*}\frac{p_{\theta}(X_i)}{p_{N_0^*,m'}^*(X_i)}\prod_{j=1}^{m'}\frac{p_{\theta}(Y_j)}{p_{N_0^*,m'}^*(Y_j)}d\pi(p_{\theta})\times \mathbbm{1}_{G_{n+m}}\mathbbm{1}_{U_{N_0^*,m'}}\mathbbm{1}_{H_{N_0^*, m'}^*} \mathbbm{1}_{H_{u,v}'}\right].
            \end{split}
        \end{equation}
        The expectation part of the last display of \eqref{step1} can be further upper bounded by
        \begin{equation}\label{step2}
            \begin{split}
            & E_{\{X_i\}_{i=1}^{N_0^*},\{Y_j\}_{j=1}^{N_0'},\{Z_k\}_{k=1}^{N_1}}\left[\int_{\|p_{\theta}-p_{u,v}^*\|_1\leq K \epsilon_{u+v}}\prod_{i=1}^u\frac{p_{\theta}(X_i)}{p_{u,v}^*(X_i)}\prod_{j=1}^v\frac{p_{\theta}(Y_j)}{p_{u,v}^*(Y_j)}d\pi(p_{\theta}\mid \{Z_k\}_{k=1}^{N_1})\right.\\
            & \left. \int \prod_{i=1}^{N_0^*}\frac{p_{\theta}(X_i)}{p_{N_0^*,m'}^*(X_i)}\prod_{j=1}^{m'}\frac{p_{\theta}(Y_j)}{p_{N_0^*,m'}^*(Y_j)}d\pi(p_{\theta})\times \mathbbm{1}_{G_{n+m}}\mathbbm{1}_{U_{N_0^*,m'}}\mathbbm{1}_{H_{N_0^*, m'}^*}\mathbbm{1}_{H_{u,v}'}\right]\\
            & \stackrel{(a)}{\lesssim} E_{\{X_i\}_{i=1}^{N_0^*},\{Y_j\}_{j=1}^{N_0'},\{Z_k\}_{k=1}^{N_1}}\left[\int_{\|p_{\theta}-p_{u,v}^*\|_1\leq K \epsilon_{u+v}}\prod_{i=1}^u\frac{p_{\theta}(X_i)}{p_{u,v}^*(X_i)}\prod_{j=1}^v\frac{p_{\theta}(Y_j)}{p_{u,v}^*(Y_j)}d\pi(p_{\theta}\mid \{Z_k\}_{k=1}^{N_1})\right.\\
            & \left. \int_{\|p_{\theta}-p_{N_0^*,m'}^*\|_1\leq K\epsilon_{N_0^*+m'}} \prod_{i=1}^{N_0^*}\frac{p_{\theta}(X_i)}{p_{N_0^*,m'}^*(X_i)}\prod_{j=1}^{m'}\frac{p_{\theta}(Y_j)}{p_{N_0^*,m'}^*(Y_j)}d\pi(p_{\theta})\times \mathbbm{1}_{U_{N_0^*,m'}}\mathbbm{1}_{H_{N_0^*, m'}^*}\mathbbm{1}_{H_{u,v}'}\right],\\
            & \lesssim r_{N_0^*,m'}^3 \times \pi\left(\left\{p_{\theta}: \|p_{\theta}-p_{N_0^*,m'}^*\|_1\leq K\epsilon_{N_0^* + m'} \right\}\right) \times \\
            & E_{\{S_i\}_{i=1}^{u+v},\{Z_k\}_{k=1}^{N_1}}\left[\int_{\|p_{\theta}-p_{u,v}^*\|_1\leq K \epsilon_{u+v}}\prod_{i=1}^u\frac{p_{\theta}(S_i)}{p_{u,v}^*(S_i)}\prod_{j=u+1}^{u+v}\frac{p_{\theta}(S_j)}{p_{u,v}^*(S_j)}d\pi(p_{\theta}\mid \{Z_k\}_{k=1}^{N_1})\right],
            \end{split}
        \end{equation}
        where $S_1,\ldots,S_{u+v}\stackrel{i.i.d.}{\sim} p_{u,v}^*$ and inequality (a) is given by Assumption (B1), the definition of $\mathbbm{1}_{G_{n+m}}$ and the definition of Bayes' theorem. Suppose $\sqrt{\frac{M_{N_1}}{N_1}} + \sqrt{\frac{M_{u+v}}{u+v}} = o\left(\|p_{u,v}^*-p_{\theta_1}\|_1\right)$, that is, $$\sqrt{\frac{M_{N_1}}{N_1}} + \sqrt{\frac{M_{u+v}}{u+v}} = o\left(\frac{v}{u+v}\|p_{\theta_1}-p_{\theta_0}\|_1\right),$$
        it implies that $\{p_{\theta}:\|p_{\theta} - p_{u,v}^*\|_1\leq K\times \epsilon_{u+v}\}\subseteq \{p_{\theta}:\|p_{\theta} - p_{\theta_1}\|_1\leq K\times \epsilon_{N_1}\}^c$. Therefore, the last display of \eqref{step2} can be further upper bounded by
        \begin{equation}
            \begin{split}
                & \exp\left\{3\log(r_{N_0^*,m'}) - \frac{d}{2}\log(N_0^* + m') - \mathbbm{1}_{\text{[1]}}\times K^* \times M_{N_1}\right\}.
            \end{split}
        \end{equation}
        The proof is finished by combining \eqref{step1} with \eqref{step2}.
    \end{proof}
\end{theorem}

It is not hard to show that the probability of observing $Q_{u,v}$, $R_{N_0^*,m'}$, $G_{n+m}$, $U_{N_0^*,m'}$, $H_{N_0^*,m'}^*$ and $H_{N_0^*,m'}'$ tend to 1 as $u+v$ and $N_0^* + m'$ approach $\infty$ using Markov's inequality as shown in the proof of Lemma \ref{lemma1.2.}. These conclude that $$\frac{\int_{\|p_{\theta}-p_{u,v}^*\|_1\leq K \epsilon_{u+v}} \prod_{k=1}^{N_1} p_{\theta}(Z_k)d\pi(p_{\theta}\mid \left\{X_i\right\}_{i=1}^{u},\left\{Y_j\right\}_{j=1}^{v})}{\int_{\|p_{\theta}-p_{N_0^*,m'}^*\|_1\leq K \epsilon_{N_0^*+m'}} \prod_{k=1}^{N_1} p_{\theta}(Z_k)d\pi(p_{\theta}\mid \left\{X_i\right\}_{i=1}^{N_0^*},\left\{Y_j\right\}_{j=1}^{m'})}$$tends to 0 in probability. For the proof when $u + v$ is finite, we can directly apply the result presented in Theorem \ref{postExample}.

\section{Details for the simulation studies}\label{appenx: detailed}
  {In this section, we also investigate the settings where both external and internal trajectories are generated without censoring.}

\subsection{Time schedules}\label{appenx: time schedule}
In addition to the parameter settings outlined in Section 4 of the main article, we assume the following:
\begin{enumerate}
    \item For DGPs 4 and 5, the number of observations per external and internal trajectory is a random integer $K$ generated by,
    $$K = \text{as.integer}(K^*),~~~~K^*\sim \text{Unif}(25, 30).$$The time schedule for each trajectory is then evenly spaced into $K$ intervals spanning from year 0 to year 6 post-treatment.
    \item For DGPs 1 and 6, the number of observations per external trajectory is a random integer $K'$ generated by,
    $$K' = \text{as.integer}(K^{'*}),~~~~K^{'*}\sim \text{Unif}(21, 26).$$The time schedule for each external trajectory is then evenly spaced into $K'$ intervals between year 0 and year 2 post-treatment. The number of observations per internal trajectory is a random integer $K*$ generated by,
    $$K^* = \text{as.integer}(K^{**}),~~~~K^{**}\sim \text{Unif}(25, 30).$$The time schedule for each internal trajectory is then evenly spaced into $K^*$ intervals into $K^*$ intervals between year 0 and year 5 post-treatment.
    \item For DGPs 2 and 3, the number of observations per external and internal trajectory is uniformly 34. The time schedule for each external and internal trajectory is evenly spaced into 34 intervals between year 0 and year 6. Furthermore, the external trajectories are censored at year 2 post-treatment, while the internal trajectories are censored at year 5 post-treatment.
\end{enumerate}

\subsection{Metrics}
\subsubsection{Settings 1 and 3}
  {To evaluate the performance of the three methods in estimating the true parameters, we adopt the following metrices,
\begin{equation}\label{ell1}
    \begin{split}
        & \ell_{\text{S}} = \sum_{k=1}^4\left|\frac{\hat{\boldsymbol{\theta}}_{k} - \boldsymbol{\theta}_k^*}{\boldsymbol{\theta}_k^*}\right|,~~~~\ell_{\text{P}} = \left|\frac{\hat{\boldsymbol{\theta}}_{5} - \boldsymbol{\theta}_5^*}{\boldsymbol{\theta}_5^*}\right|,\\
        & p_{\text{s}} = \mathbbm{1}(\ell_{\text{S}}/\ell_{\text{P}} \text{ of the current method is smaller than the SP one}),\\
    \end{split}
\end{equation}
where $\boldsymbol{\theta}_k^*$ and $\hat{\boldsymbol{\theta}}_k^*$ denote the $k$-th entry of $\boldsymbol{\theta}^*$ and $\hat{\boldsymbol{\theta}}^*$, respectively, for $k=1,\ldots,5$, and $\hat{\boldsymbol{\theta}}$ denotes the posterior mean of $\boldsymbol{\theta}$ based on the 1,000 posterior samples drawn following the procedure defined in Section 3.2. The relative error is used to assess the accuracy of both the trend estimation up to the turning point ($\ell_{\text{S}}$) and the plateau value estimation ($\ell_{\text{P}}$) for the three methods (i.e., SP, DC, and NB). These metrics mean to demonstrate whether incorporating the relevant external information, specifically, the correct external trajectories lead to improvements in early-stage and long-term estimation accuracy.}

  {We also assess the empirical coverage of the $95\%$ credible intervals provided by the three methods. Specifically, we focus on the frequency with which these credible intervals cover the true outcome value at 3 years post-treatment. For each method, the upper and lower bounds of the $95\%$ credible interval are determined by the 2.5th and 97.5th percentiles of the corresponding 1,000 posterior samples. The indicator $Cvr_3$ denotes whether the true factor levels at 3 years post-treatment is covered by the $95\%$ credible intervals, taking value 1 if the true value is covered and 0 otherwise. Additionally, we let $len_3$ represent the median length of the $95\%$ credible intervals across the 100 Monte Carlo replications, and use $p_{\text{small};3}$ to compare the interval widths between methods, defined as,
\begin{equation}\label{lens}
    \begin{split}
        & p_{\text{s};3} = \mathbbm{1}(len_3 \text{ of the current method is smaller than the SP one}).
    \end{split}
\end{equation}}
  {For our proposed approach, we also assess the proportion of correct and incorrect selections, as well as the preference level of favoring the correct external trajectories, which are defined as,
\begin{equation}\label{prop}
    \begin{split}
        & p_{[1]} = \frac{|\hat{\mathcal{C}} \cap \mathcal{C}_{[1]}|}{K_1},~~~~p_{[2]} = \frac{|\hat{\mathcal{C}} \cap \mathcal{C}_{[2]}|}{K_2},~~~~p_{\geq} = \mathbbm{1}(p_{[1]} \geq p_{[2]}),
    \end{split}
\end{equation}
where $|\cdot|$ denotes the size of an indice set, and $\hat{\mathcal{C}}$ represents the most representative subset, as defined below (8), $\mathcal{C}_{[1]}$ and $\mathcal{C}_{[2]}$ respectively denote the indices of the correct and incorrect external trajectories. The values $1 - p_{[1]}$ and $p_{[2]}$ can be interpreted as the probability of making Type \rom{1} and Type \rom{2} errors, respectively. These three metrics are designed to quantify the ability of our approach's ability to correctly select relevant external trajectories.}

  {Across the 100 Monte Carlo replications, we report the median values (and the mean $p_{\text{s}}$) for $\ell_{\text{S}}$ and $\ell_{\text{P}}$; the median values for $p_{[1]}$ and $p_{[2]}$; the mean for $p_{\geq}$; the mean for $Cvr_3$; the median value (and the mean $p_{\text{s};3}$) for $len_3$; and the median for $p_{[1]}$ and $p_{[2]}$.}

\subsubsection{Setting 2}

  {To evaluate the prediction performance of our approach and SSCM, we use the following metrics to assess the accuracy of predictions for the true outcome value at 3 years and 5 years post-treatment.
\begin{equation}\label{SSCMMetric}
    \begin{split}
        & \ell_{3} = \left|\frac{\hat{y}_3 - y_3^*}{y_3^*}\right|,~~~~\ell_{5} = \left|\frac{\hat{y}_5 - y_5^*}{y_5^*}\right|,\\
        & p_{\text{s};y} = \mathbbm{1}(\ell_{3}/\ell_{5} \text{ of the current method is smaller than the SP one}),\\
    \end{split}
\end{equation}
where $y_3^*$ and $y_5^*$ denote the true outcome values at 3 years and 5 years post-treatment. Similarly, $\hat{y}_3$ and $\hat{y}_5$ represent the estimated outcome at these time points. For our approach, the estimates are derived from the posterior mean of the predictive outcome values based on the 1,000 posterior samples. For SSCM, the estimates are obtained by averaging the external trajectories using the weight $\hat{\boldsymbol{w}}$ obtained from (10) in the main article. The estimated outcome at 3 years and 5 years post-treatment are directly accessible because time points are included in the simulated time schedule $\mathcal{T}$.}

  {For both approaches, we evaluate their proportions of correct and incorrect selections, as well as the preference level for favoring the correct external trajectories. They are defined as,
\begin{equation}\label{SSCMPrefer}
    \begin{split}
        & w_{[1]} = \sum_{i\in \hat{\mathcal{C}}\cap \mathcal{C}_{[1]}}\hat{w}_i,~~~~w_{[2]} = \sum_{i\in \hat{\mathcal{C}}\cap (\mathcal{C}_{[2]}\cup \mathcal{C}_{[3]})}\hat{w}_i,\\
        & w_{\geq} = \mathbbm{1}(\hat{w}_{[1]} \geq \hat{w}_{[2]}),
    \end{split}
\end{equation}
where $\hat{w}_i$ represents the $i$-th entry of $\hat{w}$. For our approach, $\hat{w}_i$ is equal to $\mathbbm{1}(i\in\hat{\mathcal{C}})/|\hat{\mathcal{C}}|$, $\mathcal{C}_{[1]}$, $\mathcal{C}_{[2]}$ and $\mathcal{C}_{[3]}$ respectively denote the indices set of the external trajectoris from processes [1], [2] and [3]. For SSCM, $\hat{\mathcal{C}}$ denotes the index set $\{i: \hat{w}_i\neq 0\}$, with $\hat{w}$ obtained from (10) in the main article.}

  {Across the 100 Monte Carlo replications for each DGP, we report the median values (and the mean $p_{\text{s};y}$) for $\ell_{3}$ and $\ell_{5}$; the median values for $w_{[1]}$ and $w_{[2]}$; and the mean for $w_{\geq}$.}

\subsection{Setting 1, with no censoring}

  {We consider a DGP, termed DGP 4, that uses the parameter setting of DGP 1, and the time schedule defined in Section \ref{appenx: time schedule} (without censoring). In Figure \ref{fig: Simul12}, we present 10 trajectories for each generating processes by choosing $\rho$ to be 50 and $K_1 = K_2 = 5$ for illustration.}

\begin{figure}[ht]\centering
      \includegraphics[width=\linewidth]{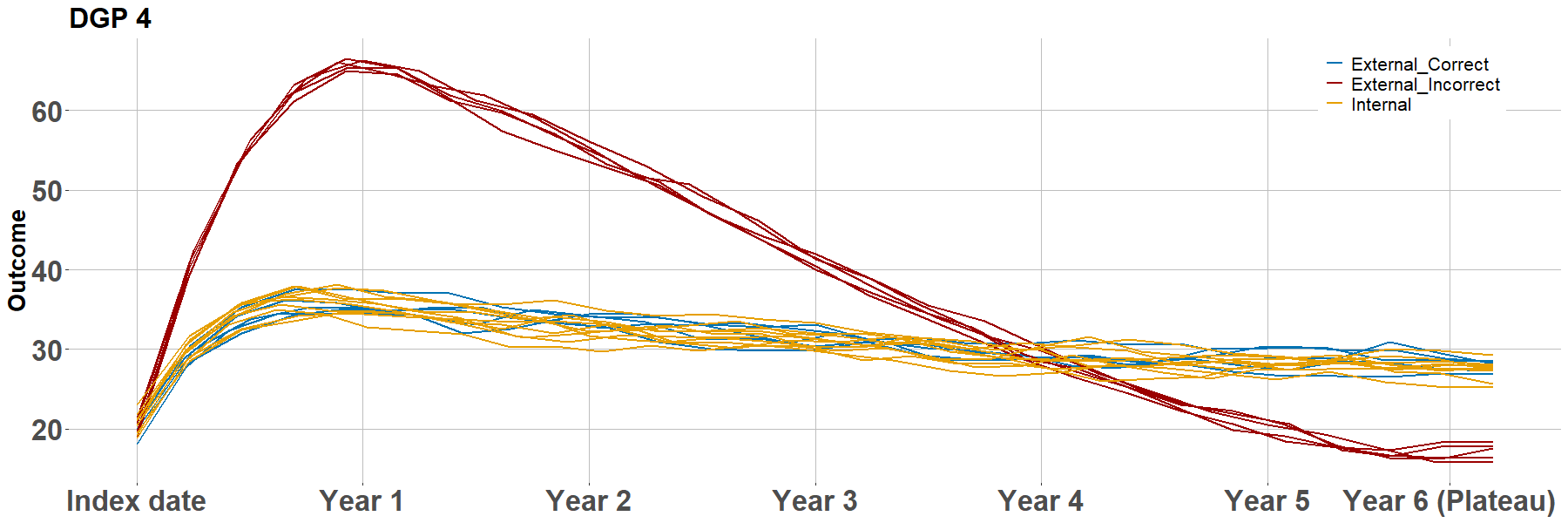}
        \caption{\label{fig: Simul12} Simulated sample data under DGP 4. The trajectories are generated without censoring.}
\end{figure}

\begin{table}[h]
    \centering
        \caption{Performance of the three methods in parameter estimation across 100 Monte Carlo replications for DGP 4, evaluated under varying values of $K_1$, $K_2$ and $\rho$.}
    \label{tab 1: setting1}
   \begin{tabular}{l|l|l|l|l|l|l|l|l}
   \toprule
    \toprule
  \multicolumn{3}{c|}{}& \multicolumn{2}{c|}{SP} & \multicolumn{2}{c|}{DC} & \multicolumn{2}{c}{NB} \\
    \midrule
 $\rho$ & $K_1$ & $K_2$  & $\ell_{\text{S}}$ & $\ell_{\text{P}}$ & $\ell_{\text{S}}$ & $\ell_{\text{P}}$ & $\ell_{\text{S}}$ & $\ell_{\text{P}}$ \\
    \midrule
 10 & 2 & 8 & 0.07 & 0.00 & 0.07 (0.41) & 0.00 (0.22) & 0.07 (0.54) & 0.00 (0.59)\\
  & 5 & 5 & 0.08 & 0.00 & 0.08 (0.46) & 0.00 (0.24) & 0.08 (0.40) & 0.00 (0.59)\\
  & 8 & 2 & 0.07 & 0.00 & 0.07 (0.48) & 0.00 (0.35) & 0.07 (0.49) & 0.00 (0.45) \\
 30 & 2 & 8 & 0.08 & 0.00 & 0.09 (0.33) & 0.00 (0.28) & 0.08 (0.46) & 0.00 (0.51) \\
 & 5 & 5 & 0.09 & 0.00 & 0.09 (0.34) & 0.00 (0.28) & 0.09 (0.40) & 0.00 (0.59)\\
  & 8 & 2 & 0.07 & 0.00 & 0.07 (0.42) & 0.00 (0.41) & 0.07 (0.45) & 0.00 (0.51)\\
50 & 2 & 8 & 0.08 & 0.00 & 0.08 (0.37) & 0.00 (0.32) & 0.07 (0.40) & 0.00 (0.50) \\
 & 5 & 5 & 0.08 & 0.00 & 0.09 (0.39) & 0.00 (0.29) & 0.09 (0.34) & 0.00 (0.56) \\
 & 8 & 2 & 0.07 & 0.00 & 0.08 (0.44) & 0.00 (0.45) & 0.07 (0.45) & 0.00 (0.50) \\
  \bottomrule
  \bottomrule
    \end{tabular}
    \newline
    \footnotesize\textsuperscript{*} Median values (with mean $p_{\text{s}}$) for $\ell_{\text{S}}$ and $\ell_{\text{P}}$.
\end{table}

  {Our approach continues to outperform the DC approach in this setting, suggested by smaller $\ell_{\text{P}}$, due to SP's ability to identify and exclude incorrect external trajectories, which would otherwise introduce substantial error. Additionally, our approach performs similarly to the NB approach in DGP 4, with $p_{\text{s}}$ values generally near 0.5, suggesting minimal improvement in estimation accuracy for both $\ell_{\text{S}}$ and $\ell_{\text{P}}$. This result is expected, as in this setting, the internal trajectories already include sufficient information for making reliable long-term inferences. Thus, the benefit of integrating external information diminishes when it does not substantially increase the effective sample size for estimation.}

\begin{table}[h]
    \centering
        \caption{Preference level of our approach (SP) and coverage level of the true outcome value at 3 years post-treatment for the three methods across 100 Monte Carlo replications for DGP 4, with varying values of $K_1$, $K_2$ and $\rho$.}
    \label{tab 2: setting1 appendx}
   \begin{tabular}{l|l|l|l|l|l|l|l|l|l|l|l}
   \toprule
    \toprule
  \multicolumn{3}{c|}{}& \multicolumn{5}{c|}{SP} & \multicolumn{2}{c|}{DC} & \multicolumn{2}{c}{NB} \\
    \midrule
 $\rho$ & $K_1$ & $K_2$ & $p_{[1]}$ & $p_{[2]}$ & $p_{\geq}$ & $Cvr_3$ & $len_3$ & $Cvr_3$ & $len_3$ & $Cvr_3$ & $len_3$ \\
    \midrule
 10 & 2 & 8 & 0.25 & 0.13 & 0.50 & 0.99 & 0.48 & 1.00 & 0.68 (0.00) & 0.99 & 0.44 (0.98)\\
 & 5 & 5 & 0.40 & 0.20 & 0.70 & 0.99 & 0.49 & 1.00 & 0.67 (0.00) & 1.00 & 0.45 (0.88)\\
 & 8 & 2 & 0.44 & 0.00 & 0.70 & 0.99 & 0.49 & 1.00 & 0.60 (0.00) & 0.98 & 0.45 (0.68) \\
30 & 2 & 8 & 0.00 & 0.13 & 0.47 & 0.96 & 0.53 & 1.00 & 0.79 (0.00) & 0.97 & 0.52 (0.73) \\
 & 5 & 5 & 0.60 & 0.00 & 0.87 & 0.98 & 0.53 & 1.00 & 0.78 (0.00) & 0.98 & 0.52 (0.58)\\
 & 8 & 2 & 0.50 & 0.00 & 0.89 & 0.97 & 0.51 & 1.00 & 0.74 (0.00) & 0.98 & 0.51 (0.36)\\
 50 & 2 & 8 & 0.00 & 0.13 & 0.49 & 0.97 & 0.57 & 0.98 & 0.74 (0.00) & 0.97 & 0.55 (0.88) \\
 & 5 & 5 & 0.60 & 0.00 & 0.87 & 0.98 & 0.74 & 0.98 & 0.74 (0.00) & 0.98 & 0.55 (0.68) \\
 & 8 & 2 & 0.50 & 0.00 & 0.93 & 0.96 & 0.55 & 0.99 & 0.72 (0.00) & 0.96 & 0.55 (0.38) \\ 
  \bottomrule
  \bottomrule
    \end{tabular}
    \newline
    \footnotesize\textsuperscript{*} Median values for $p_{[1]}$, $p_{[2]}$; Mean values for $p_{\geq}$ and $\text{Cvr}_3$; Median value (with mean $p_{\text{s};3}$) for $len_3$.
\end{table}

  {The results in Table \ref{tab 2: setting1 appendx} show that all approaches achieve higher-than-expected coverage, with our approach yielding coverage levels closer to those of the NB approach, which serves as a benchmark in DGP 4. The higher-than-expected coverage might partly be due to the limited number of Monte Carlo replications (i.e., 100). Notably, despite involving a selection procedure that probably results in under-coverage, our approach maintains uniformly satisfactory coverage in DGP 4. This provides empirical evidence that the selection procedure does not severely undermine the coverage probability of the credible intervals. Additionally, in DGP 4, the $p_{\geq}$ values exceed 0.5 in most cases, indicating that our approach effectively favors relevant external trajectories over irrelevant ones, especially when $K_1$ is reasonably large (i.e., $K_1 \geq 5$). This suggests that our approach is robust in distinguishing between relevant and irrelevant external trajectories without censoring, supporting its practical utility in scenarios similar to this setting.}

\subsection{Setting 2, SSCM comparison}\label{sec: setting2}

  {In this setting, we compare our approach to SSCM \citep{abadie2010synthetic, doudchenko2016balancing} to evaluate their performance in selecting relevant external subsets in prediction. While SSCM was originally designed to generate synthetic controls, it aligns with the objectives of our problem setting \citep{doudchenko2016balancing}. Specifically, in applying SSCM, we aim to find a weighted combination of the external trajectories from a given external subset such that this combination minimizes the mean squared error relative to the mean trend of the internal trajectories during the early-stage, i.e., up to 2 years post-treatment. This procedure, adapted from Equation (5.1) of \citet{doudchenko2016balancing} to our notation, is defined as follows,
\begin{equation}\label{SSCM}
    \begin{split}
        & \argmin_{\boldsymbol{w}} \sum_{i=1}^{N_1}\sum_{t\in\mathcal{T}, t \leq 2}(Y_{1it} - \sum_{j=1}^{N_0}\boldsymbol{w}_j\times Y_{0jt})^2,\\
        & 0\leq \boldsymbol{w}_j \leq 1,~~~~\sum_{j=1}^{N_0}\boldsymbol{w}_j = 1,\\
        & 0 < a\leq \sum_{j=1}^{N_0}\mathbbm{1}(\boldsymbol{w}_j\neq 0) \leq b \leq N_0,
    \end{split}
\end{equation}
where $\boldsymbol{w}$ is a weight vector of length $N_0$, with $\boldsymbol{w}_j$ denoting its $j$-th entry. To ensure the mean squared error in \eqref{SSCM} is well-defined, the time schedules of the external and internal trajectories are matched, that is, we let $\mathcal{T}_i^{(1)} = \mathcal{T}_j^{(0)} = \mathcal{T}$ for $i=1,\ldots,N_1$ and $j= 1,\ldots,N_0$. The bounds $a$ and $b$ allow SSCM to choose external subsets of varying sizes. Here, we set $a = K_1$ and $b = K_1 + 4$, where $K_1$ denotes the number of external trajectories generated by the same process as the internal one, namely, the process [1] in settings 1 and 2. This ensures that the selected external subset can include all external trajectories from the process [1]. The SSCM approach is realized in \textit{R} using the built-in function \textit{constrOptim}.}

\begin{figure}[ht]\centering
      \includegraphics[width=\linewidth]{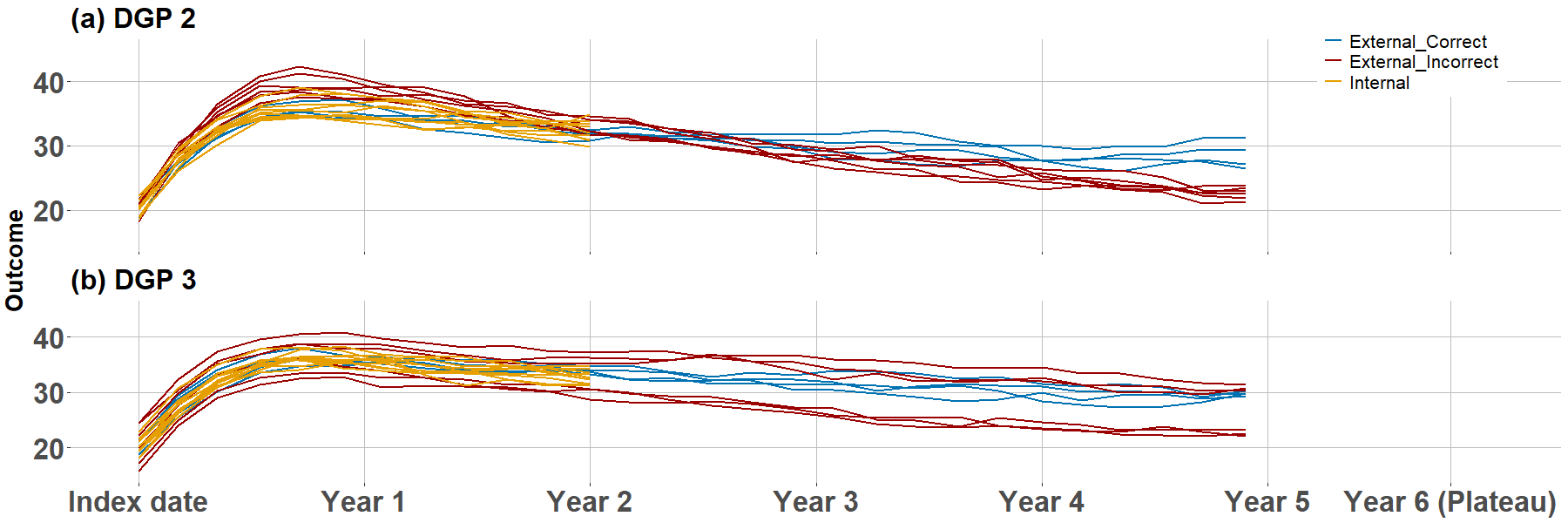}
    \caption{\label{fig: Simul23} Simulated sample data under DGPs 2 and 3. The trajectories are generated under the real-data censoring patterns.}
\end{figure}

  {Two additional DGPs, denoted as DGP 2 and 3, are introduced in this setting to explore the conditions under which our approach may outperform SSCM and vice versa. For both DGPs, we assume each trajectory has 34 longitudinal observations, with the time schedule evenly spaced over the entire 6-year study duration. After generating the external and internal trajectories, we manually censor the internal and external trajectories at 2 years and 5 years post-treatment, respectively. For the internal trajectories, the true CCHs parameter is set to $\boldsymbol{\theta}^* = (20,1,35,-0.05,28)$. The external trajectories, however, are generated from a mixture of three generating processes, labeled as processes [1], [2] and [3]. The corresponding number of external trajectories are denoted by $K_1$, $K_2$ and $K_3$. In DGP 2, the CCHs parameters $\boldsymbol{\theta}^*$'s are $(20,1,35,-0.05,28)$ for process [1] and $(20,1.2,38,-0.08,20)$ for processes [2] and [3]. In DGP 3, the CCHs parameters $\boldsymbol{\theta}^*$'s are $(20,1,35,-0.05,28)$ for process [1], $(24,1,39,-0.05,30)$ for process [2], and $(16,1,31,-0.05,22)$ for process [3]. In DGP 3, the mean trends in processes [2] and [3] are designed to approximate the vertical shifts of the mean trend in process [1], with a slight tilt in the long-term progression of process [2]. In Figure \ref{fig: Simul23}, we present 10 trajectories for each generating process by choosing $\rho$ to be 50, $K_1 = 4$, and $K_2 = K_3 = 3$.}

  {To evaluate the prediction performance of our approach and SSCM, we use relative error metrics to assess the accuracy of predictions for the true outcome values at 3 years and 5 years post-treatment, denoted as $\ell_3$ and $\ell_5$, respectively. We also define $p_{\text{s};y}$ as an indicator function, taking value 1 if the $\ell_3$ or $\ell_5$ of SSCM is smaller than that of SP. Additionally, we evaluate their proportions of correct and incorrect selections, denoted as $w_{[1]}$ and $w_{[2]}$, as well as the preference level for favoring the correct external trajectories, denoted by $w_{\geq}$. The definitions for these metrics are defined similarly as the ones introduced in Section 1 of the main article, detailed in Section 2.2 of the Supplementary File. The results for DGPs 2 and 3 are provided in Table \ref{tab 1: setting2} and \ref{tab 2: setting2}, with the reported statistics annotated in the footnotes, based on 100 Monte Carlo replications.}

\begin{table}[h]
    \centering
        \caption{Predict performance of SP and SSCM across 100 Monte Carlo replications for DGPs 2 and 3, evaulated under varying values of $K_1$, $K_2$, $K_3$ and $\rho$.}
    \label{tab 1: setting2}
   \begin{tabular}{l|l|l|l|l|l|l|l}
   \toprule
    \toprule
  \multicolumn{4}{c|}{}& \multicolumn{2}{c|}{SP} & \multicolumn{2}{c}{SSCM} \\
    \midrule
DGP & $\rho$ & $K_1$ & $K_2=K_3$  & $\ell_{3}$ & $\ell_{5}$ & $\ell_{3}$ & $\ell_{5}$ \\
\midrule
2 & 10 & 2 & 4 & 0.03 & 0.17 & 0.02 (0.79) & 0.06 (0.93) \\
  &    & 4 & 3 & 0.02 & 0.09 & 0.01 (0.65) & 0.04 (0.85) \\
  & 30 & 2 & 4 & 0.03 & 0.14 & 0.02 (0.62) & 0.07 (0.71) \\
  &    & 4 & 3 & 0.02 & 0.09 & 0.02 (0.52) & 0.05 (0.77) \\
  & 50 & 2 & 4 & 0.03 & 0.16 & 0.02 (0.61) & 0.07 (0.75) \\
  &    & 4 & 3 & 0.02 & 0.10 & 0.02 (0.56) & 0.05 (0.77) \\
  \midrule
3 & 10 & 2 & 4 & 0.03 & 0.10 & 0.01 (0.66) & 0.01 (0.78) \\
  &    & 4 & 3 & 0.02 & 0.07 & 0.01 (0.63) & 0.01 (0.86) \\
  & 30 & 2 & 4 & 0.03 & 0.14 & 0.01 (0.76) & 0.02 (0.90) \\
  &    & 4 & 3 & 0.01 & 0.05 & 0.01 (0.45) & 0.02 (0.82) \\
  & 50 & 2 & 4 & 0.03 & 0.13 & 0.01 (0.67) & 0.02 (0.90) \\
  &    & 4 & 3 & 0.01 & 0.05 & 0.01 (0.47) & 0.02 (0.83) \\
  \bottomrule
  \bottomrule
    \end{tabular}
    \newline
    \footnotesize\textsuperscript{*} Median values (with mean $p_{\text{s};y}$) for $\ell_{3}$ and $\ell_{5}$.
\end{table}

  {The main finding from Table \ref{tab 1: setting2} is that SSCM performs slightly better than our approach for 3 years outcome prediction and significantly better than our approach in 5 years prediction across DGPs 2 and 3. This is indicated by $p_{\text{s};y}$ values exceeding 0.6 in the $\ell_3$ column and greater than 0.75 in the $\ell_5$ column. We attribute this result to two main factors. First, our approach lacks the dynamic weighting flexibility inherent to SSCM. Specifically, once an external subset is selected, our approach assigns equal weight to all external trajectories within that subset, which offers less flexibility in optimizing weights for prediction than SSCM. Second, DGPs 2 and 3 present challenges for our approach, as the incorrect external trajectories are similar to the correct ones in terms of the CCHs parameter. In these scenarios, our approach is more prone to make Type \rom{2} errors, as reflected by the high $w_{[2]}$ values in Table \ref{tab 2: setting2}. In summary, SSCM generally outperforms our approach in both short-term and long-term outcome predictions under these DGPs.}

\begin{table}[h]
    \centering
        \caption{Preference level of SP and SSCM across 100 Monte Carlo replications for DGPs 2 and 3, evaulated under varying values of $K_1$, $K_2$, $K_3$ and $\rho$.}
    \label{tab 2: setting2}
   \begin{tabular}{l|l|l|l|l|l|l|l|l|l}
   \toprule
    \toprule
  \multicolumn{4}{c|}{}& \multicolumn{3}{c|}{SP} & \multicolumn{3}{c}{SSCM} \\
    \midrule
DGP & $\rho$ & $K_1$ & $K_2=K_3$  & $w_{[1]}$ & $w_{[2]}$ & $w_{\geq}$ & $w_{[1]}$ & $w_{[2]}$ & $w_{\geq}$ \\
\midrule
2 & 10 & 2 & 4 & 0.40 & 0.60 & 0.48 & 0.81 & 0.19 & 0.99 \\
  &    & 4 & 3 & 0.67 & 0.33 & 0.81 & 0.88 & 0.12 & 1.00 \\
  & 30 & 2 & 4 & 0.50 & 0.50 & 0.59 & 0.77 & 0.23 & 0.93 \\
  &    & 4 & 3 & 0.67 & 0.33 & 0.72 & 0.84 & 0.16 & 0.99\\
  & 50 & 2 & 4 & 0.50 & 0.50 & 0.51 & 0.75 & 0.25 & 0.92 \\
  &    & 4 & 3 & 0.67 & 0.33 & 0.62 & 0.83 & 0.17 & 0.99 \\
  \midrule
3 & 10 & 2 & 4 & 0.33 & 0.67 & 0.42 & 0.18 & 0.82 & 0.00\\
  &    & 4 & 3 & 0.75 & 0.25 & 0.87 & 0.42 & 0.58 & 0.34\\
  & 30 & 2 & 4 & 0.50 & 0.50 & 0.53 & 0.17 & 0.83 & 0.02\\
  &    & 4 & 3 & 0.75 & 0.25 & 0.97 & 0.43 & 0.57 & 0.38 \\
  & 50 & 2 & 4 & 0.25 & 0.75 & 0.29 & 0.17 & 0.83 & 0.02 \\
  &    & 4 & 3 & 0.67 & 0.33 & 0.78 & 0.43 & 0.57 & 0.32\\
  \bottomrule
  \bottomrule
    \end{tabular}
    \newline
    \footnotesize\textsuperscript{*} Median values for $w_{[1]}$ and $w_{[2]}$; Mean values for $w_{\geq}$.
\end{table}

  {Nevertheless, in DGP 3, where integrating processes [2] and [3] yields a mean trend closely aligns with the internal trend, our approach outperforms SSCM in selecting the correct external trajectories. This is indicated by the higher $w_{\geq}$ values for our approach in DGP 3, as shown in Table \ref{tab 2: setting2}. One possible explanation is that SSCM method selects external subsets based on minimizing the distance between the synthesized trend of the selected external trajectories and the internal trend. This, however, does not directly account for the differences in trend progression. Consequently, SSCM is more prone to selecting external trajectories that, although generated differently from the internal trajectories, resemble the internal trend, rendering potentially misleading results by incorporating non-negligible portion of irrelevant external trajectories.}

  {Another key difference between our approach and SSCM is that SSCM does not assume an explicit generating process for the internal and selected external trajectories, while our approach assumes that both sets of trajectories are generated by the same underlying process. This assumption enhances the interpretability of our method compared to SSCM. Additionally, SSCM requires pre-determined bounds $a$ and $b$ for selecting external subsets, whereas our approach infers the external subset through a valid Bayesian procedure, providing valid inferential results that could inform future studies.} 

\subsection{Setting 3, stress test}

\begin{figure}[ht]\centering
      \includegraphics[width=\linewidth]{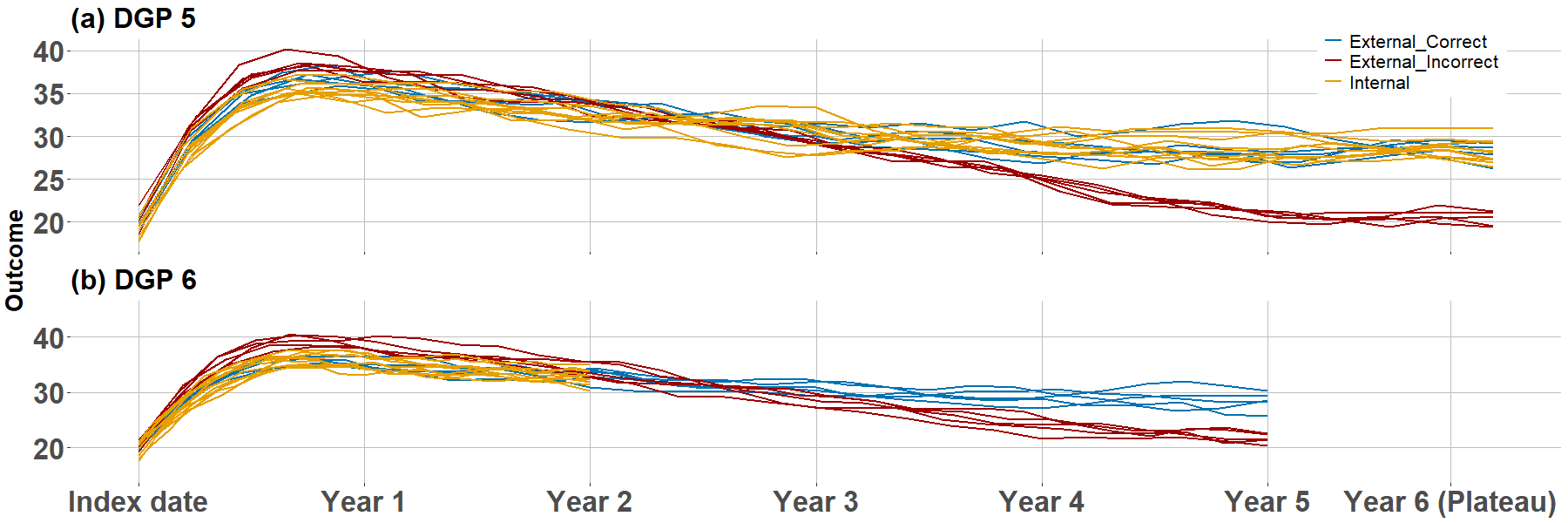}
        \caption{\label{fig: Simul56} Simulated sample data under DGPs 5 and 6. The trajectories are generated with the corresponding censoring patterns.}
\end{figure}

  {In this setting, we assume that the correct and incorrect external trajectories are generated similarly in terms of their CCHs parameters. Specifically, for the internal trajectories, we set the true CCHs parameter as $\boldsymbol{\theta}^* = (20,1,35,-0.05,28)$. For the external trajectories, the CCHs parameters $\boldsymbol{\theta}^*$'s are a mixture of $(20,1,35,-0.05,28)$ and $(20,1.2,38,-0.08,20)$. This setting is challenging for our method because the cross-sectional differences in mean outcomes between the correct and incorrect external trajectories are mostly within one standard deviation of the cross-sectional variability. We consider two DGPs in this setting, labeled DGPs 5 and 6, corresponding to scenarios without censoring and with the real-data censoring patterns. The time schedules applied are specified in Section \ref{appenx: time schedule}. In Figure \ref{fig: Simul56}, we present 10 trajectories for each generating processes given $\rho = 50$, and $K_1 = K_2 = 5$ for illustration. The simulation results are detailed in Table \ref{tab 1: setting3} and \ref{tab 2: setting3}.}

  {In cases where the incorrect and correct external trajectories are similar during the early-stage but diverge later, our approach does not exhibit a significant advantage over its competitors and is occasionally surpassed, particularly in early-stage parameter estimation ($\ell_{\text{S}}$ and $p_{\text{s}}$). This diminished performance is likely due to the increased likelihood of including incorrect external trajectories in DGPs 5 and 6, as these trajectories exhibit higher similarity to the correct external ones, reflected in the closer distance in their CCHs parameters, compared to that in DGPs 1 and 4. This is evidenced by the trend of higher $p_{[2]}$ in Table \ref{tab 2: setting3} compared to Table \ref{tab 2: setting1 appendx} and Table 2 in the main article. We need to underscore that despite the challenges posed by DGPs 5 and 6, our approach still achieves a more precise plateau value estimation compared to the DC approach, indicated by the small $p_{\text{s}}$ values under the $\ell_{\text{P}}$ column of the DC sector. This provides further empirical evidence that our approach is more robust than the DC approach in the settings where useful external information is mixed with trajectories generated by multiple irrelevant processes.}

\begin{table}[h]
    \centering
        \caption{Performance of the three methods in parameter estimation across 100 Monte Carlo replications for DGPs 5 and 6, evaluated under varying values of $K_1$, $K_2$ and $\rho$.}
    \label{tab 1: setting3}
   \begin{tabular}{l|l|l|l|l|l|l|l|l|l}
   \toprule
    \toprule
  \multicolumn{4}{c|}{}& \multicolumn{2}{c|}{SP} & \multicolumn{2}{c|}{DC} & \multicolumn{2}{c}{NB} \\
    \midrule
DGP & $\rho$ & $K_1$ & $K_2$  & $\ell_{\text{S}}$ & $\ell_{\text{P}}$ & $\ell_{\text{S}}$ & $\ell_{\text{P}}$ & $\ell_{\text{S}}$ & $\ell_{\text{P}}$ \\
    \midrule
5 & 0.19 & 2 & 8 & 0.07 & 0.00 & 0.06 (0.50) & 0.01 (0.03) & 0.07 (0.43) & 0.00 (0.66) \\
  &  & 5 & 5 & 0.08 & 0.00 & 0.08 (0.55) & 0.01 (0.06) & 0.08 (0.43) & 0.00 (0.68)\\
  &  & 8 & 2 & 0.07 & 0.00 & 0.07 (0.53) & 0.00 (0.12) & 0.07 (0.48) & 0.00 (0.60)\\
  & 0.58 & 2 & 8 & 0.08 & 0.00 & 0.08 (0.45) & 0.02 (0.01) & 0.08 (0.47) & 0.00 (0.55)\\
  &  & 5 & 5 & 0.08 & 0.00 & 0.09 (0.50) & 0.01 (0.08) & 0.09 (0.37) & 0.00 (0.63) \\
  &  & 8 & 2 & 0.08 & 0.00 & 0.07 (0.48) & 0.01 (0.16) & 0.07 (0.49) & 0.00 (0.55)\\
  & 0.96 & 2 & 8 & 0.08 & 0.00 & 0.08 (0.43) & 0.02 (0.03) & 0.07 (0.47) & 0.00 (0.65) \\
  &  & 5 & 5 & 0.08 & 0.00 & 0.08 (0.45) & 0.01 (0.12) & 0.09 (0.38) & 0.03 (0.59) \\
  &  & 8 & 2 & 0.07 & 0.00 & 0.08 (0.48) & 0.01 (0.22) & 0.07 (0.49) & 0.00 (0.53)\\ 
  \midrule
6 & 0.19 & 2 & 8 & 0.28 & 0.18 & 0.30 (0.44) & 0.23 (0.19) & 0.55 (0.11) & 0.37 (0.09) \\
  &  & 5 & 5 & 0.12 & 0.07 & 0.19 (0.34) & 0.14 (0.19) & 0.44 (0.04) & 0.34 (0.00) \\
  &  & 8 & 2 & 0.12 & 0.05 & 0.13 (0.50) & 0.06 (0.25) & 0.46 (0.05) & 0.33 (0.02) \\
  & 0.58 & 2 & 8 & 0.24 & 0.14 & 0.24 (0.49) & 0.23 (0.25) & 0.32 (0.34) & 0.21 (0.30)\\
  &  & 5 & 5 & 0.12 & 0.05 & 0.16 (0.53) & 0.14 (0.24) & 0.25 (0.26) & 0.18 (0.12)\\
  &  & 8 & 2 & 0.12 & 0.04 & 0.11 (0.63) & 0.06 (0.36) & 0.25 (0.24) & 0.18 (0.15) \\
  & 0.96 & 2 & 8 & 0.19 & 0.15 & 0.23 (0.53) & 0.23 (0.30) & 0.22 (0.51) & 0.13 (0.43)\\
  &  & 5 & 5 & 0.13 & 0.04 & 0.15 (0.57) & 0.14 (0.28) & 0.16 (0.37) & 0.12 (0.27)\\
  &  & 8 & 2 & 0.13 & 0.02 & 0.10 (0.70) & 0.06 (0.34) & 0.16 (0.41) & 0.12 (0.20)\\
  \bottomrule
  \bottomrule
    \end{tabular}
    \newline
    \footnotesize\textsuperscript{*} Median values (with mean $p_{\text{s}}$) for $\ell_{\text{S}}$ and $\ell_{\text{P}}$.
\end{table}

\begin{table}[h]
    \centering
        \caption{Preference level of our approach (SP) and coverage level of the true outcome value at 3 years post-treatment for the three methods across 100 Monte Carlo replications for DGPs 5 and 6, with varying values of $K_1$, $K_2$ and $\rho$.}
    \label{tab 2: setting3}
   \begin{tabular}{l|l|l|l|l|l|l|l|l|l|l|l|l}
   \toprule
    \toprule
  \multicolumn{4}{c|}{}& \multicolumn{5}{c|}{SP} & \multicolumn{2}{c|}{DC} & \multicolumn{2}{c}{NB} \\
    \midrule
DGP & $\rho$ & $K_1$ & $K_2$ & $p_{[1]}$ & $p_{[2]}$ & $p_{\geq}$ & $Cvr_3$ & $len_3$ & $Cvr_3$ & $len_3$ & $Cvr_3$ & $len_3$ \\
    \midrule
5 & 0.19 & 2 & 8 & 0.50 & 0.25 & 0.63 & 0.99 & 0.48 & 1.00 & 0.55 (0.00) & 0.99 & 0.44 (1.00) \\
  &  & 5 & 5 & 0.60 & 0.20 & 0.79 & 1.00 & 0.49 & 1.00 & 0.53 (0.02) & 1.00 & 0.45 (0.95)\\
  &  & 8 & 2 & 0.63 & 0.50 & 0.63 & 1.00 & 0.47 & 1.00 & 0.49 (0.13) & 0.98 & 0.45 (0.83)\\
  & 0.58 & 2 & 8 & 0.50 & 0.13 & 0.68 & 0.97 & 0.54 & 1.00 & 0.72 (0.00) & 0.97 & 0.52 (0.77) \\
  &  & 5 & 5 & 0.60 & 0.20 & 0.98 & 0.99 & 0.53 & 1.00 & 0.67 (0.00) & 0.98 & 0.52 (0.65) \\
  &  & 8 & 2 & 0.50 & 0.50 & 0.77 & 0.98 & 0.53 & 1.00 & 0.58 (0.04) & 0.98 & 0.51 (0.62)\\
  & 0.96 & 2 & 8 & 0.50 & 0.13 & 0.71 & 0.98 & 0.58 & 1.00 & 0.72 (0.00) & 0.97 & 0.55 (0.87)\\
  &  & 5 & 5 & 0.60 & 0.00 & 0.99 & 0.98 & 0.57 & 1.00 & 0.70 (0.00) & 0.98 & 0.55 (0.70) \\
  &  & 8 & 2 & 0.50 & 0.00 & 0.87 & 0.96 & 0.56 & 1.00 & 0.63 (0.01) & 0.96 & 0.55 (0.60)\\ 
  \midrule
6 & 0.19 & 2 & 8 & 1.00 & 0.38 & 0.86 & 0.14 & 0.96 & 0.00 & 0.95 (0.51) & 0.58 & 5.28 (0.00) \\
  &  & 5 & 5 & 0.80 & 0.20 & 0.83 & 0.34 & 0.86 & 0.00 & 0.96 (0.13) & 0.62 & 5.32 (0.00) \\
  &  & 8 & 2 & 0.63 & 0.50 & 0.80 & 0.73 & 0.80 & 0.66 & 0.80 (0.43) & 0.63 & 5.22 (0.00)\\
  & 0.58 & 2 & 8 & 0.50 & 0.13 & 0.81 & 0.34 & 1.06 & 0.00 & 1.00 (0.77) & 0.75 & 4.19 (0.00) \\
  &  & 5 & 5 & 0.60 & 0.20 & 0.79 & 0.56 & 0.91 & 0.00 & 1.01 (0.24) & 0.74 & 4.06 (0.00) \\
  &  & 8 & 2 & 0.50 & 0.50 & 0.69 & 0.66 & 0.87 & 0.66 & 0.90 (0.44) & 0.78 & 4.10 (0.00)\\
  & 0.96 & 2 & 8 & 0.50 & 0.13 & 0.70 & 0.38 & 1.10 & 0.00 & 0.95 (0.89) & 0.84 & 3.49 (0.00) \\
  &  & 5 & 5 & 0.60 & 0.00 & 0.75 & 0.57 & 0.89 & 0.00 & 0.95 (0.34) & 0.82 & 3.40 (0.00)\\
  &  & 8 & 2 & 0.50 & 0.00 & 0.67 & 0.70 & 0.85 & 0.61 & 0.89 (0.41) & 0.88 & 3.40 (0.00)\\
  \bottomrule
  \bottomrule
    \end{tabular}
    \newline
    \footnotesize\textsuperscript{*} Median values for $p_{[1]}$, $p_{[2]}$; Mean values for $p_{\geq}$ and $\text{Cvr}_3$; Median value (with mean $p_{\text{s};3}$) for $len_3$.
\end{table}

  {In DGP 6, it is notable that the coverage level of our approach is significantly lower than that of the NB approach, which serves as the benchmark for coverage, as shown in Table \ref{tab 2: setting3}. This discrepancy arises primarily because our approach tends to select more incorrect external trajectories, as indicated by relatively high $p_{[2]}$ values compared to DGPs 1 and 4, leading to increased estimation error. Nevertheless, our approach still provides credible intervals with relatively higher coverage levels than the DC approach, while maintaining narrower intervals than those produced by the NB approach.}

\section{A discussion on the margina likelihood approximation}\label{appenx: ml dis}

  {In this section, we revisit the issue discussed in Section 4.1 in the main article, focusing on the stochastic error in marginal likelihood approximation due to limited samples. In theory, our approach would report identical values for $Z_{\hat{\mathcal{C}}}$ in the real data analysis across the 100 chains if each chain were run for a sufficiently long duration. However, this is not the case for the results presented in Table 6 of the main article, as the entry-wise means do not converge toward 0 or 1, and certain pseudo IDs exhibit non-zero SD values. There are two potential possibilities for these findings. First, the stochastic error may come from the insufficient chain length. This possibility is less likely in our case, as each chain has been run for 10,000 iterations -- an adequate length to reach the stationary distribution. Second, the marginal likelihood approximation may have a non-negligible stochastic error, which is the more plausible cause of the non-identical $Z_{\hat{\mathcal{C}}}$ values. While increasing the number of samples would improve the accuracy of the marginal likelihood approximation, hence leading to more entires of $Z_{\hat{\mathcal{C}}}$ converging toward 0 or 1, it would also increase computation time. In our real data analysis, we trade some approximation accuracy for a shorter sampling duration. As most subjects in Table 6 of the main article exhibit entry-wise mean values close to 0 or 1 (e.g., within 0.2 of 0 or 1), we believe that the marginal likelihood approximation remains reliable in revealing the ideal $Z_{\hat{\mathcal{C}}}$, which should be invariant to across-chain initializations.}

\section{Additional sensitivity analyses}\label{appenx: sensitivity}

In this section, we provide the numeric results for the estimated trend given different hyper-parameter settings and plateau points. In Table \ref{tab:5}, we present the hyper-parameter settings used for the sensitivity analysis in Figure 5(b) of the main article.

\begin{table}[h]
    \centering
        \caption{The hyper-parameter settings of the two settings that are under investigation.}
    \label{tab:5}
   \begin{tabular}{l|l|l|l|l}
   \toprule
    \toprule
Setting & $\nu_0$ & $\Psi_0$ & $a_0$ & $b_0$\\
\midrule
\textbf{Current} & $10^{-2}$ & $10^{2}\times \text{I}_{5\times 5}$ & $10^{-2}$ & $10^{-2}$\\
\midrule
Informative & $10^{-1}$ & $10\times \text{I}_{5\times 5}$ & $10^{-1}$ & $10^{-1}$\\
\midrule
Vague & $1e^{-3}$ & $1e^{3}\times \text{I}_{5\times 5}$ & $1e^{-3}$ & $1e^{-3}$\\
  \bottomrule
  \bottomrule
    \end{tabular}
\end{table}

Additionally, to assess whether a higher plateau point ($T$) significantly impacts inference outcomes, we examine scenarios where $T$ is set to be $9$ and $10$ and compare these results with the baseline setting of $T = 8$, which is presented in Figure \ref{fig: ThreeTrends}. The result of comparing $T = 6$ with $T = 7, 8$ has been presented in Figure 5(c) of the main article. The detailed results are detailed in Table \ref{tab: CIresultsHP} and \ref{tab: CIresultsT}. 

\begin{figure}[htp]
\minipage{1\textwidth}
  \includegraphics[width=\linewidth]{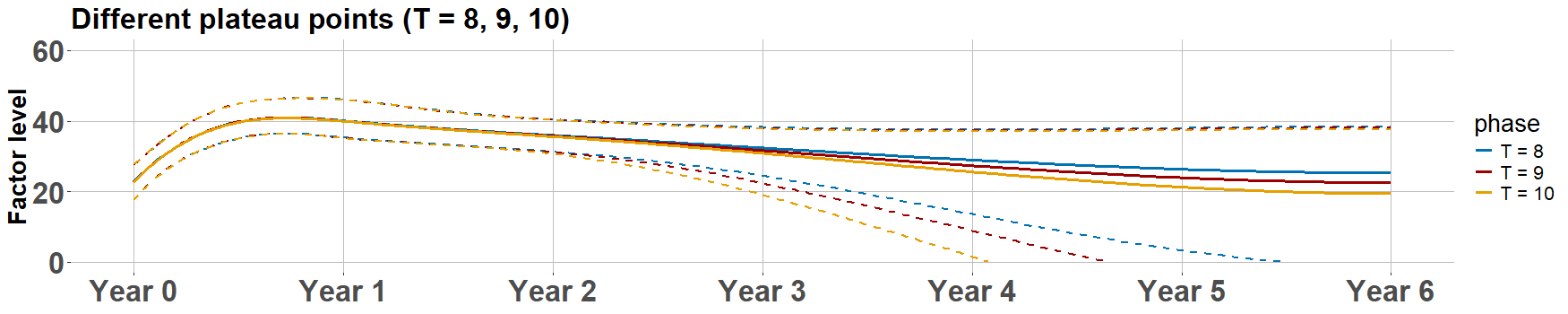}
\endminipage\hfill
\caption{\label{fig: ThreeTrends} The estimated trends (median) given by different plateau points, plotted with its $95\%$ credible interval for each.}
\end{figure}

\begin{table}[h]
    \centering
        \caption{The posterior median value (MD), the lower (LB) and upper (UB) bounds of the 95 $\%$ credible interval for the annual factor levels, the turning point $\alpha$ and the temporal correlation $\rho$ given different hyper-parameters.}
    \label{tab: CIresultsHP}
   \begin{tabular}{l|l|l|l|l|l|l|l|l|l}
   \toprule
    \toprule
\multicolumn{2}{c|}{\textbf{Method}} & Year 0 & Year 1  & Year 2 & Year 3 & Year 4 & Year 5 & $\alpha$ & $\rho$ \\
    \midrule
\textbf{Current} & MD & 22.93 & 40.10 & 36.11 & 33.42 & 31.23 & 29.89 & 1.29 & 1.12\\
                   & LB & 17.90 & 35.30 & 31.48 & 27.29 & 21.67 & 16.13 & 1.01 & 0.89\\
                   & UB & 27.79 & 46.17 & 40.46 & 38.34 & 37.66 & 38.10 & 2.08 & 1.41\\
\midrule
\textbf{Informative} & MD & 22.89 & 40.06 & 36.08 & 33.34 & 31.11 & 29.72 & 1.28 & 1.11\\
                     & LB & 17.90 & 35.30 & 31.48 & 27.24 & 21.60 & 16.12 & 1.01 & 0.88\\
                     & UB & 27.73 & 46.15 & 40.52 & 38.43 & 37.72 & 38.17 & 2.10 & 1.39\\
\midrule
\textbf{Vague} & MD & 22.88 & 40.06 & 36.03 & 33.23 & 30.96 & 29.53 & 1.29 & 1.12\\
                   & LB & 17.90 & 35.26 & 31.33 & 27.14 & 21.32 & 15.59 & 1.01 & 0.89\\
                   & UB & 27.74 & 46.12 & 40.46 & 38.32 & 37.65 & 38.03 & 2.07 & 1.49\\
  \bottomrule
  \bottomrule
    \end{tabular}
\end{table}

Our sensitivity analysis indicate that varying hyper-parameter settings has minimal impact on long-term outcome inferences, as demonstrated in Table \ref{tab: CIresultsHP}. Additionally, increasing plateau time $T$ results in wider credible intervals, as shown in Table \ref{tab: CIresultsT}, particularly beyond year 2 post-treatment, where most internal trajectories are censored. This finding supports our conjecture in the main article that information from selected external trajectories beyond $\alpha$ contributes to inferring long-term outcome over extended time frames, potentially reducing the effective sample size for these longer-term estimates. Importantly, early-stage inferences remain relatively unaffected by increases in $T$.

\begin{table}[h]
    \centering
        \caption{The posterior median value (MD), the lower (LB) and upper (UB) bounds of the 95 $\%$ credible interval for the annual factor levels, the turning point $\alpha$ and the temporal correlation $\rho$ given different $T$ values.}
    \label{tab: CIresultsT}
   \begin{tabular}{l|l|l|l|l|l|l|l|l|l}
   \toprule
    \toprule
\multicolumn{2}{c|}{\textbf{Method}} & Year 0 & Year 1  & Year 2 & Year 3 & Year 4 & Year 5 & $\alpha$ & $\rho$ \\
    \midrule
\textbf{$\boldsymbol{T}$ = 7} & MD & 22.84 & 40.01 & 36.01 & 32.83 & 30.03 & 28.05 & 1.27 & 1.11\\
                              & LB & 17.80 & 35.18 & 31.32 & 25.93 & 17.56 & 9.57 & 1.01 & 0.88\\
                              & UB & 27.72 & 45.94 & 40.45 & 38.23 & 37.66 & 38.15 & 2.11 & 1.41\\
\midrule
\textbf{$\boldsymbol{T}$ = 8} & MD & 22.87 & 40.12 & 35.96 & 32.39 & 28.95 & 26.35 & 1.28 & 1.10\\
                              & LB & 17.82 & 35.31 & 31.29 & 24.46 & 13.61 & 3.30 & 1.01 & 0.88\\
                              & UB & 27.85 & 46.05 & 40.43 & 38.24 & 37.59 & 37.98 & 2.14 & 1.40\\
\midrule
\textbf{$\boldsymbol{T}$ = 9} & MD & 22.73 & 39.99 & 35.78 & 31.67 & 27.36 & 23.94 & 1.27 & 1.10\\
                              & LB & 17.73 & 35.18 & 31.09 & 22.38 & 8.94 & -3.84 & 1.01 & 0.90\\
                              & UB & 27.80 & 45.93 & 40.32 & 37.87 & 37.30 & 37.68 & 2.13 & 1.42\\
                   \midrule
\textbf{$\boldsymbol{T}$ = 10} & MD & 22.67 & 39.96 & 35.60 & 30.88 & 25.61 & 21.31 & 1.28 & 1.09\\
                               & LB & 17.72 & 35.14 & 30.66 & 19.10 & 1.57 & -14.51 & 1.01 & 0.89\\
                               & UB & 27.84 & 45.96 & 40.21 & 37.82 & 37.08 & 37.39 & 2.14 & 1.39\\
  \bottomrule
  \bottomrule
    \end{tabular}
\end{table}

\newpage

\section{Additional simulations for validity check}\label{appen: additional simul}

In this section, we provide additional simulation settings to check our model's validity from two perspectives, (1) i.i.d. samples are generated from parametric distributions, (2) mixed effect models are considered for generating data. 

\subsection{Samples generated i.i.d. from parametric distributions}\label{appen: iid}

Suppose $X_1,\ldots,X_{N_0}\stackrel{i.i.d.}{\sim} p_{\theta_0}$ are external samples, and $Y_1,\ldots,Y_{N_1}\stackrel{i.i.d.}{\sim} p_{\theta_1}$ are internal samples. We use the \textbf{uniform} prior defined in the main article. Next, we compare the performance of our selection procedure (SP) using the coverage probability of $\theta_1$ given by $95\%$ credible intervals with two alternative strategies:
\begin{itemize}
    \item A baseline strategy (BSL) that only uses internal samples, and
    \item A direct combination (DC) strategy that merges both external and internal datasets.
\end{itemize} 
To conduct this comparison, we perform 1,000 Monte Carlo replications to compute the coverage probability, and collect 1,000 posterior samples to obtain the $95\%$ credible intervals given the most representative external subset $\hat{\mathcal{C}}$, defined as follows, 
\begin{equation}
    \begin{split}
        & \hat{\mathcal{C}} = \argmin_{\mathcal{C}\in \text{Post}(\mathcal{C})}\|\boldsymbol{Z_{\mathcal{C}}} - \boldsymbol{\bar{Z}_{\mathcal{C}}}\|_2, \\
        & \boldsymbol{\bar{Z}_{\mathcal{C}}} = \frac{1}{|\text{Post}(\mathcal{C})|}\sum_{\mathcal{C}\in \text{Post}(\mathcal{C})}\boldsymbol{Z_{\mathcal{C}}}, \\
    \end{split}
\end{equation}
where $\|\cdot\|_2$ denotes the $\ell^2$ vector norm, $\boldsymbol{Z_{\mathcal{C}}}$ denotes a vector of length $N_0$, with the $i$-th entry being 1 if $X_i$ is selected ($i\in\mathcal{C}$) and 0 otherwise, $\text{Post}(\mathcal{C})$ represents the collection of the posterior samples of $\mathcal{C}$. To improve the efficiency of the sampling algorithm, we randomly initialize the external subset and employ the following Metropolis-Hastings proposal $q(\mathcal{C}_{\text{new}}\mid \mathcal{C}_{\text{old}},\mathcal{I}_{\text{prop}},\mathcal{C}_{\text{max}})$, which assigns positive mass to the best external subset $\mathcal{C}_{\text{max}}$ identified thus far based on the posterior marginal likelihood,
\begin{equation}
    \begin{split}\label{MHNew}
        & q(\mathcal{C}_{\text{new}}\mid \mathcal{C}_{\text{old}},\mathcal{I}_{\text{old}},\mathcal{C}_{\text{max}}) = 0.1 \times \mathbbm{1}(\mathcal{C} = \mathcal{C}_{\text{max}}) + 0.9 \times q^*(\mathcal{C}\mid \mathcal{C}_{\text{old}}, \mathcal{I}_{\text{prop}}),\\
        & \mathcal{I}_{\text{prop}} = [(\mathcal{C}_{\text{max}} \cup \mathcal{C}_{\text{old}})\setminus(\mathcal{C}_{\text{max}} \cap \mathcal{C}_{\text{old}})] \cup \mathcal{C}_{\text{prop}},
    \end{split}
\end{equation}
where $\mathcal{C}_{\text{prop}}$ is a subset from $(\mathcal{C}_{\text{max}} \cap \mathcal{C}_{\text{old}})$ that is randomly selected in each MCMC iteration. To define $q^*(\mathcal{C}\mid \mathcal{C}_{\text{old}}, \mathcal{I}_{\text{prop}})$, we introduce $\boldsymbol{Z}$ and $\boldsymbol{Z_{\text{old}}}$, which are vectors of length $N_0$, where the $i$-th entry $Z_i$ ($Z_{\text{old},i}$) is 1 if $X_i$ is selected in $\mathcal{C}$ ($\mathcal{C}_{\text{old}}$) and 0 otherwise. A new subset $\mathcal{C}$ generated from the proposal $q^*(\mathcal{C}\mid \mathcal{C}_{\text{old}}, \mathcal{I}_{\text{prop}})$ is then defined as follows,
\begin{equation}
    \begin{split}\label{prop}
        & \mathcal{C} = \{i:~Z_i = 1,~ i=1,\ldots,N_0\},\\
        & Z_i\stackrel{i.i.d.}{\sim} \text{Bernoulli}(0.5),~~ i\in \mathcal{I}_{prop}.
    \end{split}
\end{equation}
For the external and internal sample sizes, we set $N_1 = 6 \times N_0$ given $N_0 = 20,~50,~100$ to mimic the larger amount of internal data observed in our real-world application. Additionally, we consider the extreme case $N_1 = N_0 = 100$ to challenge our model. We let $p_{\theta}$ denote the probability mass function of a Bernoulli distribution parameterized by $\theta$ and choose $\theta_0 = 0.2,~\theta_1 = 0.8$ and $\theta_0 = 0.5,~\theta_1 = 0.8$ to evaluate our model's performance across different distances between $\theta_0$ and $\theta_1$.

The results from Bernoulli distributions are presented in Table \ref{tab: BernoulliCoverSpec}. 

\begin{table}[h]
    \centering
        \caption{Data are generated using Bernoulli distributions. The coverage probability of $\theta_1$ is evaluated over 1,000 Monte Carlo replications.}
    \label{tab: BernoulliCoverSpec}
   \begin{tabular}{l|l|l|l|l|l|l}
   \toprule
    \toprule
$N_0$ & $N_1$ & $\theta_0$ & $\theta_1$  & SP & BSL & DC\\
    \midrule
20 & 120 & 0.2 & 0.8 & 0.933 & 0.936 & 0.284\\
50 & 300 &  &  & 0.934 & 0.921 & 0.017\\
100 & 600 &  &  & 0.934 & 0.933 & 0.000\\
100 & 100 &  &  & 0.895 & 0.923 & 0.000\\
\midrule
20 & 120 & 0.5 & 0.8 & 0.912 & 0.927 & 0.729\\
50 & 300 &  &  & 0.925 & 0.928 & 0.470\\
100 & 600 &  &  & 0.917 & 0.943 & 0.194 \\
100 & 100 &  &  & 0.848 & 0.921 & 0.001 \\
  \bottomrule
  \bottomrule
    \end{tabular}
\end{table}

Our main finding is that, the coverage probability of our method (SP) is satisfactory (between $90\%$ and $93\%$) and comparable to that of the baseline (BSL) except for the last row, which is the most difficult scenario when $\theta_0$ and $\theta_1$ taking similar values and the sample sizes are the same. Even when non-trivial external subsets included, with expected sizes of $N_0 \times 0.5/0.8$ when $\theta_0 = 0.5$ and $\theta_1 = 0.8$, the coverage probability of our model is not severely compromised. The selected external subset may introduce slight bias, because it typically centers around the MLE derived from the internal data other than the internal truth. However, this bias can be controlled by the larger internal data size, which has a greater impact on the coverage probability than the selected external subsets. In contrast, the direct combination method fails to maintain the coverage probability because of not ruling out irrelevant external data. 

Additionally, when $\theta_0$ and $\theta_1$ are more separate, i.e., $\theta_0 = 0.2$ and $\theta_1 = 0.8$, our selection procedure can identify most external data are irrelevant, including only external subsets with expected sizes of $N_0 \times 0.2/0.8$. 

In addition to the Bernoulli distributions, we also consider the normal distribution, that is,  $p_{\theta}$ is the density function of $N(\theta,1)$, with designs similar to those used in the Bernoulli case, detailed in Table \ref{tab: NormalCoverSpec}. It can be seen that our proposed method (SP) manages to achieve a satisfactory coverage probability when $\theta_0 = 0$ and $\theta_1=1$. When the difference between $\theta_0$ and $\theta_1$ increases, the coverage increases closer to the nominal level of $95\%$ as expected. 
\begin{table}[h]
    \centering
        \caption{Data are generated using normal distributions. The coverage probability of $\theta_1$ is evaluated over 1,000 Monte Carlo replications.}
    \label{tab: NormalCoverSpec}
   \begin{tabular}{l|l|l|l|l|l|l}
   \toprule
    \toprule
$N_0$ & $N_1$ & $\theta_0$ & $\theta_1$  & SP & BSL & DC \\
    \midrule
20 & 120 & -1 & 1 & 0.926 & 0.931 & 0.153\\
50 & 300 &  &  & 0.933 & 0.925 & 0.003\\
100 & 600 &  &  & 0.931 & 0.918 & 0.000\\ 
100 & 100 &  &  & 0.932 & 0.944 & 0.000\\
\midrule
20 & 120 & 0 & 1 & 0.919 & 0.930 & 0.626\\
50 & 300 &  &  & 0.928 & 0.920 & 0.243\\
100 & 600 &  &  & 0.925 & 0.928 & 0.051\\
100 & 100 &  &  & 0.900 & 0.944 & 0.000\\
  \bottomrule
  \bottomrule
    \end{tabular}
\end{table}

\subsection{Samples from linear mixed effect models}

We generate data using linear mixed-effects models for simplicity. Specifically, we assume the external data follow
    \begin{equation}\label{LME_ext}
        \begin{split}
            & X^{(j)}_{i1},\ldots,X^{(j)}_{iN_i^{(j)}}\mid \begin{pmatrix} \mu^{(j)}_{i1} \\ \mu^{(j)}_{i2}\end{pmatrix} \stackrel{i.i.d.}{\sim} N\left(
            \begin{pmatrix} \mu^{(j)}_{i1} \\ \mu^{(j)}_{i2}\end{pmatrix}, \begin{pmatrix} 
            1, & \rho \\ 
            \rho, & 1
            \end{pmatrix}\right),\\
            & \begin{pmatrix} \mu^{(j)}_{i1} \\ \mu^{(j)}_{i2}\end{pmatrix} \stackrel{i.i.d.}{\sim} N\left(
            \begin{pmatrix} \mu^{(j)}_{1} \\ \mu^{(j)}_{2}\end{pmatrix}, \sigma_0^2 \cdot \begin{pmatrix} 
            1, & \rho \\ 
            \rho, & 1
            \end{pmatrix}\right),~~\text{for}~i=1,\ldots,K^{(j)},\\
            & N_i^{(j)} = \left\lceil R_i^{(j)} \right\rceil,~ R_i^{(j)}\sim \text{Unif}(15,45),~~\text{for}~j = 1, 2,
        \end{split}
    \end{equation}
    and the internal data follow
    \begin{equation}\label{LME_int}
        \begin{split}
            & Y_{i1},\ldots,Y_{iN_i}\mid \begin{pmatrix} \mu_{i1} \\ \mu_{i2}\end{pmatrix} \stackrel{i.i.d.}{\sim} N\left(
            \begin{pmatrix} 
            \mu_{i1} \\ 
            \mu_{i2}
            \end{pmatrix}, \begin{pmatrix} 
            1, & \rho \\ 
            \rho, & 1
            \end{pmatrix}\right),\\
            & \begin{pmatrix} 
            \mu_{i1} \\ 
            \mu_{i2}\end{pmatrix} \stackrel{i.i.d.}{\sim} N\left(
            \begin{pmatrix} 
            \mu_1 \\ 
            \mu_2\end{pmatrix}, \sigma_0^2 \cdot \begin{pmatrix} 
            1, & \rho \\ 
            \rho, & 1
            \end{pmatrix}\right),~~\text{for}~i=1,\ldots,K,\\
            & N_i = \left\lceil R_i \right\rceil,~ R_i\sim \text{Unif}(15,45).
        \end{split}
    \end{equation}    
Here, we select subsets of external trajectories, with each trajectory being bi-variate, denoted by $\{X_{il}^{(j)}\}_{l = 1}^{N_i^{(j)}}$. The selection is based on the first-entry of the bi-variate using a linear mixed-effects model. Our goal is to investigate the coverage probability of $\mu_1$ and $\mu_2$ using $95\%$ credible intervals.
\begin{itemize}
    \item For $\mu_1$, we combine the first entry from both the internal and the selected external trajectories, using the linear mixed-effects model to obtain the posterior samples.
    \item For $\mu_2$, we use only the second entry of the selected external trajectories to mimic the scenario where the internal trajectories are censored after a specific time point in our real application.
\end{itemize} 
In all simulations under this response, we assume $\mu_2^{(1)} = 2$, $\mu_1^{(2)} = \mu_2^{(2)} = \mu_1 = \mu_2 = 1$, $\rho = 0.5$ and $K = 60$. 

\textbf{Our simulation designs correspond to the two key assumptions of our method, as introduced in Section 2.1 in the main article.} 
\begin{itemize}
    \item We assume that the external trajectories that match the internal trajectories before the censoring point will continue to match afterward. That is, for the second part of the external trajectories, when it holds that $\mu_1^{(2)} = \mu_1$, we have $\mu_2^{(2)} = \mu_2$.
    \item We assume the external trajectories differing from the internal trajectories before the censoring point will continue to differ afterward. Specifically, for the first part of the external trajectories, when it holds that $\mu_1^{(1)} \neq \mu_1$, it follows that $\mu_2^{(1)} \neq \mu_2$.
\end{itemize}
These simulation designs allow for information gain, reflected in shorter credible interval lengths, when the data selection method is effective. We evaluate the performance of our method using two main settings: (1) longitudinally independence ($\sigma_0^2 = 0$), and (2) a longitudinal correlation of 0.5 ($\sigma_0^2 = 1$) before and after the censoring point.

Using similar sampling settings introduced in Section \ref{appen: iid}, we present our results in Table \ref{tab: LMECoverSpec10} and \ref{tab: LMECoverSpec20}, under different parameter settings using the \textbf{uniform} prior. The results indicate our method (SP) achieves information augmentation by reducing the median credible interval length for $\mu_1$, particularly when $K^{(2)} \geq K^{(1)}$, though with a slight decrease in coverage probability compared to the baseline method, which does not incorporate external information. For $\mu_2$, the baseline method (BSL) directly samples from the prior due to the censoring of internal trajectories, leading to poor control of the credible interval length. By leveraging external information, our method approximately achieves more than $80\%$ coverage using $95\%$ credible intervals in most cases, particularly when $K^{(2)} \geq K^{(1)}$. We should emphasize that directly combining (DC) the external information, which is systematically different from the internal truth in our simulation designs, results in obvious under-coverage across all the settings. 

\begin{table}[h]
    \centering
        \caption{10 external trajectories are generated using linear mixed-effects models. The coverage probabilities (the median length of $95\%$ credible intervals) of $\mu_1$ and $\mu_2$ are evaluated over 1,000 Monte Carlo replications.}
    \label{tab: LMECoverSpec10}
   \begin{tabular}{l|l|l|l|l|l|l|l|l|l}
   \toprule
    \toprule
\multicolumn{4}{c|}{} & \multicolumn{3}{c}{$\mu_1$} & \multicolumn{3}{|c}{$\mu_2$} \\
    \midrule
$K^{(1)}$ & $K^{(2)}$ & $\mu_1^{(1)}$  & $\sigma_0^2$ & SP & BSL & DC & SP & BSL & DC \\
    \midrule
8 & 2 & -1 & 0 & 0.97 (0.11) & 0.95 (0.10) & 0.00 (0.27) & 1.00 (2.24) & 1.00 ($3e^7$) & 0.00 (0.54)\\
  &   &  & 1 & 0.91 (0.59) & 0.94 (0.51) & 0.61 (0.55) & 0.79 (2.55) & 1.00 ($3e^7$) & 0.31 (1.27)\\
  &   & 0 & 0 & 0.97 (0.11) & 0.95 (0.10) & 0.00 (0.27) & 1.00 (2.24) & 1.00 ($3e^7$) & 0.00 (0.54)\\
  &   &  & 1 & 0.91 (0.50) & 0.94 (0.51) & 0.61 (0.55) & 0.79 (2.55) & 1.00 ($3e^7$) & 0.31 (1.27)\\
  \midrule
5 & 5 & -1 & 0 & 0.95 (0.09) & 0.97 (0.11) & 0.14 (0.26) & 0.98 (0.45) & 1.00 ($3e^7$) & 0.03 (0.66)\\
  &   &  & 1 & 0.91 (0.46) & 0.94 (0.54) & 0.82 (0.55) & 0.89 (2.09) & 1.00 ($3e^7$) & 0.72 (1.32)\\
  &   & 0 & 0 & 0.95 (0.09) & 0.97 (0.11) & 0.14 (0.26) & 0.98 (0.45) & 1.00 ($3e^7$) & 0.03 (0.66)\\
  &   &  & 1 & 0.91 (0.46) & 0.94 (0.54) & 0.82 (0.55) & 0.89 (2.09) & 1.00 ($3e^7$) & 0.72 (1.33)\\
  \midrule
2 & 8 & -1 & 0 & 0.95 (0.09) & 0.97 (0.11) & 0.92 (0.17) & 0.99 (0.38) & 1.00 ($3e^7$) & 0.88 (0.57)\\
  &   &   & 1 & 0.93 (0.45) & 0.94 (0.50) & 0.90 (0.46) & 0.93 (1.70) & 1.00 ($3e^7$) & 0.91 (1.38)\\
  &   & 0 & 0 & 0.95 (0.09) & 0.97 (0.11) & 0.92 (0.17) & 0.99 (0.38) & 1.00 ($3e^7$) & 0.88 (0.57)\\
  &   &   & 1 & 0.93 (0.45) & 0.94 (0.50) & 0.90 (0.46) & 0.93 (1.70) & 1.00 ($3e^7$) & 0.91 (1.38)\\
  \bottomrule
  \bottomrule
    \end{tabular}
\end{table}

\begin{table}[h]
    \centering
        \caption{20 external trajectories are generated using linear mixed-effects models. The coverage probabilities (the median length of $95\%$ credible intervals) of $\mu_1$ and $\mu_2$ are evaluated over 1,000 Monte Carlo replications.}
    \label{tab: LMECoverSpec20}
   \begin{tabular}{l|l|l|l|l|l|l|l|l|l}
   \toprule
    \toprule
\multicolumn{4}{c|}{} & \multicolumn{3}{c}{$\mu_1$} & \multicolumn{3}{|c}{$\mu_2$} \\
    \midrule
$K^{(1)}$ & $K^{(2)}$ & $\mu_1^{(1)}$  & $\sigma_0^2$ & SP & BSL & DC & SP & BSL & DC \\
    \midrule
16 & 4 & -1 & 0 & 0.96 (0.10) & 0.95 (0.10) & 0.00 (0.38) & 1.00 (0.65) & 1.00 ($3e^7$) & 0.00 (0.37)\\
  &   &  & 1 & 0.91 (0.46) & 0.93 (0.49) & 0.17 (0.56) & 0.72 (2.06) & 1.00 ($3e^7$) & 0.07 (0.91)\\
  &   & 0 & 0 & 0.95 (0.10) & 0.96 (0.10) & 0.00 (0.20) & 0.98 (0.80) & 1.00 ($3e^7$) & 0.00 (0.36)\\
  &   &  & 1 & 0.90 (0.45) & 0.94 (0.50) & 0.66 (0.47) & 0.56 (1.69) & 1.00 ($3e^7$) & 0.07 (0.91)\\
  \midrule
10 & 10 & -1 & 0 & 0.95 (0.09) & 0.96 (0.10) & 0.00 (0.28) & 0.99 (0.37) & 1.00 ($3e^7$) & 0.00 (0.49)\\
  &   &  & 1 & 0.91 (0.45) & 0.94 (0.51) & 0.58 (0.52) & 0.86 (1.52) & 1.00 ($3e^7$) & 0.46 (0.97)\\
  &   & 0 & 0 & 0.95 (0.09) & 0.96 (0.10) & 0.05 (0.16) & 0.99 (0.37) & 1.00 ($3e^7$) & 0.00 (0.48)\\
  &   &  & 1 & 0.91 (0.45) & 0.94 (0.51) & 0.84 (0.46) & 0.79 (1.42) & 1.00 ($3e^7$) & 0.45 (0.96)\\
  \midrule
4 & 16 & -1 & 0 & 0.97 (0.10) & 0.97 (0.11) & 0.59 (0.20) & 0.99 (0.26) & 1.00 ($3e^7$) & 0.36 (0.36)\\
  &   &   & 1 & 0.90 (0.45) & 0.94 (0.52) & 0.90 (0.48) & 0.93 (1.18) & 1.00 ($3e^7$) & 0.85 (0.93)\\
  &   & 0 & 0 & 0.97 (0.10) & 0.96 (0.11) & 0.71 (0.12) & 0.99 (0.27) & 1.00 ($3e^7$) & 0.36 (0.36)\\
  &   &   & 1 & 0.91 (0.45) & 0.94 (0.51) & 0.94 (0.45) & 0.91 (1.17) & 1.00 ($3e^7$) & 0.85 (0.93)\\
  \bottomrule
  \bottomrule
    \end{tabular}
\end{table}

\section{External links}\label{appenx: ext link}
The figure given in the oral presentation by \citet{samelson2021follow} is available at \url{https://genetherapy.isth.org/follow-up-of-more-than-5-years-in-a-cohort-of-patients-with-hemophilia-b-treated-with-fidanacogene-elaparvovec-adeno-associated-virus-gene-therapy}.

\spacingset{0.7}
\bibliographystyle{chicago}
\bibliography{ref.bib}

\end{document}